\documentclass[prl,aps,twocolumn,amsmath,footinbib]{revtex4-2}
\usepackage{amssymb}
\usepackage{graphicx}

\usepackage{color} 

\definecolor{darkgreen}{rgb}{0,0.7,0}

\usepackage[caption=false]{subfig}
\newcommand{\phantomsubfloat}[1]{
    {
        \captionsetup[subfigure]{labelformat=empty}
        \subfloat[][]{#1}
    }%
}

\begin{document}

\title{Extension of the Generalized Hydrodynamics to the Dimensional Crossover Regime}

\author{Frederik M{\o}ller\textsuperscript{1}, 
Chen Li\textsuperscript{1,2},
Igor Mazets\textsuperscript{1,3},
Hans-Peter Stimming\textsuperscript{3},
Tianwei Zhou\textsuperscript{2,5},\\
Zijie Zhu\textsuperscript{6},
Xuzong Chen\textsuperscript{2},
and J\"{o}rg Schmiedmayer\textsuperscript{1}
}

\affiliation{
\textsuperscript{ 1} Vienna Center for Quantum Science and Technology (VCQ), Atominstitut, TU Wien, Vienna, Austria
\\
\textsuperscript{ 2} School of Electronics Engineering and Computer Science, Peking University, Beijing 100871, China
\\
\textsuperscript{ 3} Research Platform MMM "Mathematics--Magnetism--Materials",
c/o Fakult{\"a}t f{\"ur} Mathematik, Universit{\"a}t Wien, 1090 Vienna, Austria
\\
\textsuperscript{ 5} INO-CNR Istituto Nazionale di Ottica del CNR, Sezione di Sesto Fiorentino, I-50019 Sesto Fiorentino, Italy
\\
\textsuperscript{ 6} Institute for Quantum Electronics, ETH Zurich, 8093 Zurich, Switzerland
}

\date{\today}

\begin{abstract} 
In an effort to address integrability breaking in cold gas experiments, we extend the integrable hydrodynamics of the 1d Lieb-Liniger model with two additional components representing the population of atoms in the first and second transverse excited states, thus enabling a description of quasi-1d condensates. Collisions between different components are accounted for through the inclusion of a Boltzmann-type collision integral in the hydrodynamic equation. Contrary to standard generalized hydrodynamics, our extended model captures thermalization of the condensate at a rate consistent with experimental observations from a quantum Newton's cradle setup.  
\end{abstract} 

\maketitle 

Over the last decades, the advances in experimentally realizing and manipulating quantum many-body systems in low dimensions have increased the demand for theoretical methods capable of describing their complex dynamics~\cite{RevModPhys.77.259, francesco2012conformal,PhysRevA.67.053615,caux2009correlation,PhysRevLett.98.147205,PhysRevA.89.033605,Caux_2016}. Arguably one of the most prominent experimental platforms for studying out-of-equilibrium phenomena is ultracold Bose gases~\cite{kinoshita2006quantum, RevModPhys.80.885, greiner2002collapse, langen2013local, Gring1318,paredes2004tonks, PhysRevLett.115.085301, PhysRevLett.111.053003, schweigler2017experimental, Fabbri2015Dynamical, Langen207, schemmer2019generalized, Polkovnikov2011, schweigler2020decay, Kinoshita1125,Li2018, Haller2009}, which upon confinement to one dimension exhibit integrability. Integrable systems abide by an extended set of conservation laws, strongly constraining their dynamics and inhibiting thermalization~\cite{rigol2007relaxation,rigol2009breakdown,Rigol2008,Gogolin2016, PhysRevLett.110.257203}. Within the integrable limit, the recent theory of Generalized Hydrodynamics (GHD) has established itself as a powerful and flexible framework by capturing both the transport of all the conserved charges and the Wigner delay time in elastic scattering of particles~\cite{wigner1955scattering,boldrighini1983one,mazets2011integrability} within a single continuity equation~\cite{castro2016emergent,bertini2016transport,bastianello2019inhomogeneous}. For the 1d Bose gas, GHD has the added benefit of being valid across the entire phase diagram of the Lieb-Liniger model. Building upon the framework of GHD, a wide array of extensions have enabled the study of correlations~\cite{doyon2018exact,bastianello2018exact,bastianello2018sinh,doyon2020fluctuations,mller2020eulerscale}, Drude weights~\cite{PhysRevLett.119.020602,PhysRevB.97.045407,SciPostPhys.3.6.039,PhysRevB.96.081118}, diffusion constants~\cite{PhysRevLett.121.160603,PhysRevB.98.220303,10.21468/SciPostPhys.6.4.049,PhysRevLett.122.127202}, and more. 

However, real systems realized in even very controlled environments are only approximately integrable, as various mechanisms can break the integrability of the system, thus changing its dynamics and over time driving it towards thermalization. Among such mechanisms are small experimental imperfections like atom losses~\cite{bouchoule2020effect} or noise~\cite{bastianello2020generalised}, diffusive effects in the presence of an external potential~\cite{bastianello2020thermalisation}, and processes outside the realm of GHD specifically related to the physical realization of the system~\cite{Mazets2008,Gerbier2010,Pichler2010,Mazets2011,Riou2012,Riou2014,tang2018cradle,Zundel2019,caux2019cradle}. 
The breaking of integrability is perhaps best demonstrated in the seminal quantum Newton's cradle experiment~\cite{kinoshita2006quantum}. In a fully integrable system, the oscillating motion of the cradle would persist indefinitely, whereas the presence of any of the aforementioned mechanisms would eventually lead to equilibration. Further, the rate of thermalization depends on the severity of the integrability breaking~\cite{tang2018cradle,Li2018}. Thus, the quantum Newton's cradle is the ideal setup for studying weakly broken integrability.
Unfortunately, due to the many possible mechanisms, formulating a generally applicable theory for thermalization appears intractable, whereby each mechanism must be considered separately~\cite{PhysRevX.9.021027,PhysRevB.101.180302,bastianello2020generalised,durnin2020non,lopez2020hydrodynamics,bastianello2020thermalisation,Bland_2018}.

\begin{figure}
\center
\includegraphics[width = 0.8\columnwidth]{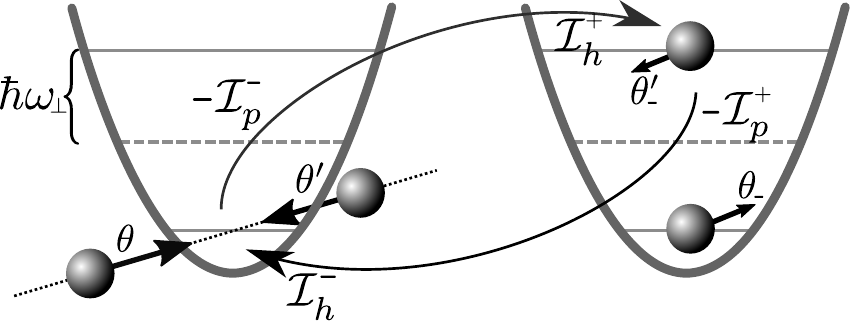}
\caption{\label{fig:drawing} Mechanism for thermalization in quasi-1d Bose gas. Two atoms in the transverse ground state collide with large opposite momenta, exciting one of the atoms to the second excited state. The excited atom can decay to the ground state through collisions with ground state atoms.}
\end{figure}

In this Letter, we seek to extend the applicability of GHD to the dimensional crossover regime, which is accessed when the collisional energy of atoms exceeds the level spacing of the transverse confinement. Thus, based on heuristic considerations, we introduce two additional components to the Lieb-Liniger model, representing atoms in the first and second transverse excited states. The coupling between components is accounted for by introducing a Boltzmann-type collision integral~\cite{pitaevskii2012physical} to the GHD equation.
We then study the role of the transverse states during the evolution of a Bose gas in a quantum Newton's cradle-type setup by comparing our extended model to standard GHD. We further demonstrate its applicability by comparing to experimental observations from Ref. \cite{Li2018}.

The degenerate gas of  $N$  bosonic atoms of mass $m$ is described by the second-quantized Hamiltonian
\begin{equation} 
\begin{aligned}
\hat H=\, & \int d\mathbf{r}\, \Big{ \{ } \frac{\hbar ^2}{2m} (\nabla \hat \Psi ^\dag )(\nabla \hat \Psi )+
[U(z)+V_\perp (x,y)]\hat \Psi ^\dag \hat \Psi + \\
& \frac{2\pi \hbar ^2 a_s}m \hat \Psi ^\dag \hat \Psi ^\dag \hat \Psi \hat \Psi \Big{ \} }, \label{Hmt} 
\end{aligned} 
\end{equation}
where $\hat \Psi =\hat \Psi (\mathbf{r})$ is the atom annihilation operator, $a_s$ is the \textit{s}-wave scattering length, 
$U(z)$ is the loose trapping potential in the longitudinal direction, $V_\perp (x,y)$ is the tight transverse trapping potential. 
We assume that $V_\perp (x,y)$ is harmonic and axially symmetric, $\omega _\perp $ being its fundamental frequency and 
$l_\perp =\sqrt{\hbar /(m\omega _\perp )}$ being the corresponding length scale.  

We treat the motion of atoms in the longitudinal direction within the GHD framework, while the transverse motion is 
accounted for via a collision integral. 
The GHD provides a coarse grained theory for the dynamics of systems close to an integrability point~\cite{castro2016emergent,bertini2016transport}. Just like the thermodynamics Bethe ansatz, the theory encodes the thermodynamic properties of a local equilibrium macrostate in a distribution of quasiparticles~\cite{PhysRevLett.19.1312,doi:10.1063/1.1664947}. Each quasiparticle is uniquely labeled by its rapidity, $\theta$, expressed in inverse length units~\cite{lieb1963exact,lieb2004exact}. In the thermodynamics limit, the rapidity becomes a continuous variable, with the density of occupied rapidities in the phase $(z,\theta )$-space given by the time-dependent quasiparticle density, $\rho_p (z,\theta ,t)$. Similarly, one can introduce a density of holes, $\rho_h (z,\theta ,t)$, describing the density of unoccupied rapidities~\cite{takahashi2005thermodynamics}. Together these two densities describe the density of states and obey the relation 
$\rho_p (\theta )+\rho_h (\theta )=(2\pi )^{-1}+ 
\pi ^{-1}\int _{-\infty }^\infty d\theta ^\prime \, \{ c/[ c^2 +(\theta ^\prime -\theta )^2]\} \rho_p (\theta ^\prime )$, where $c=2a_s/l_\perp ^2$ is the interaction parameter of the Lieb-Liniger model. Here we omit the coordinate and time arguments when appearing the same in all terms.
A quasiparticle with rapidity $\theta $ propagates at velocity $v^{\mathrm{eff}}$, which obeys the integral equation $v^{\mathrm{eff}}(\theta )=\hbar \theta /m + \int _{-\infty }^\infty d\theta ^\prime \, \{ 2c/[c^2 +(\theta ^\prime -\theta )^2]\}\rho_p (\theta ^\prime )[v^{\mathrm{eff}}(\theta ^\prime )-v^{\mathrm{eff}}(\theta )]$ and encodes the Wigner delay time associated with the phase shifts occurring under elastic collisions in integrable systems~\cite{PhysRevB.97.045407,PhysRevLett.120.045301}. In an external potential, a force $F^{\mathrm{eff}}=-\partial _z U(z)$ acts on the quasiparticles~\cite{Doyon2017note}. 

By considering parity conservation and using multicomponent extensions~\cite{klauser2011, Sutherland1968} of Yang's theory~\cite{PhysRevLett.19.1312}, we develop a simple model that accounts for collisional population of the first and second transverse excited states, denoted by indices $n=1$ and $n=2$, respectively. In the multicomponent case, $\rho _p (z, \theta , t)$ comprises all components and excitations are accounted for using a pseudospin
degree of freedom~\footnote{See Supplemental Material for further details.}.
If two atoms with rapidities $\theta $ and $\theta ^\prime $ collide and the collision energy exceeds $2\hbar \omega _\perp $, their transverse states can change. Neglecting degeneracy, two collision outcomes are equally probable: 
(i) one atom remains in the transverse ground state and the other one occupies the second excited state; 
(ii) both of the atoms are transferred to the first excited state. Similar selection rules exist for deexciting collisions.
In the presence of these processes, our extended model yields 
\begin{equation}
\partial_t \rho_p + \partial_z (v^{\mathrm{eff}} \rho_p) +\hbar ^{-1} \partial_\theta (F^{\mathrm{eff}} \rho_p) = \mathcal{I}(\theta)\; .       \label{main}
\end{equation}
Eq. (\ref{main}) differs from the conventional GHD equation by the Boltzmann-type collision integral $\mathcal{I}(\theta)$~\cite{pitaevskii2012physical}, which encodes the state-changing collisions and reads~\cite{Note1}
\begin{equation} 
\mathcal{I}(\theta)= \sum_{n = 1}^{2} \zeta_n \left[  \mathcal{I}_h^{-}(\theta)\nu_{n}^{\beta_n} -\mathcal{I}_p^{-}(\theta) -\mathcal{I}_p^{+}(\theta)\nu_{n}^{\beta_n} +\mathcal{I}_h^{+}(\theta) \right]
\label{coll.I.1}
\end{equation}
where $\nu_n$ is the probability for an atom to be in the $n$'th transverse excited state, $\zeta_n$ is the relative transition strength accounting for neglected degeneracy of the excited states, and $\beta_{1} = 2$ and $\beta_{2} = 1$ are the number of atoms changing state via the collision.
We have assumed that $\nu_{n} \ll 1$ and is uniform, and set $\zeta_{1} = \zeta_2 = 0.5$.

If two atoms in different transverse states collide, a rapidity exchange is a relatively highly probable outcome. Therefore, the transverse excitations rapidly spread over the entire phase space~\cite{Note1}, whereby we neglect correlations between transverse excitations and rapidities and let $\nu_{n} (t)$ be uniform and obey the simple equation 
\begin{equation} 
\frac {d\nu_n }{dt}= \zeta_n\beta_n \left[ \Gamma _h^+ - \Gamma _p^+ \nu_{n}^{\beta_n} \right] + \gamma_n, 
\label{nu12} 
\end{equation}
where $\Gamma _\alpha ^+=(2N)^{-1}\int _{-\infty }^\infty  dz \int _{-\infty}^\infty d\theta \, 
\mathcal{I}_\alpha ^+(\theta )$, $\alpha =p,\, h$, and $\gamma_n$ accounts for any heating rate caused by experimental imperfections. We assume $\gamma_2 = \gamma_1 \nu_1$~\footnote{Unlike for collisions, transitions caused by external heating do not abide to parity conservation. Thus, the heating will typically cause an atom to jump one transverse level~\cite{Li2018}. It is therefore reasonable to assume that the rate of atoms transferred via heating to the second excited state is proportional to the population of the first one, hence $\gamma_2 = \gamma_1 \nu_1$.}.
The terms in Eqs. (\ref{coll.I.1}) and (\ref{nu12}) are defined as 
\begin{equation} 
\begin{aligned} 
\mathcal{I}_\alpha ^\pm (\theta)=\, &\frac{(2\pi )^2\hbar}{m}\int _{\mathcal{R}_\pm }d\theta ^\prime \, 
|\theta -\theta ^\prime |P_\updownarrow (|\theta -\theta ^\prime |,\, |\theta _\pm -\theta ^\prime _\pm |)\times \\
&\rho _\alpha (\theta )\rho_\alpha (\theta ^\prime )\rho _{\bar \alpha}(\theta _\pm )\rho _{\bar \alpha }(\theta ^\prime _\pm ), \label{coll.I.2}   
\end{aligned}    
\end{equation} 
where $\bar \alpha  = h$ for $\alpha =p$ and vice versa, 
$P_\updownarrow (\theta _1,\, \theta _2)=4 c^2\theta _1\theta_2 /[\theta _1^2\theta_2^2 +c^2(\theta _1+\theta_2 )^2]$ is the scattering probability, while
$\theta _\pm = \frac 12 (\theta +\theta ^\prime )+\frac 12 (\theta -\theta ^\prime )
\sqrt{1\pm 8/[(\theta -\theta ^\prime )l_\perp ]^2}$ 
and $\theta _\pm ^\prime = \frac 12 (\theta +\theta ^\prime )-\frac 12 (\theta -\theta ^\prime )
\sqrt{1\pm 8/[(\theta -\theta ^\prime )l_\perp ]^2}$ are the rapidities after a collision leading to excitation ('-') or deexcitation ('+') of the transverse states (see figure \ref{fig:drawing}). 
The integration ranges in Eq. (\ref{coll.I.2}) are the following: 
$\mathcal{R}_+$ is the whole real axis, and $\mathcal{R}_-$ is comprised of those real values of $\theta ^\prime $, which yield real $\theta _- $ and $\theta _-^\prime $, i.e. 
$\mathcal{R}_- =\{ \theta ^\prime : \theta ^\prime < \theta -2\sqrt 2/l_\perp \} \cup \{ \theta ^\prime : \theta ^\prime > \theta +2\sqrt 2/l_\perp \}$~\footnote{See Supplemental Material for a detailed construction of the collision integral, which includes Refs.~\cite{Olshanii1998}}.
Further mechanisms of thermalization present, such as virtual quantum excitations~\cite{PhysRevA.65.043614, PhysRevA.77.013617, adhikari2009gap, Mazets_2010}, are neglected, as these are too slow for the parameters in this work~\cite{Note1}. 

\begin{figure*}
\center
\includegraphics[width = 0.9\textwidth]{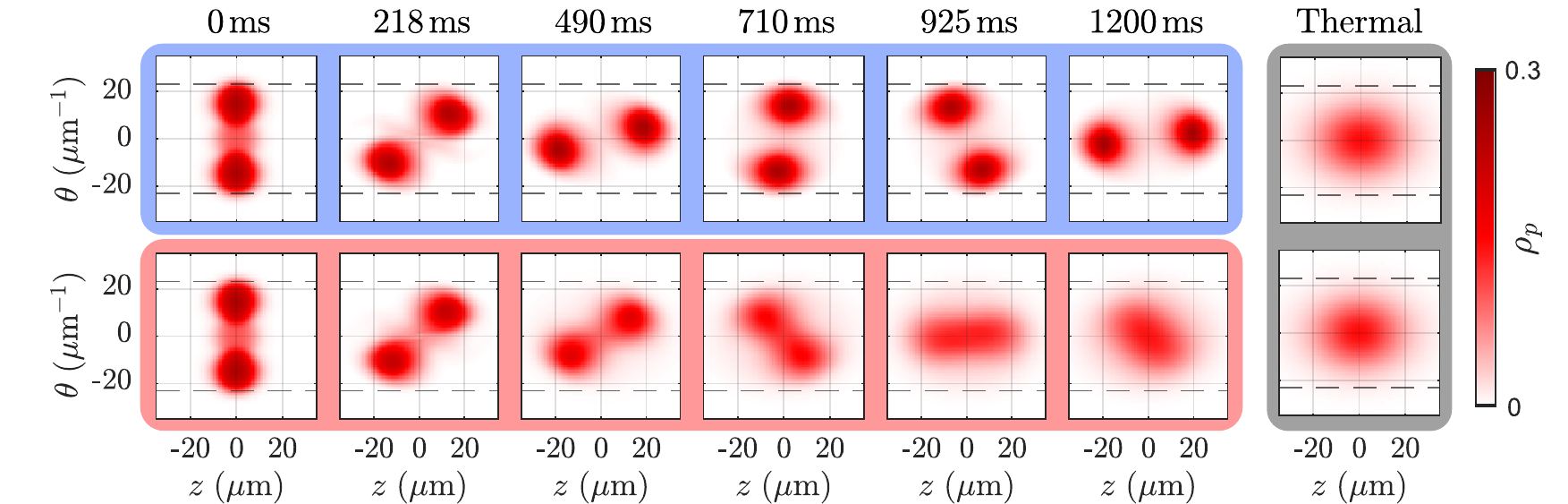}
\caption{\label{fig:theta} Evolution of a single tube containing 130 atoms at 94nK during the first 100 oscillation periods. The top row displays the quasiparticle density evolved using the standard GHD equation, while the bottom row is computed via the extended model. The dashed lines mark the excitation threshold at $\pm \sqrt{2}/l_\perp$, while the final panels show the best fitted thermal state at 660nK.}
\end{figure*}

Owing to their parity, the two transverse excited states play very different roles in the thermalization of the gas. While their excitation rate is practically identical, the deexcitation from the first excited state is very slow, as it requires a collision of two excited atoms, seen by the quadratic dependence on $\nu_1$ in Eq. \eqref{coll.I.1}. Thus, atoms accumulate in the first excited state, effectively reducing the number of atoms available for further thermalization driven by repeated transitions between the ground and second excited state.

In the following, we demonstrate how the presence of transverse excited states influence the dynamics of the otherwise integrable Lieb-Liniger model. As stated earlier, the dynamics of the quantum Newton's cradle~\cite{kinoshita2006quantum} are particularly sensitive to integrability breaking mechanisms. Therefore, we study a particular experimental realization of the setup presented in Ref. \cite{Li2018}, where transverse state-changing collisions were shown to drive the observed thermalization. 

To briefly summarize the experiment~\cite{Note1, Li2018}, it studied the dynamics of $^{87}$Rb Bose-Einstein condensates in a 2d lattice of independent 1d tubes with a tight transverse confinement of $\omega_\perp/2\pi=31\,\mathrm{kHz}$ and weak longitudinal confinement of $\omega _\Vert/2\pi=83.3\,\mathrm{Hz}$ (oscillation period $\mathcal{P} = 12\,\mathrm{ms}$). Owing to the Gaussian profile of the trapping beams, the longitudinal potential was slightly anharmonic, $U(z) = m \omega _\Vert^2 \sigma^2 \left( 1 - e^{-2 z^2 / \sigma^2} \right)/ 4$, where $\sigma = 145\,\mathrm{\mu m}$ being the beam-waist. The dynamics in the longitudinal direction were initiated by two Bragg pulses, imparting opposite momenta of $\pm 2 \hbar k_\mathrm{Bragg}$ to the atomic cloud, with $k_\mathrm{Bragg} = 2 \pi / 852\,\mathrm{nm}$. Following Refs. \cite{caux2019cradle, berg2016separation}, we assume the pre-pulse quasiparticle density to be a thermal state $\rho_{p}^{\mathrm{th.}}(\theta)$, while the pulse sequence simply shifts the distribution along the rapidity axis, yielding $\rho_{p}^{\mathrm{init}}(\theta) = \frac{1}{2}(1- \eta)\rho_{p}^{\mathrm{th.}}(\theta + 2 k_{\mathrm{Bragg}}) + \frac{1}{2}(1- \eta)\rho_{p}^{\mathrm{th.}}(\theta - 2 k_{\mathrm{Bragg}}) + \eta \rho_{p}^{\mathrm{th.}}(\theta)$. The parameter $\eta$ is the fraction of atoms unaffected by the Bragg pulses.

First, we consider a single tube containing $N=130$ atoms at a temperature of 94nK. Based on independent measurement, we set $\eta = 0.17$ and $\gamma_1 = 0.035 \: \mathrm{s}^{-1}$ corresponding to a heating rate of 55nK/s~\footnote{See Supplemental Material for a detailed description of the measurement of each parameter, which includes Refs.~\cite{Cazalilla2004, PhysRevLett.87.050404, PhysRevLett.91.010405, PhysRevA.67.051602, PhysRevA.83.031604, PhysRevLett.121.220402}}. This leaves an initial fraction 0.02 of atoms at rapidities above the excitation threshold $\sqrt{2}/l_\perp$, namely the minimum required rapidity of at least one quasiparticle for a state-changing collision. Employing both standard and our extended GHD, we simulate the dynamics for the first 100 oscillation periods of the cradle~\cite{iFluid}.
In figure~\ref{fig:theta} we plot the resulting quasiparticle distributions for various times throughout the evolution. Comparing the two theories side by side clearly demonstrates the influence of the additional transverse components; while the Bragg peaks of the initial state persist throughout the evolution when propagated using the standard GHD equation, the inclusion of the collision integral enables quasiparticles to distribute across the phase space. Hence, the additional components initially accelerate the dephasing of the gas and eventually cause it to thermalize. 
After 100 oscillation periods (1.2s), the quasiparticle density of the extended model resembles that of a thermal state, although the dynamics have not yet completely subsided. For comparison, we plot a thermal state with the same number of atoms and total energy as the initial, post-Bragg pulse state, yielding a final temperature of 660nK. The large difference in temperature between the initial and final thermal state is due to the large amount of kinetic energy pumped into the system during the Bragg pulse sequence.
Notably, in this setup no dephasing is observed when evolving the system according to the standard GHD equation. Further, the initial population of atoms at low rapidities is rapidly depleted, as it is transferred to the Bragg peaks. This is contrary to Ref. \cite{caux2019cradle}, where an anharmonic trapping potential was sufficient to induce dephasing, although the dephased state was distinctly different from thermal. However, in our setup, the interactions between atoms manifested in the effective velocity protect the gas against dephasing~\footnote{A similar phenomenon can also be seen in Ref. \cite{caux2019cradle} at long times, albeit not as clearly.}.

\begin{figure}
    \centering
    \includegraphics[width=0.49\textwidth]{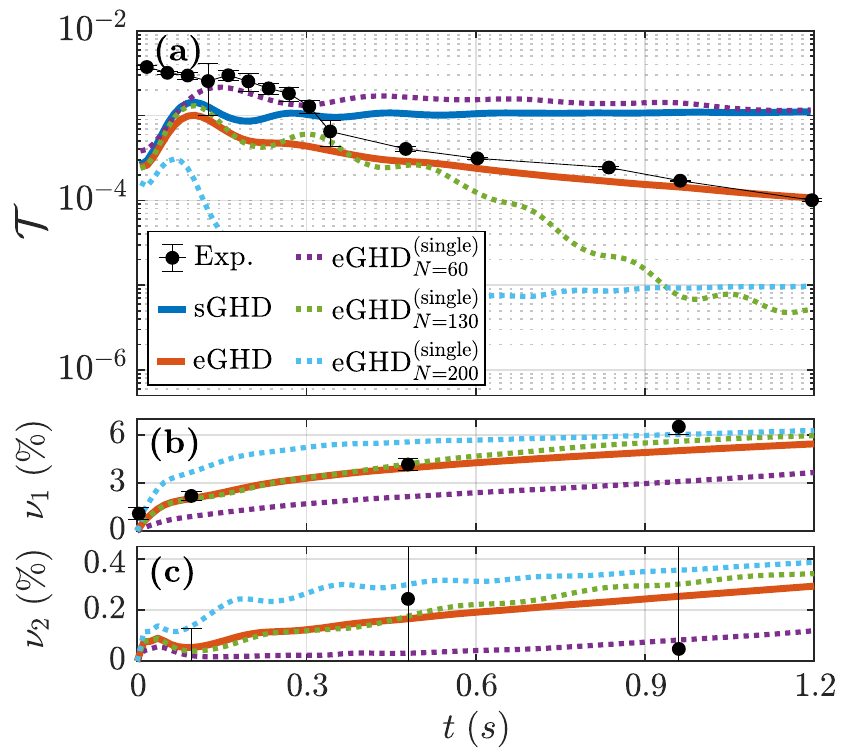}%
    \phantomsubfloat{\label{fig:thermalization_a}}
    \phantomsubfloat{\label{fig:thermalization_b}}
    \phantomsubfloat{\label{fig:thermalization_c}}

    \vspace{-2\baselineskip}

    \caption{Thermalization in the quantum Newton's cradle with $N_{tot} = 1 \times 10^5$ atoms at 94nK. \textbf{(a)} Comparison of measure $\mathcal{T}$ between standard GHD, extended GHD, and experimental observations for a weighted sum over the full lattice. Additionally, 3 single tubes of the lattice computed with extended GHD are plotted. \textbf{(b, c)} Percentage of atoms in first and second transverse excited states. }%
    \label{fig:thermalization}%
\end{figure}

Next, we wish to more quantitatively compare the two theories and demonstrate how the thermalization rate is dependent on the degree of integrability breaking. Further, we also compare our findings to the experimental observations of Ref. \cite{Li2018} to see whether our relatively simple model applies to realistic scenarios. 
Importantly, the experimental system consists of $N_{tot} = 1 \times 10^5$ atoms distributed over many 1d tubes, with each individual tube containing up to 200 atoms~\cite{Note1}. To emulate the lattice in our GHD simulation, we bin the tubes according to their atom numbers and solve the dynamics for a representative system for each bin. The overall results are obtained by summing up the contributions from all the bins, weighing each by its underlying number of tubes. Although several quantities besides the atom number vary slightly across the lattice, we employ the same parameters, namely those used for the previous simulation featured in figure \ref{fig:theta}, for all tubes in order to keep matters simple and transparent.

For the quantitative comparison, we consider a metric of distance to thermalization $\mathcal{T}(t) = \int \mathrm{d}\theta [ F(t,\theta) - \hat{F}(t,\theta) ]^ 2$, where $F(t,\theta) = \int_{t - \mathcal{P}/2}^{t + \mathcal{P}/2} \mathrm{d}t' \int \mathrm{d}z \: \rho_p(\theta, z, t') / \mathcal{N}$ is the normalized period-mean rapidity distribution (RDF) of the quasiparticles, and $\hat{F}(t,\theta)$ is its best fit to a Gaussian~\cite{tang2018cradle,Li2018}. We consider a period averaged quantity to eliminate contributions from any potential dephasing between individual tubes owing to variance in oscillator frequency across the lattice~\footnote{Averaging the profiles over one periods also reduces the difference between the RDFs and MDFs.}. Importantly, the experiment measures the momentum distribution function (MDF) rather than the RDF~\cite{Note1, Li2018, caux2019cradle, Wilson2020}. In the degenerate regime the two distributions differ, however, as the density of the gas drops due to the intra-tube dephasing~\cite{tang2018cradle} induced by the state-changing collisions (as seen in figure \ref{fig:theta}), the gas becomes increasingly nondegenerate and the two distributions start to coincide~\cite{Note1, giamarchi2004quantum}.

Figure \ref{fig:thermalization_a} shows the values of $\mathcal{T}(t)$ obtained when considering the RDFs obtained via the weighted average over all tubes. The figure contains results from both standard and our extended GHD, and compares them to experimental measurements.
As already illustrated in our previous demonstration, the integrability of the standard GHD prohibits thermalization, whereby its associated $\mathcal{T}(t)$ remains constant even at long timescales. Meanwhile, the inclusion of the additional components in the extended GHD enables it to thermalize causing its $\mathcal{T}(t)$ to decrease~\footnote{Note, the initial increase of $\mathcal{T}(t)$ observed in figure \ref{fig:thermalization_a} for the GHD simulations stems from the initial depletion of quasiparticles at low rapidity (which can also be seen in figure \ref{fig:theta}). This is caused by the interactions in standard GHD and is not a product of our extended model.}.
Initially, we observe a large discrepancy between simulations and experiment owing to the difference between the MDF and RDF. However, after roughly 30 oscillation periods (0.36s) the gas is practically nondegenerate, whereby our extended model exhibits a thermalization rate comparable with the experiment.

Further understanding of the thermalization process can be gained from figures \ref{fig:thermalization_b} and \ref{fig:thermalization_c}, which depict the excitation probabilities $\nu_1$ and $\nu_2$, respectively. Despite their equal collision probability, we observe a much larger fraction of atoms occupying the $n=1$ state due to its very low deexcitation rate.
Additionally, these measures are unaffected by discrepancy between the MDF and RDF, whereby they provide a valid comparison to the experiment even at short timescales~\footnote{Note, the experimental band-mapping technique used to extract $\nu_1$ and $\nu_2$ requires a sufficient number of atoms in the transverse excited states in order to overcome the measurement noise. Since the second transverse state is only sparsely populated, the resulting errorbars on $\nu_2$ are quite large. See Ref. \cite{Li2018} for more details.}. Here, we find a decent agreement, although at long timescales we do observe a small discrepancy between the measured and simulated population of the first excited state. We attribute this to an underestimation of the heating $\gamma_1$.

To understand the thermalization in the lattice, it is instructive to examine the contribution from different subsystems. Therefore, figure \ref{fig:thermalization} also includes plots for three single tubes with $N=$ 60, 130, and 200 atoms. 
For higher atom number and temperature, a larger fraction of the atoms occupy rapidities large enough to cause excitation upon collision, i.e. the system is deeper in the dimensional crossover~\cite{Note1}. Thus, we observe the most rapid thermalization in the $N=200$ case, whose transverse populations quickly reach a dynamic equilibrium from the many collision events, whereafter they increase slowly due to the continuous heating. Meanwhile, in the $N=60$ subsystem, very few atoms can partake in state-changing collisions. Thus, the thermalization is almost entirely driven by external heating, leading to a very slow and steady increase of the transverse populations in addition to almost no change in the corresponding $\mathcal{T}(t)$ value.

Lastly, we emphasize that both the underlying theory of GHD and the extension to the crossover regime are valid across the entire Lieb-Liniger phase diagram~\footnote{See Supplemental Material for a detailed discussion of the various methods applicable in different regimes of the Lieb-Liniger phase diagram, which includes Refs.~\cite{PhysRevA.67.053615,Bland_2018,Tonks1936,Girardeau1960,Collura2014,Krauth2006,PhysRevLett.120.045301,Li2018,schemmer2019generalized,pitaevskii2012physical}}. Thus, our method should be applicable to a wide range of setups~\cite{Amerongen2008yang,davis2012thermometry,jacqmin2011fluctuations,jacqmin2012momentum,paredes2004tonks, PhysRevLett.115.085301,Wilson2020,PhysRevA.83.021605}. For instance, the first experimental demonstration of GHD~\cite{schemmer2019generalized} mimicked a quantum Newton's cradle in an atom chip trap with a Bose gas mainly in its quasicondensate phase. However, the high temperature and chemical potential combined with the lower transverse confinement of the chip trap setup, places it even deeper within the dimensional crossover than the lattice setup. Further, unlike the typical quantum Newton's cradle where the system is brought into the dimensional crossover via a Bragg pulse sequence, the system of Ref.~\cite{schemmer2019generalized} is initialized directly in the crossover regime. In the Supplemental Material we simulate the dynamics on the chip trap using both standard and extended GHD. Indeed, after just a single period the predictions of standard GHD start deviating from experimental observations, just as in Ref.~\cite{schemmer2019generalized}. Meanwhile, extended GHD reproduces the experimental observations to a high degree, thus once again highlighting the importance of considering transverse excitations when studying realistic setups.

In conclusion, we have extended the theory of Generalized Hydrodynamics with a multicomponent Lieb-Liniger model in order to address the question of thermalization in the dimensional crossover. Our model takes into account collisions with transverse excited atoms through a Boltzmann-type collision integral and can be readily applied to most realizations of quasi-1d condensates. Through comparisons between standard GHD and our extended model, we have demonstrated the large influence on dynamics a small fraction of transverse excited atoms can have. Furthermore, comparing predictions of our model to experimental data from a quantum Newton's cradle setup yields good agreement of the thermalization rate, despite the simplicity of our model. Thus, our results demonstrate that accounting for transverse excited states is necessary when applying GHD to quasi-1d Bose gases.

This work was supported by the Austrian Science Fund (FWF) via the 
SFB 1225 ISOQUANT (I~3010-N27). We further acknowledge financial support by the ESQ (Erwin Schr\"{o}dinger Center for Quantum Science and Technology) Discovery programme, hosted by the Austrian Academy of Sciences (\"{O}AW). I.M. and H.-P. S. acknowledge the support by the Wiener Wissenschafts- und Technologiefonds (WWTF)
via Grant No. MA16-066 (SEQUEX)
and by the Austrian Science Fund (FWF) via Grant SFB F65 (Complexity in PDE systems). 
X.C. acknowledges the support from the National Natural Science Foundation of China (Grant No. 11920101004, 91736208).
We thank Vincenzo Alba, Alvise Bastianello, Jean-S{\'{e}}bastien Caux, Marcos Rigol, David Weiss, Wei Xiong, Hepeng Yao and Xiaoji Zhou for enlightening discussions. 

\clearpage

\appendix
\begin{widetext}
\section{SUPPLEMENTAL MATERIAL}

\section{Standard GHD equations}
We report here the equations for the standard GHD of the Lieb-Liniger model. Note, we omit all spacial and temporal arguments, as they will remain the same on either side of the equations.

The standard GHD propagation equation reads
\begin{equation}
    \partial_{t} \rho_{p}+\partial_{z}\left(v^{\mathrm{eff}} \rho_{p}\right)+\hbar^{-1} \partial_{\theta}\left(F^{\mathrm{eff}} \rho_{p}\right)= 0 \; ,
    \label{eq:GHDstandard}
\end{equation}
where the effective force on the quasiparticles $F^{\mathrm{eff}}$ describes changes in the rapidity distribution in the presence of inhomogeneous interactions, while the effective velocity is given by
\begin{equation}
   v^{\mathrm{eff}}(\theta)= \frac{\hbar \theta}{m} + \int_{-\infty}^{\infty} d \theta^{\prime} \Phi(\theta , \theta')  \rho_p (\theta') \left[v^{\mathrm{eff}}\left(\theta^{\prime}\right)-v^{\mathrm{eff}}(\theta)\right] \, . 
\end{equation}
Here, $\Phi(\theta , \theta') = \frac{2 c}{c^2 + (\theta - \theta')^2}$ is the Lieb-Liniger two-body scattering kernel. From the quasiparticle density, one can extract the expectation values of the  conserved charges and their associated current, respectively, via
\begin{align}
    \mathrm{q}_{i} &= \int \mathrm{d} \theta \: h_{i}(\lambda) \rho(\lambda) \\
    \mathrm{j}_{i} &= \int \mathrm{d} \theta \: h_{i}(\lambda) v^{\mathrm{eff}}(\lambda) \rho(\lambda) \; ,
\end{align}
with $h_{i}(\lambda)$ being the one-particle eigenvalue of the $i$'th conserved charge.

As an alternative to the quasiparticle density, one can encode the thermodynamic properties of the system in the filling function
\begin{equation}
    \vartheta(\theta) = \frac{\rho_p (\theta)}{\rho_p (\theta) + \rho_h (\theta)} \; ,
\end{equation}
where the density of states is given by
\begin{equation}
    \rho_p (\theta) + \rho_h (\theta) = \frac{1}{2 \pi} + \frac{1}{2 \pi} \int_{-\infty}^{\infty} \mathrm{d}\theta' \: \Phi(\theta , \theta') \rho_p (\theta ') \; .
\end{equation}
The quasiparticles of the Lieb-Liniger model follow Fermionic statistics. Thus, a thermal state can be calculated from
\begin{equation}
    \vartheta (\theta) = \frac{1}{1 + e^{ \epsilon(\theta) \beta }} \; ,
\end{equation}
where $\beta$ is the inverse temperature and the pseudoenergy $\epsilon(\theta)$ is acquired from solving the equation
\begin{equation}
    \epsilon(\theta) = \frac{\hbar^2 \theta^2}{2 m} - \mu - \frac{1}{2 \pi \beta} \int_{\infty}^{\infty} \mathrm{d}\theta' \: \Phi(\theta , \theta') \ln \left( 1 + e^{ \epsilon(\theta') \beta } \right) \; .
\end{equation}
The chemical potential $\mu (z) = \mu_0 - U(z)$ accounts for the external potential.

\section{Numerically solving the propagation equation}
For the numerical GHD computations we employ the iFluid library~\cite{iFluid}. In order to solve the hydrodynamic equation with collision integral we employ a first order split step propagation scheme.

First, we evaluate the collision integral, which requires quantities readily available from GHD.  Throughout the entire calculation we maintain the same rapidity and collision grids. Consider the rapidity discretized on a grid $\theta_i$ with $i = 1, \ldots , i_{max}$. The collision grids then read
\begin{equation}
    \theta_{\pm}[i ; j]=\frac{1}{2}\left(\theta_{i}+\theta_{j}\right)+\operatorname{sgn}\left(\theta_{i}-\theta_{j}\right) \sqrt{ 1 \pm 8 / \left[ \left(\theta_{i}-\theta_{j}\right) l_{\perp} \right]^2}
\end{equation}
where $\theta_{\pm} ^\prime  [i ; j] = \theta_{\pm}[j ; i]$.
To obtain the particle and hole densities on the collision grids we use interpolation, which can be expressed in matrix form as
\begin{equation}
    \rho_{p, h}\left(\theta_{\pm}[i ; j]\right)=\sum_{k} \Xi_{\pm}([i ; j], k) \rho_{p, h}\left(\theta_{k}\right) \; .
\end{equation}
Throughout the simulation we maintain constant rapidity and collision grids. Thus, the interpolation matrix $\Xi_{\pm}$ can be calculated beforehand, greatly reducing the computational time needed.
For linear interpolation, the interpolation matrix can be constructed as follows
\begin{align}
    \Xi_{\pm}([i ; j], k-1) &= \frac{\theta_k - \theta_{\pm}[i ; j]}{\theta_k - \theta_{k-1}} \\
    \Xi_{\pm}([i ; j], k) &= 1 - \frac{\theta_k - \theta_{\pm}[i ; j]}{\theta_k - \theta_{k-1}} \; ,
\end{align}
where $k$ is the index minimizing $\min_{k} | \theta_k - \theta_{\pm}[i ; j] |$. This matrix structure is sparse, allowing for very fast interpolation.

Once the collision integral has been obtained, we solve the equations
\begin{align}
    &\frac{\mathrm{d} }{\mathrm{d}t} \rho_p (\theta, z, t) = \mathcal{I}(\theta, z, t) \\
   &\frac{\mathrm{d} }{\mathrm{d}t} \nu_n (t) = \zeta_n\beta_n \left[ \Gamma _h^+ - \Gamma _p^+ \nu_{n}^{\beta_n} \right] + \gamma_n
\end{align}
using the two-step Adams–Bashforth method. Next, we solve the standard GHD equation (\ref{eq:GHDstandard}) without collision integral using the method of characteristics. Here we employ the second order scheme detailed in Ref. \cite{bastianello2019inhomogeneous}. This method has proved itself very stable, exhibiting a loss of less than $3\%$ of atoms (and total energy) throughout the full evolution of 100 periods with 120 propagation steps taken per period. 
The numerical inaccuracies leading to the loss most likely stem from small errors in the interpolation used when propagating the state according to the characteristics. However, we found that employing finer spacial and rapidity grids lead to only a small gain in accuracy while substantially increasing the runtime of the simulation. Increasing the order of the integration scheme would allow fewer, but larger propagation steps to be used, thus reducing the number of interpolations used.
However, we found that simulations using a worse resolution produced results very similar to the settings employed for the results of this work. Thus, we are confident that the small numerical errors have no substantial effect on the results presented.

\section{Comparing GHD to other methods}

\begin{figure*}
\center
\includegraphics[width = 0.7\textwidth]{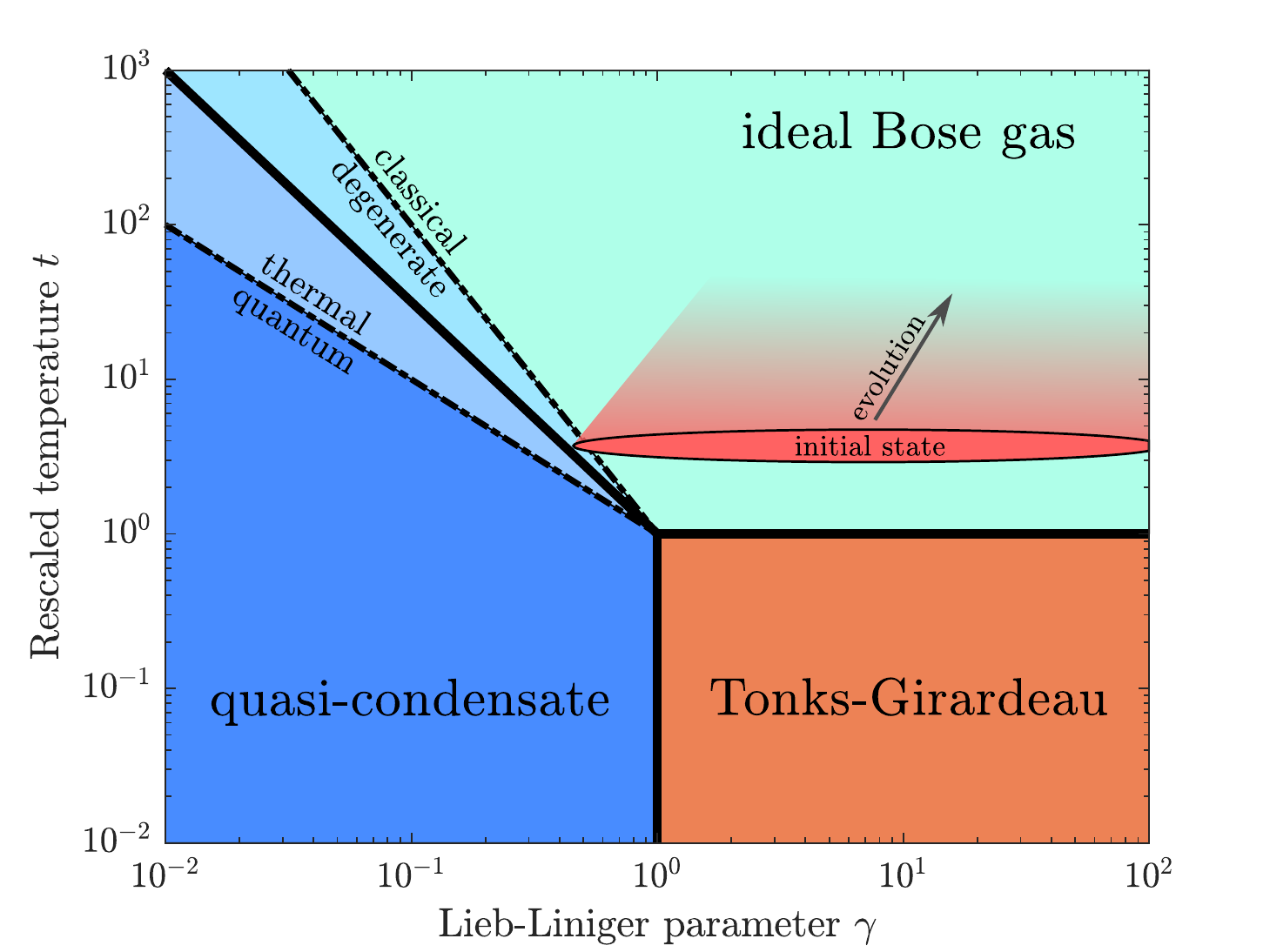}
\caption{\label{fig:PhaseDiagram} Phase diagram of the repulsive Lieb-Liniger model. The solid lines mark the boundaries between the three main regimes, namely the quasicondensate, the ideal Bose gas and the Tonks-Girardeau gas. Note, the transitions between the phases are continuous rather than abrupt. This is further illustrated by the dot-dashed lines indicating the degeneracy transition ($t \gamma^2 =1$) and the thermal-quantum boundary ($t \gamma = 1$). In red, the parameter regime of the quantum Newton's cradle setup ($N = 130$ atoms, $T = $ 94nK) is indicated.}
\end{figure*}

One particular feature of the thermodynamic Bethe ansatz (and by extension GHD), which makes it desirable for describing cold gas experiments, is its applicability throughout the entire Lieb-Liniger phase diagram. As illustrated in figure \ref{fig:PhaseDiagram}, the one-dimensional Bose gas with repulsive interactions can exist in one (or more) of three different phases, namely the quasicondensate, the ideal Bose gas or the Tonks-Girardeau gas. The phase of the system is determined by the interaction strength encoded in the Lieb-Liniger parameter $\gamma = c/n_{1d}$, where $n_{1d}$ is the linear atomic density, and the thermal occupation of excitations parametrized by $t = 2 m k_{\mathrm{B}} T / \hbar^2 c^2$, with $k_{\mathrm{B}}$ being the Boltzmann constant and $T$ the temperature.

Each phase has a different equation of state and exhibits different properties, whereby various methods for treating separate regimes exist. For low temperatures and weak interactions the gas is in the quasicondensate state and can be treated using mean-field theories through the Gross-Pitaevskii equation~\cite{PhysRevA.67.053615,Bland_2018}. If temperatures are sufficiently low and the interaction is increased, the gas enters the Tonks-Girardeau. In this regime the atoms act like impenetrable spheres, and the dynamics can be treated by mapping the particles onto spinless Fermions~\cite{Tonks1936,Girardeau1960,Collura2014}. Lastly, at high temperatures, the interactions are negligible compared to the thermal energy and the system can be described as a nearly ideal Bose gas. In this regime, (semi-)classical molecular-dynamics approaches can be employed to treat the dynamics of the gas~\cite{Krauth2006,PhysRevLett.120.045301,Li2018}.  

In practice, an experimentally realized gas will often occupy multiple regimes of the phase diagram, thus limiting the aforementioned methods to only a fraction of the gas. Further, experimental defects can cause heating of the gas, while dephasing of the gas will cause a drop in density. In quantum Newton's cradle setups, the initial Bragg pulse sequence pumps a lot of energy into the system, whereby the thermalized state will be much hotter than the pre-pulse thermal state. Therefore, the gas will eventually tend towards the ideal Bose gas regime. While this limits the overall applicability of most methods, GHD remains valid throughout the entire evolution, thus greatly simplifying the considerations needed when describing dynamics. For a direct comparison between GHD and other methods for a gas in the transition regions of the phase diagram, see Ref. \cite{schemmer2019generalized}.

In figure \ref{fig:PhaseDiagram} the parameter regime of the Newton's cradle setup (originally from Ref. \cite{Li2018}) described in the manuscript is indicated. Already the initial state is partially in the ideal Bose gas phase, with only the high density regions (the peaks containing most of the atoms) being degenerate. In the main manuscript, we compare the state obtained via the extended model after 100 oscillation periods to a thermal state. For the tube of 130 atoms at 94nK, the estimated final thermal state has a temperature of $T = 660\mathrm{nK}$, which has been used to draw the shaded region in figure \ref{fig:PhaseDiagram}.
Hence, the dynamical evolution brings the gas deep into the ideal Bose gas regime, resulting in the bosonic MDF and the rapidity distribution coinciding. 

Finally, it should be noted that the Boltzmann-type collision integral implies the possibility to factorize higher-order correlations~\cite{pitaevskii2012physical}. This is the case also for strongly interacting system, when we can describe them by introducing weakly interacting quasiparticles~\cite{pitaevskii2012physical}. The quantum correlation properties of the system are then manifested via the Pauli blocking factors in the collision integral. Thus, our extended model should apply in all regimes of the Lieb-Liniger phase diagram.

\section{Additional experimental details and simulation parameters}
This section details how the simulation parameters are obtained. For further details about the experiment and extraction of parameters, we refer to Ref. \cite{Li2018}.

The quantum Newton's cradle is realized experimentally in a red detuned optical lattice consisting of many 1d tubes. The lattice loading procedure is considered by following the method proposed in Ref.~\cite{Haller2009, Fabbri2015Dynamical}. Before the lattice is ramped up, a 3d BEC is prepared in a crossed dipole trap. The 1d atom numbers vary from tube to tube due to the inhomogeneous density distribution of the 3d BEC.
In order to more faithfully capture the experimental setup in the GHD simulations, we calculate the atomic distribution across the entire lattice, bin tubes according to their occupation, and weigh their contribution according to their occurrence.
The atomic distribution is determined by the Thomas-Fermi profile just before the tunneling is suppressed. Hence the number of atoms in the tube located at position ($i$,$j$) in the 2d array is given by 
\begin{equation}
N_{i,j}=N_{0,0}\Bigg[1-\bigg(i\frac{\lambda_L}{2 R_x}\bigg)^2-\bigg(j\frac{\lambda_L}{2 R_y}\bigg)^2\Bigg]^{3/2}\,,
\end{equation} 
where $N_{0,0}=5N_{tot}\lambda_L^2/(8\pi R_xR_y)$ is the atom number in the central tube, $\lambda_L = 1064 \mathrm{nm}$ is the wavelength of the lattice beams, $R_{x/y}$ is the Thomas-Fermi radius in the transverse direction $x/y$.  Within the scope of this manuscript, we study the dynamics with $N_{tot}=1.5\times 10^4$, $5\times 10^4$, and $1\times 10^5$. Assuming zero offset between the cloud and the lattice beam center, the distribution of atoms in the 1d tubes can be seen in figure \ref{fig:LatticeDistribution}, where tubes with similar atom numbers have been binned. In the main manuscript, the dynamics of subsystems with the parameters of each bin are solved. 

\begin{figure*}
    \centering
    \includegraphics[width=0.9\textwidth]{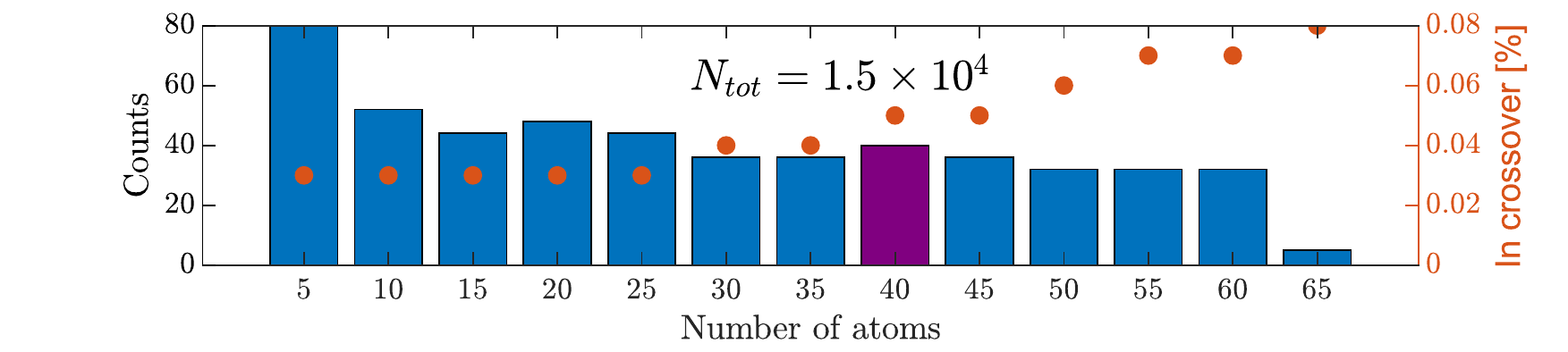}%
    \hfill
    \includegraphics[width=0.9\textwidth]{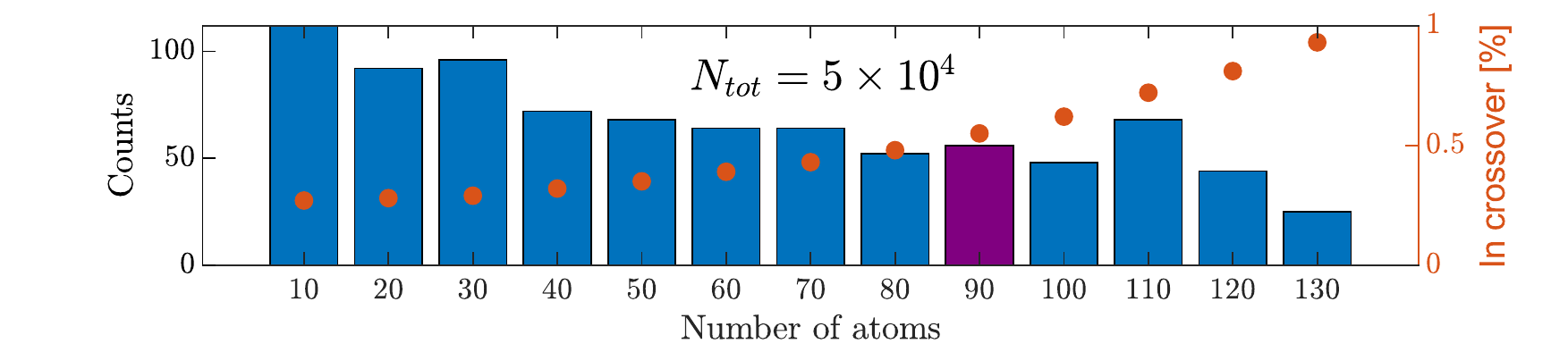}%
    \hfill
    \includegraphics[width=0.9\textwidth]{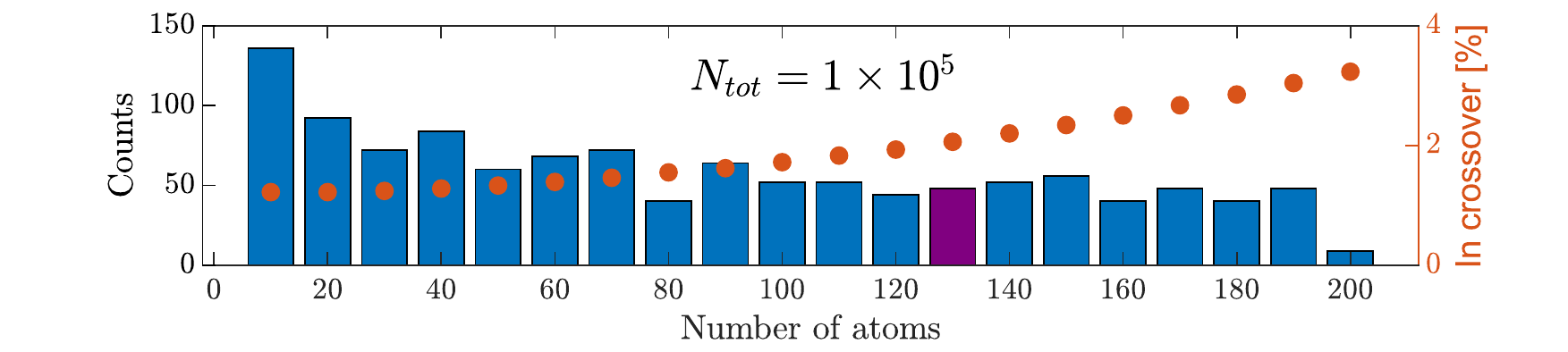}%
    \caption{Distribution of atoms in the optical lattice. The 1d tubes have been binned and their occurrence is plotted. The purple bar for each realization denoted the weighted average number of atoms per tube. The red dots mark the percentage of atoms in the post-pulse state, whose kinetic energy exceeds $\hbar \omega_\perp$, i.e. whose corresponding rapidities exceeds $\sqrt(2)/l_\perp$.}%
    \label{fig:LatticeDistribution}%
\end{figure*}

Due to the Gaussian profile of the lattice beam, the longitudinal potential in each tube is slightly anharmonic 
\begin{equation}
    U(z) = \frac{m \omega _\Vert^2 \sigma^2}{4} \left( 1 - e^{-2 z^2 / \sigma^2} \right) \; .
\end{equation}
Here, $\omega _\Vert/2\pi=83.3\,\mathrm{Hz}$ is the longitudinal trapping frequency, and $\sigma = 145 \mu \mathrm{m}$ is the beam-waist of the lattice beams. Notably, the trapping potentials in the occupied tubes are not identical. The variations of $\omega_\Vert$ and $\omega_\perp$ are subject to the shift of the cloud from the lattice beam center, which occurs as an experimental imperfection. In our experimental setup, we expect that $\omega_\Vert$ and $\omega_\perp$ have a variance of $0.5\%$ throughout the lattice, although any offset between the cloud and lattice center would increase this variance. The inhomogeneity results in a dephasing of oscillation in different tubes with a time scale of about 50 to 100 periods. Since determining the offset between the cloud from the lattice beam center is very difficult, we have chosen to employ the same trapping frequencies for all tubes in the simulation. In return, we only compared period-averaged measures to the experiment, as these are not sensitive to potential dephasing between individual tubes. 

The temperature of the system is obtained by studying the momentum distribution function (MDF) of the pre-pulse thermal state in the lattice. If the rapidity distribution is not far from a Galilean-boosted thermal distribution of a degenerate gas, then the respective momentum peak is Lorentian with the width $\lambda _T =2\hbar ^2 n_{1d}/(m k_B T)$, where $n_{1d}$ is the linear atomic density~\cite{Cazalilla2004}. Thus, the temperature of the gas is estimated from the half-width at half-maximum of the bosonic MDF~\cite{PhysRevLett.87.050404, PhysRevLett.91.010405, PhysRevA.67.051602, PhysRevA.83.031604, PhysRevLett.121.220402}, providing the initial temperatures of $T=34\mathrm{nK}$, $T=60\mathrm{nK}$ and $T=94\mathrm{nK}$ for the three realizations, respectively. Note, we assume the temperature to be the same in all tubes. Knowing the temperature of the gas, we can use the TBA for the Lieb-Liniger model to generate a thermal state for each of the 1d subsystems featured in figure \ref{fig:LatticeDistribution}. For each bin we have plotted the fraction of atoms in the post-pulse state whose kinetic energy succeeds $\hbar \omega_\perp$.

The heating process is studied experimentally by observing the evolution of a cloud held in the optical lattice without the Bragg-pulse excitation. Over time, heating effects from the trapping laser will cause the momentum peak of the cloud to expand. From the MDF we can compute the kinetic energy of the gas, and any increase in kinetic energy is attributed to heating effects. However, since we measure an average over all tubes, it is unclear whether each tube heats evenly. To reduce the number of variables and keep matters simple, we have therefore chosen to employ a heating rate of 55nK/s for all systems, which is close to the measured values (see Ref. \cite{Li2018} for more details).

The fraction of atoms leftover by the Bragg pulse $\eta$ is dependent on the number of atoms, as the efficiency of the pulse sequence decreases for higher atom numbers. When measuring the post-pulse MDF in the lattice, we found the weighted average fraction of atoms of 0.09, 0.15 and 0.2 for the three realizations. Since the exact relation between the atom number and $\eta$ is unknown, we opted for employing the same value for all tubes in order to keep matters simple. Given the values of the measured central fractions have a bias towards the more populated tubes, choosing these values for $\eta$ would yield a total central population too large when taking the weighted average over the entire ensemble. Instead, selecting the true mean value of $\eta$ for the simulation should yield a better result. We find that for the $N_{tot}=1\times 10^5$ realization, the weighted mean population of a tube is 130, while the true mean is about $35\%$ lower. From the measurements, it appears that $\eta$ scales with the atom number to the power of less than one. Thus, the true mean of $\eta$ should be within the interval $[ 0.13, 0.2 ]$. We find that picking the central, rounded value of $\eta = 0.17$ yields a central population fairly close to experimental observations when taking the weighted average over the ensemble.

Finally, the atomic losses in the experiment were around $8.4\%$ within the time scale concerned in the manuscript. In the simulations, however, these losses are not considered. Further, any tunneling between the tubes of the lattice is negligible.

\section{Setting up the extended model}
To construct a numerically tractable extension of the GHD, we need to assume several simplifications and approximations. One possible path towards thermalization is through collisions with transverse excited atoms.
Parity conserving collisions of atoms in the transverse ground state with sufficiently high collision energy can lead to excitation of either one atom into the second transverse excited state or two atoms into the first excited state. We will neglect this distinction and assume that the system contains only three components, namely atoms in the ground state and first and second excited states of the transverse confinement, denoted by the index $n = 0,1,2$, respectively. When treating collisions between transverse states, it is important to consider parity conservation. The inversion symmetry of the transverse trapping potential ensures that the $n = 1$ state is odd (has a negative parity), in contrast to the $n = 0$ and $n = 2$ states, which are even. Therefore, transitions of only one atom to the first transverse excited state are forbidden, or in the case of weak symmetry breaking of $V_\perp$, highly suppressed.
Likewise, deexcitations occur through collisions between either a second excited and a ground state atoms or two first excited ones. Since the population of the $n = 1$ state is small compared to the ground state, the $n = 1$ state is de-populated at a much slower rate than the $n = 2$.
In addition to state changing collisions, quasimomentum-exchange collisions between atoms of any state can occur. Thus, this type of collision occurs more frequently than those of state changing nature, causing a rapid distribution of the excitations among all occupied rapidities. Therefore, the excitation probability $\nu_n$ quickly becomes uniform over the phase space.

For the purpose of this work, we only consider the two lowest transverse excited states, as the population of higher excited states is negligible throughout the evolution~\cite{Li2018}. However, these states may become relevant in the presence of higher temperatures, more atoms, lower transverse trapping frequency or larger momentum transfers during the Bragg pulse sequence.
In this case, the extension of Yang's theory~\cite{PhysRevLett.19.1312} to an arbitrary number of components of a 1d Bose gas is possible~\cite{klauser2011,Sutherland1968}.
Furthermore, one could take into account the transverse degrees of freedom within the adiabatic approximation (see Refs. \cite{PhysRevA.65.043614,PhysRevA.77.013617} for low-dimensional degenerate bosonic systems or Ref. \cite{adhikari2009gap} for their fermionic counterparts. This has been shown to be equivalent to the treatment of the \textit{virtual} quantum excitations by the second-order perturbation theory~\cite{Mazets_2010}, which leads to the thermalization in 1d gases via effective three-body collisions~\cite{Mazets2008}. However, this mechanism is too slow for the parameters of the experiment (because of the low density). The quasimomenta can be redistributed fast enough only due to the \textit{real} excitations of the transverse degrees of freedom, which require kinetic treatment using a Boltzmann-type equation.

For the sake of clearness, in the following derivation we will consider only two components: the transverse ground state (denoted as the pseudospin state $| \downarrow \rangle$) and the transverse second excited state (denoted as $| \uparrow \rangle$). Analogously, we could consider only the ground and first excited states. In principle, both excited states are degenerate, and a full treatment including all possible states would require considerations of orbital momentum conservation as well. However, this falls outside the scope of the present work. Instead, we simply re-scale the transition probabilities of the considered collision channels by an empirical factor $\zeta$ and neglect quasimomentum exchanging collisions between the $n=1$ and $n=2$ states. Hence, we consider two separate subsystems ($n = \{0,1\}$ and $n = \{0,2\}$), and by combining their collision channels we obtain the effective model presented in the main text. The collision integrals of the two subsystems are very similar, as the only real difference being both atoms in the $n = \{0,1\}$ deexciting collisions carry excitations. Therefore, for simplicity, the following derivations only concern the $n = \{0,2\}$ subsystem and we drop all state-indicating subscripts.

The Bethe-ansatz solution for an integrable two-component 1d Bose gas was first proposed by Yang~\cite{PhysRevLett.19.1312}\footnote{Note, that the cited paper deals mainly with 1D Fermi systems; the result for bosons is only briefly presented}. Eigenstates of the two-component 1d Bose gas are characterized not only by the quasiparticle rapidities, but also by rapidities $\lambda$ of pseudospin waves. The bosonic wave function of $N$ bosons is symmetric with respect to permutations of atoms. For an eigenstate, it can be written as an irreducible tensor product of the pseudospin and co-ordinate parts, each of them belonging to the same irreducible representation of the symmetric group $S_N$. An irreducible representation of $S_N$ is denoted by the corresponding Young diagram. Since only two pseudospin states are present, the Young diagram can contain maximally 2 rows, i.e., has the form $\{ N-M, M \}$ where $M$ is an integer from $0$ to the integer part of $N/2$. Note that $M$ is not the number of atoms in the state $|\uparrow \rangle$; the latter number is larger than or equal to $M$.

In the general case, pseudospin rapidities can be complex, forming so-called Bethe strings. However, since the fraction of atoms in the $|\uparrow \rangle$ state is small, we can assume $\mathrm{Im} \lambda = 0$. Thus, we can introduce quasiparticle (p) and hole (h) distributions $\sigma_{p,h} (\lambda)$ for the pseudospin rapidities as well. Because $M \ll N$, the contribution of the pseudospin component to the quasimomenta density of states is negligible. Therefore, we can roughly estimate
\begin{equation}
    \sigma_{p}(\lambda)+\left.\sigma_{h}(\lambda) \approx \rho_{p}\right|_{\theta=\lambda} \; .
    \label{eq:pseudospin_density}
\end{equation}
Eq. (\ref{eq:pseudospin_density}) has a clear physical meaning: each atom can bear, additionally to its quasimomentum, a pseudospin excitation. Thus, we denote the probability of an atom bearing a pseudospin excitation by 
\begin{equation}
    \nu(\theta)=\left.\left.\frac{\sigma_{p}}{\sigma_{p}+\sigma _h}\right|_{\lambda=\theta} \approx \frac{\sigma_{p}}{\rho_{p}}\right|_{\lambda=\theta} 
\end{equation}
and assume in the following that $\nu \ll 1$.

Finally, we can formally account for the excitation component within the framework of GHD by introducing a Boltzmann-type collision integral to the hydrodynamic equation
\begin{equation}
    \partial_t \rho_p + \partial_x (v^{\mathrm{eff}} \rho_p) + \hbar^{-1}\partial_\theta (F^{\mathrm{eff}} \rho_p) = \mathcal{I}(\theta)\; .
    \label{eq:EXhydro}
\end{equation}
In the following, we will derive an expression for the collision integral for the two-level subsystem.

\section{\label{sec:collision}Collision integral}
We consider atoms in a waveguide under assumption that the collision energy may exceed $2 \hbar \omega_\perp$, but is certainly below $4 \hbar \omega_\perp$. 
We extend Olshanii's treatment~\cite{Olshanii1998} 
to collision energies high enough 
to excite the transverse degrees of freedom. The renormalized coupling strength is then 
\begin{equation} 
\tilde c = \frac c{1 -\frac 12 cl_\perp {\cal C} (\epsilon )}     ,     \label{eghd.1.1}
\end{equation}
where 
\begin{equation} 
{\cal C} (\epsilon )\approx 2 \sqrt{1-\tfrac 12 \epsilon } -\frac 1{4\sqrt{1-\frac 12 \epsilon}}-\frac 1{2\sqrt 2 \sqrt{1-\epsilon}}    ,  
\label{eghd.1.2} 
\end{equation} 
$\epsilon =\frac 18(k_1-k_2)^2l_\perp ^2$, and $\hbar k_1,\, \hbar k_2$ are the momenta of colliding bosonic atoms.  Eq. (\ref{eghd.1.2}) is derived 
using the simplest approximation to the sum 
$    
\sum _{n=2}^\infty \frac {\exp (-\sqrt 2 \sqrt{n-\epsilon }|z|/l_\perp )}{\sqrt 2 \sqrt{n-\epsilon }} \approx  
\int _2^\infty dn^\prime \, \frac {\exp (-\sqrt 2 \sqrt{n^\prime -\epsilon }|z|/l_\perp )}{\sqrt 2 \sqrt{n^\prime -\epsilon }}+
\frac 12 \frac {\exp (-\sqrt 2 \sqrt{2-\epsilon }|z|/l_\perp )}{\sqrt 2 \sqrt{2-\epsilon }} 
$ 
according to the Euler--Maclaurin formula. This approximation is quite good, since Eq. (\ref{eghd.1.2}) yields 
${\cal C} (0) \approx 1.04$, while the exact result   
is ${\cal C} (0)=1.06  \dots ~$. 

For $cl_\perp \ll 1$ the real part of $\tilde c$ is close to $c$ for almost all collision energies, except of a narrow interval near 
the excitation threshold $\epsilon = 1$. If $\epsilon >1$, the imaginary part of $\tilde c$ is non-zero, which corresponds to the 
probability of a collisional excitation of the transverse degrees of freedom 
\begin{equation} 
\mathcal{P}_{\updownarrow} (k,q) = \frac {4c^2 kq}{k^2q^2+  c^2(k+q)^2}   ,   \label{eghd.1.3}
\end{equation} 
where 
\begin{equation} 
k= |k_1-k_2|, \qquad q=\sqrt{   |k_1-k_2|^2 -8l_\perp ^{-2}}      ,    \label{eghd.1.4} 
\end{equation}
Depending on the exact transition, we may scale $\mathcal{P}_{\updownarrow}$ by its relative transition strength $\zeta$ to account for the neglected degeneracy of the excited states. 
We will use the dimensional coupling constant $c$ and the excitation probability (\ref{eghd.1.3}) as basic building blocks for our extended GHD.

\begin{figure*}
\center
\includegraphics[width = 0.98\textwidth]{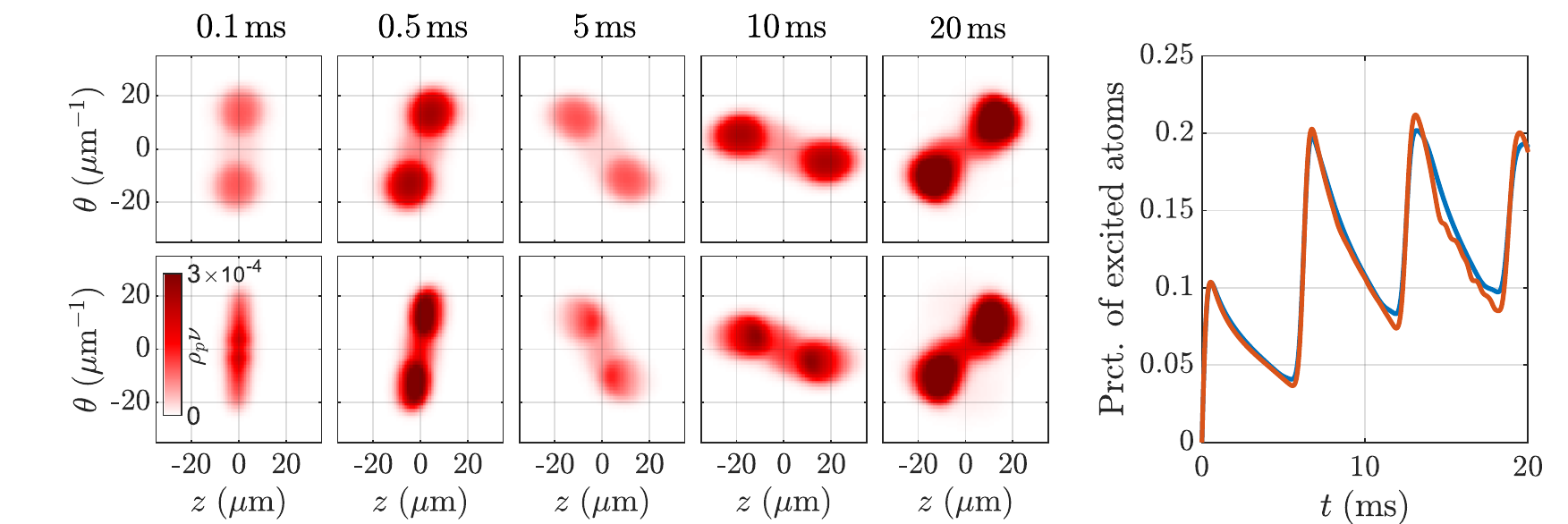}
\caption{\label{fig:nu} Comparison of uniform vs non-uniform excitation probability $\nu$ for the $n = \{0,2\}$ subsystem with 130 atoms at 94nK without heating. The top row shows $\rho_p \nu$ evaluated using the equations for uniform $\nu$, while the bottom row is evolved using eq. (\ref{eq:NonuniformNu}). The last panel shows the percentage of the total number of atoms excited to the second transverse state. The blue curve is the uniform case, while the red one is the non-uniform.}
\end{figure*}
First, we also introduce the quasimomenta of the atoms after a collisional excitation (denoted by subscript "-") or deexcitation (denoted by "+") of the transverse state as $\theta _\pm = \frac 12 (\theta +\theta ^\prime )+\frac 12 (\theta -\theta ^\prime )
\sqrt{1\pm 8/[(\theta -\theta ^\prime )l_\perp ]^2}$, 
and $\theta _\pm ^\prime = \frac 12 (\theta +\theta ^\prime )-\frac 12 (\theta -\theta ^\prime )
\sqrt{1\pm 8/[(\theta -\theta ^\prime )l_\perp ]^2}$. 
The microscopic collision velocity is $\hbar |\theta -\theta ^\prime |/m$. 
Knowing the scattering probability $P_\updownarrow $, we can write the Boltzmann-type collision integral (for the$n = \{0,2\}$ subsystem) as 
\begin{equation}
\begin{split} 
{\cal I}_{\{0,2\}}(\theta )=&\, (2\pi )^2 \frac \hbar m \int _{-\infty }^\infty d\theta ^\prime \, 
\zeta P_\updownarrow ( |\theta -\theta ^\prime |, |\theta _-  -\theta ^\prime _-|) 
|\theta -\theta ^\prime |\Theta (|\theta -\theta ^\prime |l_\perp -2\sqrt 2)
\Big{ \{ }-\rho _p (\theta )\rho _p(\theta ^\prime )\rho _h(\theta _-)\rho _h(\theta _-^\prime )+ \\
&\qquad \qquad \qquad \frac 12 \rho _h(\theta )\rho _h(\theta ^\prime ) \rho _p(\theta _-) \rho _p(\theta _-^\prime )
\left[\nu (\theta _-) + \nu(\theta _-^\prime )\right]\Big{ \} } +  \\
&\, (2\pi )^2\frac \hbar m \int _{-\infty }^\infty d\theta ^\prime \, 
\zeta P_\updownarrow ( |\theta -\theta ^\prime |, |\theta _+  -\theta ^\prime _+|)  |\theta -\theta ^\prime | 
\Big{ \{ }-\frac 12\rho _h(\theta _+)\rho _h(\theta _+^\prime )\rho _p (\theta )\rho _p(\theta ^\prime )
\left[ \nu (\theta )+\nu(\theta ^\prime )\right] +  \\ 
&\qquad \qquad \qquad 
\rho _h (\theta )\rho _h(\theta ^\prime )\rho _p(\theta _+)\rho _p(\theta _+^\prime )  \Big{ \} } ,
\end{split} 
\end{equation}
where $\Theta (x)$ is the Heaviside step function. Again, note that the collision integral for the $n = (0,1)$ subsystem looks slightly different, as both atoms in the deexciting collision carry excitations.

The key idea behind the expression for the collision integral is that in the quantum degenerate regime the scattering is affected by the Pauli blocking: scattered atoms can acquire only those values of quasimomentum, which were not occupied before the collision. Therefore, the collision integrals must contain not only particle distribution functions, but also hole distribution functions. Here the fermionic nature of particles and holes in the Lieb-Liniger model is manifested. Note that $\rho _h (\theta _\pm)$ is the Pauli blocking factor (1 minus the population) times the density of states for the scattering products. Factor $(2\pi )^2$ arises from the normalization. One factor $2\pi $ arises from 
$\int dt\, \exp [-i(E_i -E_f)t/\hbar = 2\pi \hbar \delta (E_i-E_f)$, where $E_i$, $E_f$ are the energies of the initial and final states, respectively. Another factor $2\pi $ appears when we switch from summation over discrete rapidities defined by the periodic boundary conditions over the length $L$ to the integration over continuous $\theta _\pm $: the Kronecker delta-symbol for discretized total momentum, 
$\delta _{P_i,P_f}=\mathrm{sinc}\, [(\theta _\pm +\theta ^\prime _\pm -\theta -\theta ^\prime )L/2]$, where 
$\mathrm{sinc}\, x =\sin x/x$, transforms to a $2\pi \delta (\theta _\pm +\theta ^\prime _\pm -\theta -\theta ^\prime )/L$, when we replace the discrete sum by $L\int d\theta _\pm \, \dots $, recall the normalization $L\int d\theta \, \rho _p(\theta )=N$.

\begin{figure*}
    \centering
    \includegraphics[width=0.49\textwidth]{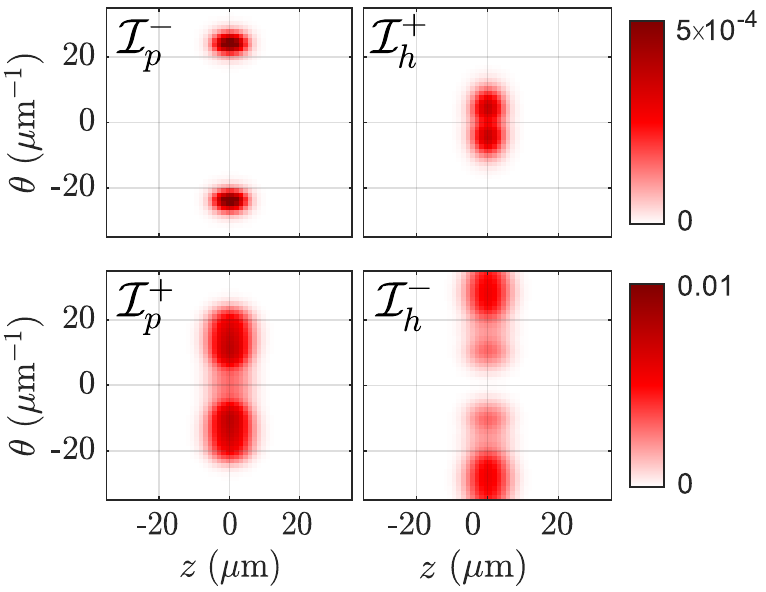}%
    \hfill
    \includegraphics[width=0.49\textwidth]{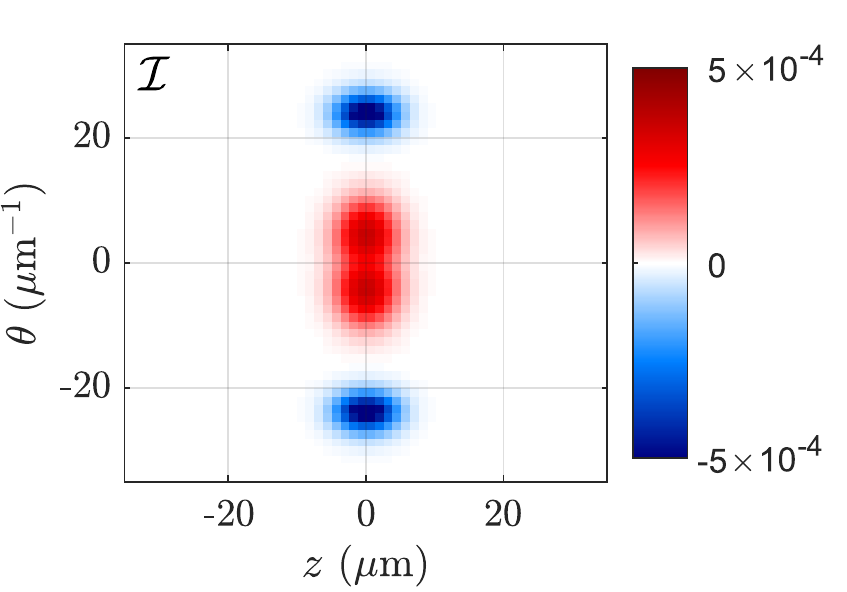}%
    \caption{Collision integral and its components for the post Bragg-pulse state with 130 atoms at 94nK. $\mathcal{I}_p^{-}$ and $\mathcal{I}_h^{+}$ describe the redistribution of quasiparticles in rapidity space due to ground state collisions leading to transverse excitations. Meanwhile, $\mathcal{I}_p^{+}$ and $\mathcal{I}_h^{-}$ describe the redistribution via deexcitation through collisions between a ground state and excited atom. The final panel displays the total collision integral. Since $\nu_n = 0$ at $t=0$, no deexcitations can occur.  }%
    \label{fig:CollisionIntegrals}%
\end{figure*}

If we assume that the pseudospin waves propagate at the same velocity of the atoms (remember $M \ll N$), then their kinetics are given by an equation similar to ground state atoms
\begin{equation}
\begin{aligned}
\partial_{t}\left(\rho_{p} \nu \right)+\partial_{z}\left(v^{\mathrm{eff}} \rho_{p} \nu \right)+\hbar^{-1} \partial_{\theta}\left(F^{\mathrm{eff}} \rho_{p} \nu \right)=& \frac{1}{2} \zeta \mathcal{I}_{h}^{+}-\frac{1}{2} \zeta \mathcal{I}_{p}^{+} \nu+\rho_{p} \gamma+\\
& \frac{2 \pi \hbar}{m} \int_{-\infty}^{\infty} d \theta^{\prime} \frac{c^{2}\left|\theta-\theta^{\prime}\right|}{c^{2}+\left(\theta-\theta^{\prime}\right)^{2}} \rho_{p}\left(\theta^{\prime}\right) \rho_{p}(\theta)\left[\nu\left(\theta^{\prime}\right)-\nu(\theta)\right] \; .
\label{eq:NonuniformNu}
\end{aligned}
\end{equation}
The two first terms are also found in the Boltzmann-type collision integral and describe the creation and deexcitations due to collisions. The final term encodes the rapidity-exchanging collisions between quasiparticles, which conserves the number of excitations. This process occurs on time scales much shorter than excitation of transverse modes, whereby $\nu$ is quickly distributed among all rapidities. Accounting for both the first and second transverse excited state simultaneously adds further possible rapidity-exchanging collisions, thus distributing $\nu$ uniformly even more rapidly.
Propagating $\rho_p \nu$ and $\rho_p$ for the $n = \{0,2\}$ subsystem using the equations above produces the results seen in figure \ref{fig:nu}. The figure also compares the results to the uniform $\nu$ employed in the main manuscript. Initially, the two atomic clouds overlap in the cradle, whereby collisions between ground state atoms can exceed $2 \hbar \omega_\perp$ in energy, exciting one of the atoms to the second transverse state. Immediately after the collisions the partaking quasiparticles are found at low rapidities, as most of their kinetic energy has been converted into transverse potential energy. However, due to quasimomentum exchange processes, the excitation probability $\nu$ is quickly spread to higher rapidities. The transversely excited atoms propagate at the same velocity as the ground state atoms, whereby they throughout the oscillation gets distributed in space as well. Hence, after less than two periods, we find $\rho_p (\theta , z) \: \nu (\theta , z)$ to be practically identical to $\rho_p (\theta, z)$ times a constant, i.e. $\nu$ is practically uniform in the $(\theta, z)$-space.
Additionally, we observe almost no difference in the percentage of excited atoms when evolving with the uniform $\nu$ as compared with the non-uniform one. 
Therefore, it is justified making the simplification that $\nu$ only depends on time, whereby its kinetics and the Boltzmann-type collision integral reduce to the much simpler equations found in the main text.

Figure \ref{fig:CollisionIntegrals} displays the collision integral calculated for the post Bragg-pulse state for 130 atoms at 94nk (a system also discussed in the main manuscript). The figure also shows the contribution to the collision integral from each component, which makes the different processes of excitation and deexcitation obvious.
The component $\mathcal{I}_{p}^{-}$ plotted in the top row describes the quasiparticles partaking in ground state collisions leading to excitations, whereby only quasiparticles occupying large rapidities contribute. Meanwhile,  $\mathcal{I}_{h}^{+}$ is the distribution of the quasiparticles immediately after a state-changing collision. Since much of their kinetic energy has been converted into potential energy during the excitation, the quasiparticles now occupy mostly low rapidities. In the collision integral, subtracting $\mathcal{I}_{p}^{-}$ while adding  $\mathcal{I}_{h}^{+}$ captures the redistribution of quasiparticles due to excitations. 
Analogously, the bottom row of figure \ref{fig:CollisionIntegrals} plots the components of the collision integral encoding the deexciting collisions. Since quasiparticles at any rapidity can partake in deexciting collisions, we find that the corresponding component $\mathcal{I}_{p}^{+}$ has roughly the same shape as the quasiparticle density. Upon deexcitation, the $2\hbar\omega_\perp$ worth of potential energy is converted to kinetic energy, whereby the quasiparticles after the collision have much higher rapidities, which is reflected in the component $\mathcal{I}_{h}^{-}$. 
Since $\nu_n = 0$ at $t = 0$, only exciting collisions occur initially. 

The collision integral $\mathcal{I}(\theta )$ is identically zero when rapidities obey the Fermi--Dirac distribution and the classical (Boltzmann) statistics hold for transverse excitations (recall that $\nu_n \ll 1 $ by assumption). The temperature-dependent collective correction to the quasiparticle energy that appears in the thermodynamic Bethe ansatz is assumed to be negligibly small in our treatment, since we consider temperatures well below  $\hbar \omega _\perp /k_\mathrm{B}$. 
For a nondegenerate 1d Bose gas, $\rho _p(\theta )\ll \rho _h(\theta )\approx 1/(2\pi )$, the collision integral takes the limit (Boltzmann) limit 
\begin{equation}
\begin{split} 
{\cal I}_{\mathrm{cl}}(\theta )=&\, \frac \hbar m \int _{-\infty }^\infty d\theta ^\prime \, 
P_\updownarrow ( |\theta -\theta ^\prime |, |\theta _-  -\theta ^\prime _-|) 
|\theta -\theta ^\prime |  \Theta (|\theta -\theta ^\prime |l_\perp -2\sqrt 2) 
\Big{ [ }-\rho _p (\theta )\rho _p(\theta ^\prime )+ 
\rho _p(\theta _-)\rho _p(\theta _-^\prime )\nu \Big{ ] } +  \\
&\, \frac \hbar m \int _{-\infty }^\infty d\theta ^\prime \, 
P_\updownarrow ( |\theta -\theta ^\prime |, |\theta _+  -\theta ^\prime _+|) |\theta -\theta ^\prime |
\Big{ [ }- \rho _p (\theta )\rho _p(\theta ^\prime )\nu +\rho _p(\theta _+)\rho _p(\theta _+^\prime )  \Big{ ] } .
\end{split} 
\end{equation}

\section{Comparing excitation rates between extended GHD and molecular dynamics}
\begin{figure*}
    \centering
    \includegraphics[width=0.95\textwidth]{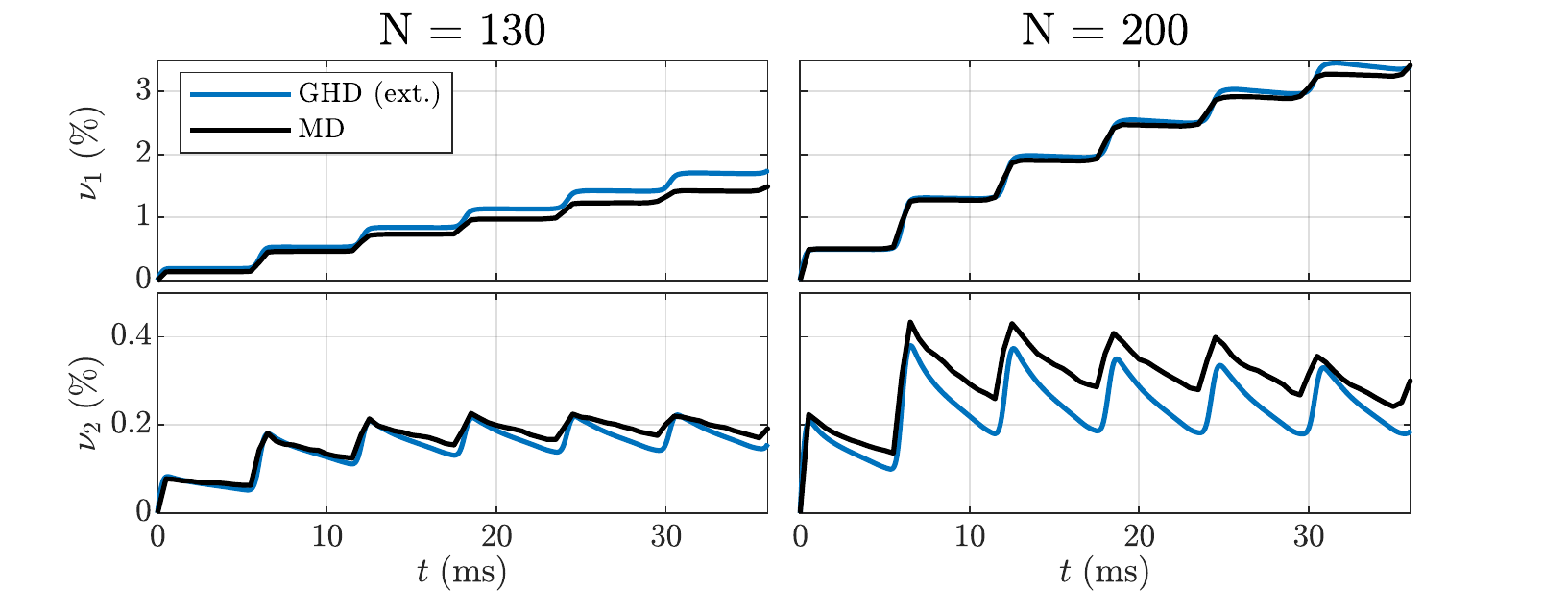}%
    \caption{Population of transverse excited states for two different realizations with $N=130$ and $N=200$ atoms at 90nK during the first 3 oscillation periods of the cradle. The calculation was performed using the extended GHD and the molecular dynamics approach featured in Ref. \cite{Li2018}.}%
    \label{fig:MDcomparison}%
\end{figure*}

In Ref. \cite{Li2018}, a molecular dynamics approach was used to describe the same sets of data as presented in the main text. The method propagated the atoms using the classical equations of motion in a harmonic confinement. Whenever two atoms collided, a random number was drawn from a uniform distribution from 0 to 1. If said number was below the probability of a state-changing collision, the transverse states of the atoms were changed accordingly. Thus, an arbitrary number of excited states could be accounted for, although the populations of the $n>3$ states were entirely negligible. Finally, many realizations with different stochastic outcomes were averaged over to obtain the final results. 

It is interesting to compare this stochastic approach to excitations with the Boltzmann-type collision integral used in our extended model. Thus, using both GHD and molecular dynamics, we simulated the first 3 periods of the Newton's cradle for 2 different systems with atom numbers $N=130$ and $N=200$ at a temperature of 90nK with $\eta = 0.17$. Since the external heating works rather differently between the two approaches, we let $\gamma = 0$. Finally, we employed a harmonic longitudinal confinement in the GHD simulations.

The results are seen in figure \ref{fig:MDcomparison}, where the population of the first and second transverse excited states are plotted. Generally, we observe a good agreement between the two methods. The populations of the transverse states exhibit the same behavior; a rapid increase during overlap of the Bragg peaks, followed by a slower decrease during the peak separation. Additionally, both methods clearly exhibit a deexcitation rate from the $n=1$ state relatively much slower than that from the $n=2$ state. We do observe some discrepancies (especially in the population of the $n=2$ state), which we attribute to the difference in dynamics; while the molecular dynamics employs classical equations of motion, the GHD takes into account the Wigner time delay associated with interactions between atoms. The interactions combined with the inhomogeneous longitudinal potential enables mixing of the quasiparticle trajectories (single particle energy not conserved)~\cite{caux2019cradle}, thus enabling the shape of the peaks to change during evolution, even without any influence from transverse excited states, through so called "many-body dephasing"~\cite{tang2018cradle}. 

Lastly, it should be noted that the thermalization rate $\mathcal{T}$ presented in this work is slightly different to those in Ref. \cite{Li2018}. Whereas in Ref. \cite{Li2018} an excellent agreement between theory and experiment was found when studying a single 1d tube containing the weighted average number of atoms, we observe here that GHD yields the best comparison with the experiment when considering an ensemble of 1d systems. Interestingly though, we observe roughly the same population of excited states, even at long time-scales, when comparing the results in the main text with Ref. \cite{Li2018}.  

\section{Comparing period-averaged distributions between GHD and experiment}
\begin{figure*}
    \centering
    \includegraphics[width=0.95\textwidth]{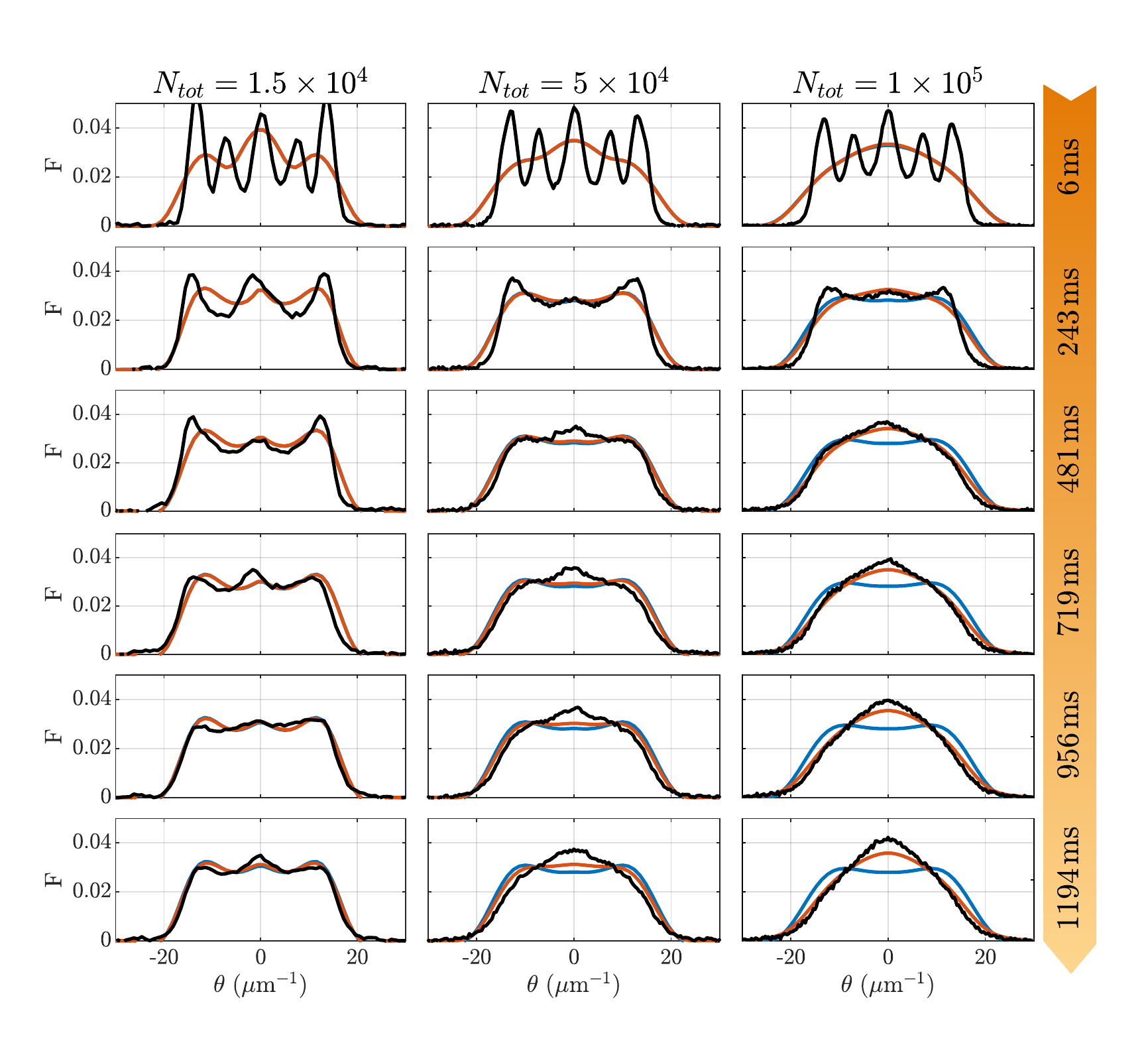}%
    \caption{Normalized period-mean rapidity distribution functions (momentum distribution functions) for the GHD simulations (experiment). The standard GHD is marked in blue, the extended GHD in red, and the experiment in black. The simulated profiles were obtained by summing over a number of 1d systems and weighing their contribution according to their occurrence in the lattice.}%
    \label{fig:MeanMDFs}%
\end{figure*}
In the main manuscript we observe good agreement in the rate of thermalization between the experiment and our extended model. However, it is important to note that the measure of thermalization $\mathcal{T}$ quantifies how close the period-mean profile is to its best Gaussian fit. Thus, the closest Gaussian for the experiment might not equal that of the GHD theory.

For a direct comparison between GHD and experiment, we plot in figure \ref{fig:MeanMDFs} the total period-mean rapidity and momentum distributions for the theory and experiment, respectively. Similar to what is shown in the main manuscript, the profiles of standard GHD exhibit only small change throughout the duration: Initially, the population of quasiparticles at low rapidities is depleted due to the non-linear dynamics emerging from interactions. This leads to the initial increase in $\mathcal{T}$. Thereafter, the mean profile remains constant. Meanwhile, in the extended model for high atom numbers, we observe a change in the mean profile over time, as it indeed tends towards a Gaussian. Comparing the two theories to the experimental observations, two main differences in the evolution are apparent. Firstly, the initial profile of the experiment looks very different from the simulations. As mentioned previously, the experimentally measured profiles are the momentum distribution functions (MDF), which in the degenerate regime are much narrower than the RDF~\cite{Cazalilla2004}. Therefore, the peaks of the experimental period-mean profiles are very pronounced. Further, measurements are made at 1ms intervals. Since the oscillation period is 12ms, the position of the peaks are similar for many of the measurements during the first periods. Hence, the observed 5-peak structure emerges. However as the density of the gas decreases through dephasing during the evolution, the MDF and RDF become increasingly alike~\cite{giamarchi2004quantum}, whereby GHD and experimental observations become comparable. After the initial stage, a relaxation towards Gaussian also occurs in the experiment. However, the observed relaxation rate is greater than in the GHD simulations, in particular for the lower atom numbers. For these realizations, many tubes have only a tiny population above the excitation threshold, whereby state-changing collisions are very rare. Instead, the relaxation is dominated by the heating from the lattice, something also seen in Ref. \cite{Li2018}.

We attribute the discrepancies in the results to three main factors: (i) An underestimation of the heating rate $\gamma$. The heating rate employed in the manuscript was obtained from observations detailed in Ref. \cite{Li2018} of the condensate held in the lattice without initiating the Bragg pulse sequence. However, the heating during the cradle dynamics appears greater than estimated in the static case. Nevertheless, since our main objective is describing the thermalization via collisions, we leave this for future studies. (ii) Inhomogeneous atom losses. In the GHD simulations, we do not account for atom losses, and when taking the period mean of the profiles we also normalize them. However, if there is an experimental bias towards losing atoms with high kinetic energy (which is often the case), the tails of the MDF will appear smaller. (iii) Poor knowledge of the initial state. Due to the nature of the experiment, it is extremely difficult to estimate the initial state: First of all, we do not measure the full quasiparticle density, but rather a projection of it onto the rapidity axis. Furthermore, the measured distribution is the MDF rather than the RDF. Thus, it is unknown exactly how the post-pulse state looks like in the full ($\theta$,$z$)-phase space. We therefore rely on the approach from Ref. \cite{caux2019cradle, berg2016separation} to construct the post-pulse state, although imperfections in the pulse-sequence clearly lead to a different distribution. We accommodate for this by introducing the parameter $\eta$ to encompass any atoms unaffected by the pulses, however, the value of $\eta$ appears atom number dependent and thus differs from tube to tube. The exact dependency is unknown, whereby we simply employ the same value of $\eta$ for all simulations. Thus, accurately predicting the full thermalization using our extended model would require a dedicated experiment addressing the issues highlighted above.

\section{Comparing standard and extended GHD for atom chip experiment}
\begin{figure*}
    \centering
    \includegraphics[width=0.95\textwidth]{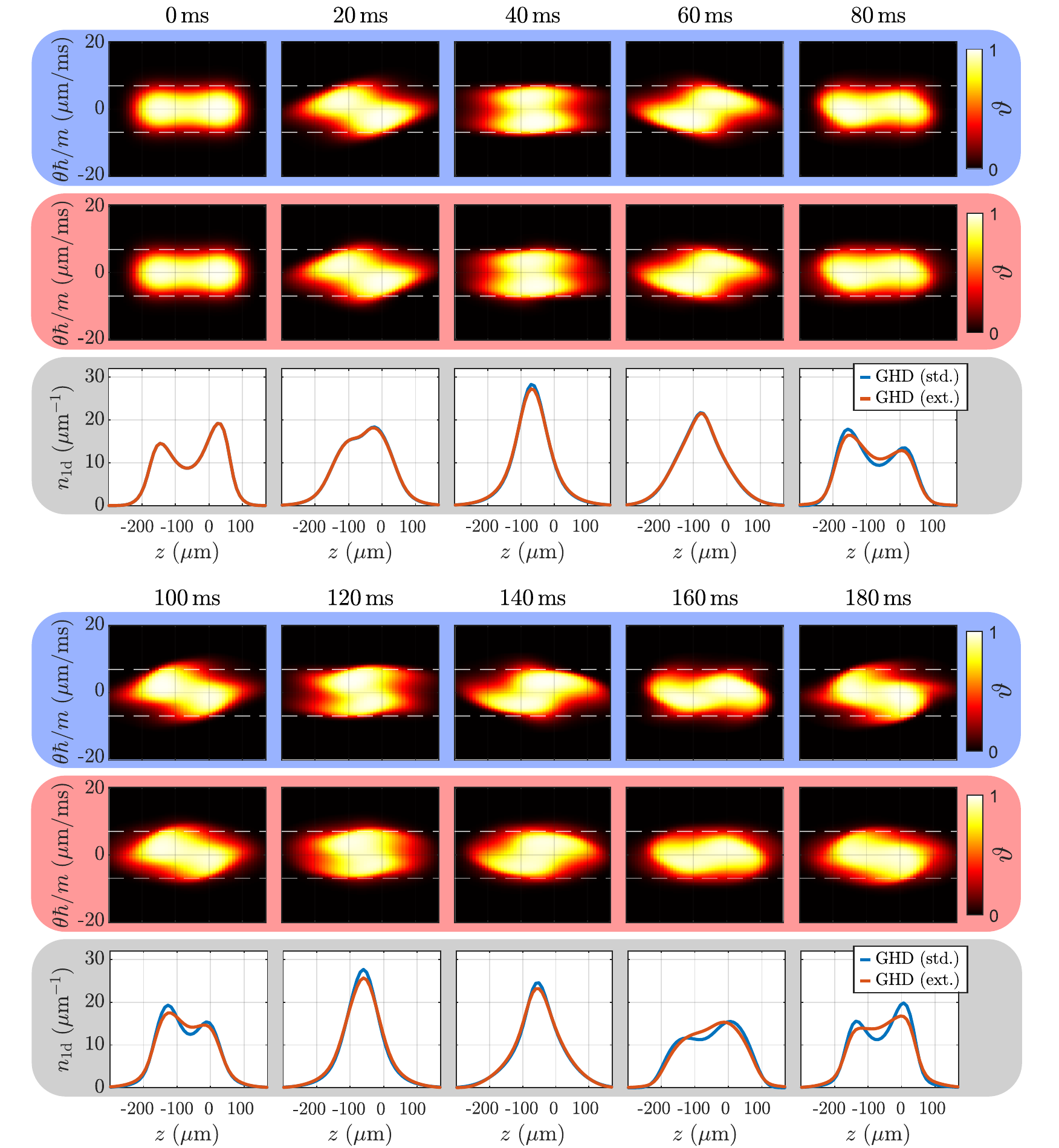}%
    \caption{Simulation of the evolution of the system described in Ref. \cite{schemmer2019generalized}, where a longitudinal double-well potential is quenched to a harmonic trap. The top (blue) rows show the filling function computed using standard GHD, while the center (red) rows depict the the filling function computed via extended GHD. The dashed white lines indicate the excitation threshold. In the bottom (grey) rows, the linear atomic density is plotted for the two approaches.}%
    \label{fig:chipexp_compar}%
\end{figure*}
In the first experimental demonstration of the applicability of GHD (see Ref. \cite{schemmer2019generalized}), a quasi-1d Bose gas mainly in the quasicondensate regime was realized on an atom chip. The setup sought to mimic a quantum Newton's cradle by quenching the longitudinal potential from an initial double-well to a harmonic trap.
After just a single oscillation period, the experimentally observed linear density started deviating from the GHD predictions. The discrepancy was attributed to atom losses~\cite{bouchoule2020effect}, however, given the parameters of the experiment, plus the fact that said setup has previously been used to explore the crossover regime using a Yang-Yang model with additional transverse states~\cite{PhysRevA.83.021605, Amerongen2008yang}, it is very reasonable to assume that transverse excitations may have influenced the observed dynamics. Therefore, in this section, we simulate the dynamics of Ref. \cite{schemmer2019generalized} using both standard and extended GHD and compare the two results. This demonstrates both the applicability of extended GHD in the quasicondensate regime and how it without any modifications can be used to describe other setups.

The system of Ref. \cite{schemmer2019generalized} is a single Bose gas consisting of $N = 3500$ $^{87}$Rb atoms at $T \approx 0.15\mu\mathrm{K}$ with a transverse trapping frequency of $\omega_\perp = 2 \pi \times 5.4$kHz. The system is realized in a longitudinal double-well, and the dynamics are initiated by quenching to a longitudinal harmonic confinement of $\omega_\Vert = 2 \pi \times 6.5$Hz. There are no indications of the heating rate in Ref. \cite{schemmer2019generalized}, whereby we set the external heating rate to $\gamma = 0$. However, atom chip setups with their magnetic trapping typically have rather low heating, so neglecting the heating for this setup should not pose a problem.

Finding the initial state of Ref. \cite{schemmer2019generalized} is unfortunately rather difficult, since no expression for the double-well is given. Instead, the authors directly constructed the quasiparticle density $\rho_p(z, \theta, t=0)$ based on the observed linear density and used the subsequent evolution to fix the temperature. Given that standard GHD was used for this sort of fitting, their approach of retroactive fitting hardly seems valid, as the extended GHD, rather than the standard GHD, should describe the dynamics. Nevertheless, the influence of transverse states is still relatively small during the initial part of the evolution (as demonstrated in the main text), hence we assume that the temperature estimated in Ref. \cite{schemmer2019generalized} is not too far from the true temperature. 
Thus, given the reported temperature of $T \approx 0.15\mu\mathrm{K}$ we have tried reconstructing the double-well potential, yielding $V_{dw} (z) \approx \hbar \omega_\perp \left( (a z)^4 + (b z)^3 + (c z)^2 \right)$ with $a = 6.7 \times 10^{3} \: \mathrm{m}^{-1}$, $b = 3.4 \times 10^{3} \: \mathrm{m}^{-1}$, and $c = 5.5 \times 10^{3} \: \mathrm{m}^{-1}$.
When treating collisionally driven transverse excitations, one of the most important quantities is the number of atoms above the excitation threshold, i.e. how far on the rapidity axis the quasiparticle density stretches. Fortunately, the supplemental material of Ref. \cite{schemmer2019generalized} features phase-space plots of the filling function $\vartheta(z, \theta, t)$ for several times during the evolution (see figure 6 of the SM in Ref. \cite{schemmer2019generalized}). Comparing our results to those, we see that our reconstructed state is fairly similar to that of Ref. \cite{schemmer2019generalized}.

Compared to the optical lattice setup treated in the main text, the atom chip setup of Ref. \cite{schemmer2019generalized} featured a high temperature and chemical potential combined with a smaller transverse trapping frequency. Further the system was initialized directly in the dimensional crossover regime rather than being brought there by a Bragg pulse sequence. Therefore, we find an initial, non-negligible population of the transverse excited states by fitting a multicomponent Lieb-Liniger model given a fixed atom number, temperature and potential. Each transverse state is treated as a separate Lieb-Lininger system, whose chemical potential is offset by $n \times \hbar \omega_\perp$, with $n$ being the transverse level (a technique fairly similar to the one of Ref. \cite{Amerongen2008yang}). Using this approach we find $\nu_1 (t=0) = 0.0462$ and $\nu_2 (t=0) = 0.0073$ with higher levels having a negligible population.

\begin{figure*}
    \centering
    \includegraphics[width=0.75\textwidth]{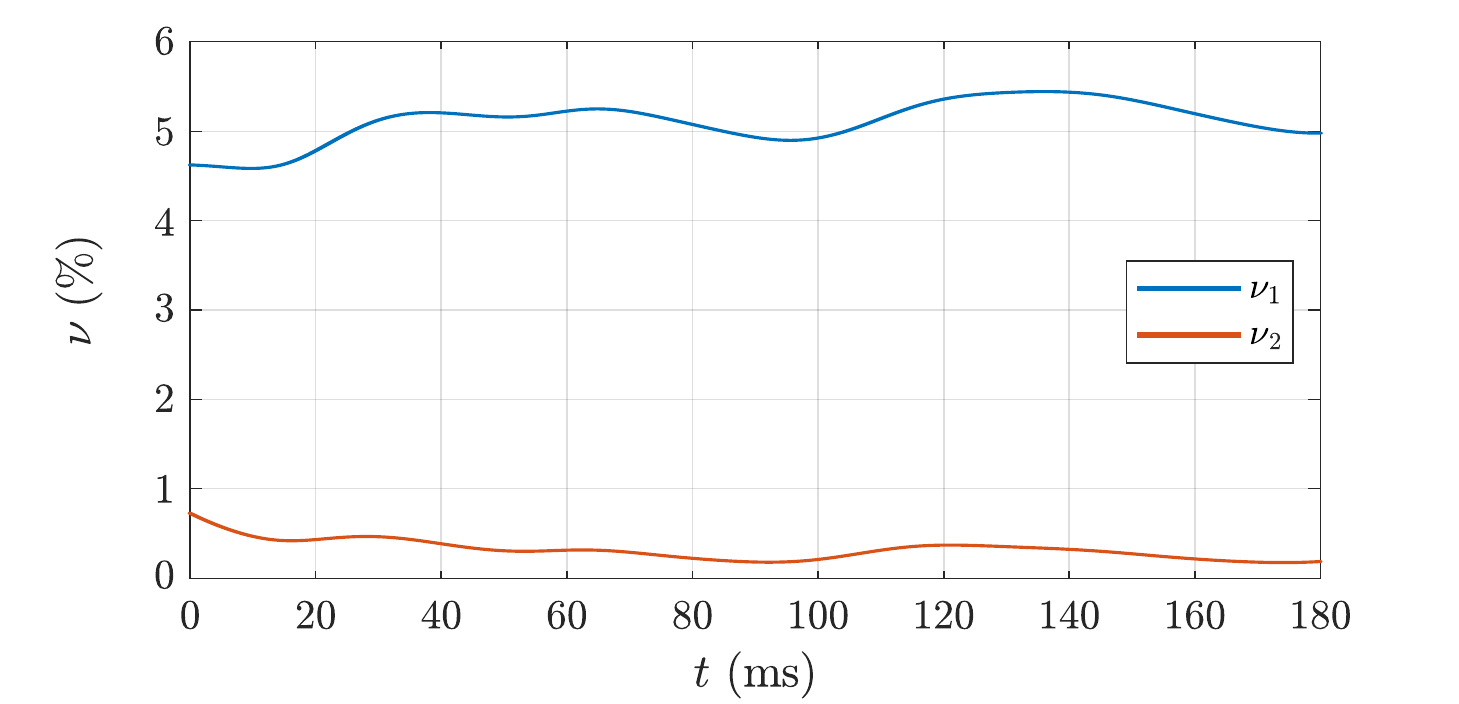}%
    \caption{Excitation probabilities of the first and second transverse excited states of the experimental setup described in Ref. \cite{schemmer2019generalized} calculated using extended GHD. The system is initialized in the dimensional crossover regime, whereby the excitation probabilities are non-zero at $t=0$.}%
    \label{fig:chipexp_nu}%
\end{figure*}

Figure \ref{fig:chipexp_compar} shows the results of propagating the reconstructed initial state using both standard and extended GHD. Starting with standard GHD, we observe rather similar results to those presented in Ref. \cite{schemmer2019generalized}, thus proving that our estimate of the initial state is fairly accurate. After about one oscillation period ($\sim 160$ms) the two peaks of the double-well are still clearly separable. Meanwhile, for the extended GHD, we observe a tendency similar to the Newton's cradle setup discussed in the main text, namely an accumulation of quasiparticles at low energy (low rapidity, close to center of trap) and an overall dephasing of the cloud (spreading of the cloud over the ($z, \theta$)-phase-space. Hence, after one oscillation period, the two density peaks are no longer separable, exactly as observed in the experiment of Ref. \cite{schemmer2019generalized} (depicted in their figure 4 and their figure 5 of the SM). Of course atom losses would contribute to the difference between the measured profile and the one predicted by standard GHD, however, given the likeness of the profiles between extended GHD and the experimental observations plus the relatively high population of the transverse excited states (see figure \ref{fig:chipexp_nu}), it is undeniable that the dimensional crossover absolutely must be taken into account when treating such systems. We also ran the simulation for $\nu_1 (t=0) = \nu_2 (t=0) = 0$ and observed only a small change in the linear density, namely the results of extended GHD being slightly closer to standard GHD, although still clearly different after 180ms. Since the influence of the transverse states accumulate over time, we believe a significant difference between having a finite or vanishing initial transverse population would emerge only at long timescales.

Examining the evolution of the excitation probabilities $\nu_1 (t)$ and $\nu_2 (t)$, plotted in figure \ref{fig:chipexp_nu}, we observe a very different tendency to those found in the setup of the main text; rather than increasing over time, the population of the transverse excited states is practically constant throughout the evolution. The reason therefore is simple, and we have already discussed it: The quantum Newton's cradle is initialized far from the dimensional crossover but is transferred there via the Bragg pulse sequence. Thus, the population of the transverse excited states are far from equilibrium, causing them increase rapidly during the evolution. Meanwhile, the transverse populations in the setup of Ref. \cite{schemmer2019generalized} start in equilibrium, and the quench of the longitudinal trapping potential barely brings the system any deeper into the dimensional crossover (see the dashed lines of figure \ref{fig:chipexp_compar}, indicating the excitation threshold). Therefore, the transverse dynamics are practically in equilibrium, whereby the components of the Boltzmann-type collision integral corresponding to excitation and deexcitation having roughly equal magnitude. This serves to illustrate a very important point of the integrability breaking collisions found in the extended GHD: It is not the \textit{population} of atoms in the transverse excited states that causes the relaxation of dynamics but rather the continuously occurring \textit{transitions} between the transverse states which drives it.

\newpage
\newpage
\section{Additional datasets and figures}
\begin{figure*}[h]
    \centering
    \includegraphics[width=0.49\textwidth]{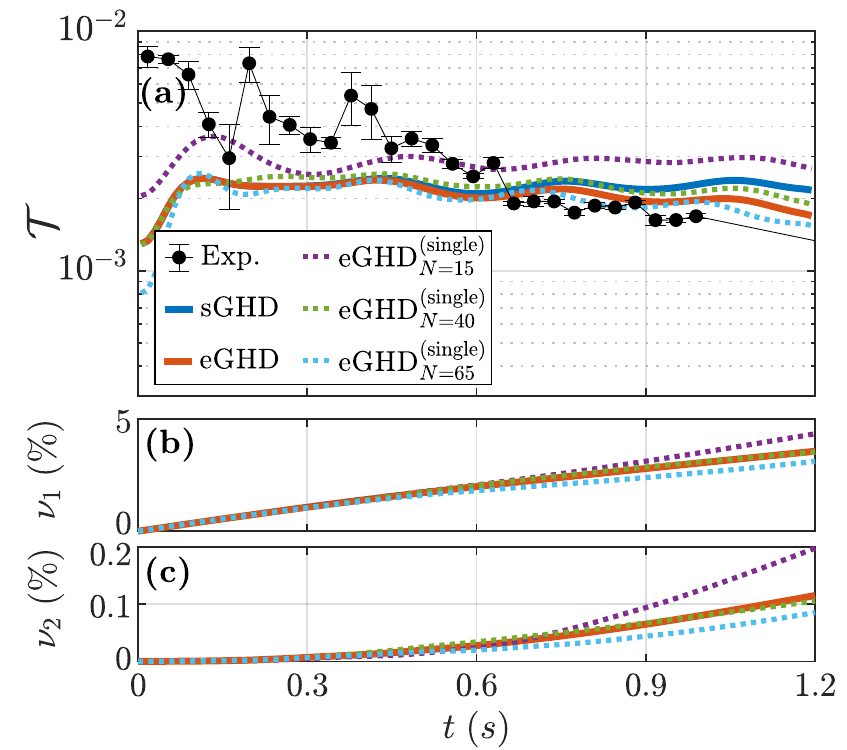}%
    \hfill
    \includegraphics[width=0.49\textwidth]{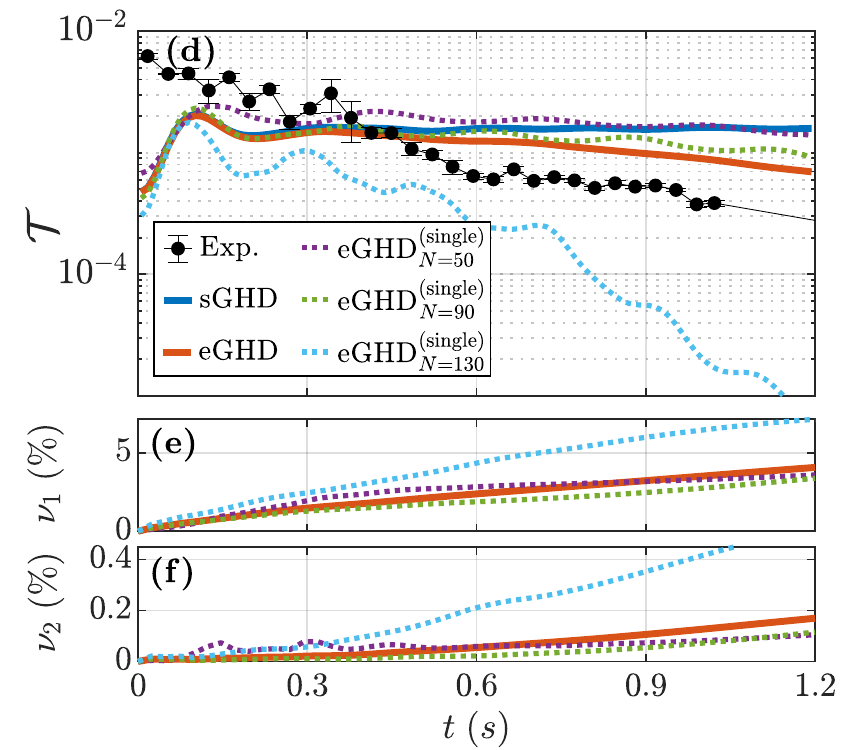}%

    \caption{Thermalization in the quantum Newton's cradle with \textbf{(a-c)} $N_{tot} = 1.5 \times 10^4$ atoms at 34nK, and \textbf{(d-f)} $N_{tot} = 5 \times 10^4$ atoms at 60nK. \textbf{(a,d)} Comparison of measure $\mathcal{T}$ between standard GHD, extended GHD, and experimental observations. Additionally, $\mathcal{T}$ is plotted for a series of single tubes evolved using extended GHD. \textbf{(b,e)} Percentage of atoms in first transverse excited states. \textbf{(c,f)} Percentage of atoms in second transverse excited states. Compared to the higher atom number realization plotted in the main text, the agreement between theory and experiment is worse here. We attribute this to an underestimation of the external heating, which for low atom numbers dominates the thermalization.}%
    \label{fig:thermalization_SM}%
    \vspace{-1\baselineskip}%
\end{figure*}

\begin{figure*}[h]
    \centering
    \includegraphics[width=0.81\textwidth]{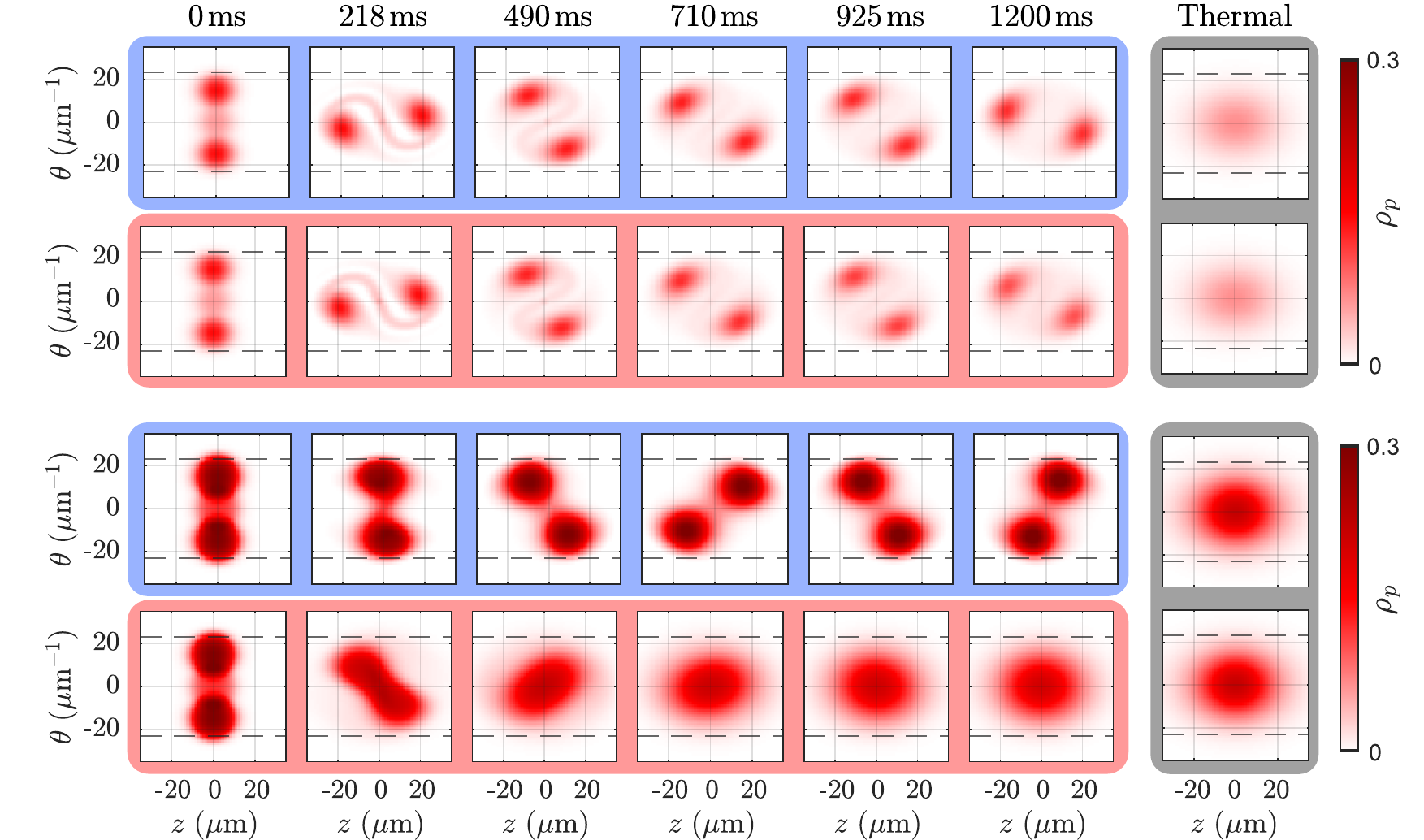}%
    \caption{Evolution of a single tube containing 60 atoms (top figure) and 200 atoms (bottom figure) at 94nK. The blue rows display the quasiparticle density evolved using the standard GHD equation, while the red rows are computed via the extended model. The final panels show the best fitted thermal state at 620nK (top) and 710nK (bottom) for comparison.}%
    \label{fig:theta_figure_200}%
    \vspace{-1\baselineskip}%
\end{figure*}


\end{widetext}

\newpage
\newpage
\bibliography{references}

\providecommand{\noopsort}[1]{}\providecommand{\singleletter}[1]{#1}%
\begin{thebibliography}{109}%
\makeatletter
\providecommand \@ifxundefined [1]{%
 \@ifx{#1\undefined}
}%
\providecommand \@ifnum [1]{%
 \ifnum #1\expandafter \@firstoftwo
 \else \expandafter \@secondoftwo
 \fi
}%
\providecommand \@ifx [1]{%
 \ifx #1\expandafter \@firstoftwo
 \else \expandafter \@secondoftwo
 \fi
}%
\providecommand \natexlab [1]{#1}%
\providecommand \enquote  [1]{``#1''}%
\providecommand \bibnamefont  [1]{#1}%
\providecommand \bibfnamefont [1]{#1}%
\providecommand \citenamefont [1]{#1}%
\providecommand \href@noop [0]{\@secondoftwo}%
\providecommand \href [0]{\begingroup \@sanitize@url \@href}%
\providecommand \@href[1]{\@@startlink{#1}\@@href}%
\providecommand \@@href[1]{\endgroup#1\@@endlink}%
\providecommand \@sanitize@url [0]{\catcode `\\12\catcode `\$12\catcode
  `\&12\catcode `\#12\catcode `\^12\catcode `\_12\catcode `\%12\relax}%
\providecommand \@@startlink[1]{}%
\providecommand \@@endlink[0]{}%
\providecommand \url  [0]{\begingroup\@sanitize@url \@url }%
\providecommand \@url [1]{\endgroup\@href {#1}{\urlprefix }}%
\providecommand \urlprefix  [0]{URL }%
\providecommand \Eprint [0]{\href }%
\providecommand \doibase [0]{https://doi.org/}%
\providecommand \selectlanguage [0]{\@gobble}%
\providecommand \bibinfo  [0]{\@secondoftwo}%
\providecommand \bibfield  [0]{\@secondoftwo}%
\providecommand \translation [1]{[#1]}%
\providecommand \BibitemOpen [0]{}%
\providecommand \bibitemStop [0]{}%
\providecommand \bibitemNoStop [0]{.\EOS\space}%
\providecommand \EOS [0]{\spacefactor3000\relax}%
\providecommand \BibitemShut  [1]{\csname bibitem#1\endcsname}%
\let\auto@bib@innerbib\@empty
\bibitem [{\citenamefont {Schollw\"ock}(2005)}]{RevModPhys.77.259}%
  \BibitemOpen
  \bibfield  {author} {\bibinfo {author} {\bibfnamefont {U.}~\bibnamefont
  {Schollw\"ock}},\ }\bibfield  {title} {\bibinfo {title} {The density-matrix
  renormalization group},\ }\href {https://doi.org/10.1103/RevModPhys.77.259}
  {\bibfield  {journal} {\bibinfo  {journal} {Rev. Mod. Phys.}\ }\textbf
  {\bibinfo {volume} {77}},\ \bibinfo {pages} {259} (\bibinfo {year}
  {2005})}\BibitemShut {NoStop}%
\bibitem [{\citenamefont {Francesco}\ \emph {et~al.}(2012)\citenamefont
  {Francesco}, \citenamefont {Mathieu},\ and\ \citenamefont
  {S{\'e}n{\'e}chal}}]{francesco2012conformal}%
  \BibitemOpen
  \bibfield  {author} {\bibinfo {author} {\bibfnamefont {P.}~\bibnamefont
  {Francesco}}, \bibinfo {author} {\bibfnamefont {P.}~\bibnamefont {Mathieu}},\
  and\ \bibinfo {author} {\bibfnamefont {D.}~\bibnamefont {S{\'e}n{\'e}chal}},\
  }\href@noop {} {\emph {\bibinfo {title} {Conformal field theory}}}\ (\bibinfo
   {publisher} {Springer Science \& Business Media},\ \bibinfo {year}
  {2012})\BibitemShut {NoStop}%
\bibitem [{\citenamefont {Mora}\ and\ \citenamefont
  {Castin}(2003)}]{PhysRevA.67.053615}%
  \BibitemOpen
  \bibfield  {author} {\bibinfo {author} {\bibfnamefont {C.}~\bibnamefont
  {Mora}}\ and\ \bibinfo {author} {\bibfnamefont {Y.}~\bibnamefont {Castin}},\
  }\bibfield  {title} {\bibinfo {title} {Extension of bogoliubov theory to
  quasicondensates},\ }\href {https://doi.org/10.1103/PhysRevA.67.053615}
  {\bibfield  {journal} {\bibinfo  {journal} {Phys. Rev. A}\ }\textbf {\bibinfo
  {volume} {67}},\ \bibinfo {pages} {053615} (\bibinfo {year}
  {2003})}\BibitemShut {NoStop}%
\bibitem [{\citenamefont {Caux}(2009)}]{caux2009correlation}%
  \BibitemOpen
  \bibfield  {author} {\bibinfo {author} {\bibfnamefont {J.-S.}\ \bibnamefont
  {Caux}},\ }\bibfield  {title} {\bibinfo {title} {Correlation functions of
  integrable models: A description of the abacus algorithm},\ }\href@noop {}
  {\bibfield  {journal} {\bibinfo  {journal} {Journal of mathematical physics}\
  }\textbf {\bibinfo {volume} {50}},\ \bibinfo {pages} {095214} (\bibinfo
  {year} {2009})}\BibitemShut {NoStop}%
\bibitem [{\citenamefont {Konik}\ and\ \citenamefont
  {Adamov}(2007)}]{PhysRevLett.98.147205}%
  \BibitemOpen
  \bibfield  {author} {\bibinfo {author} {\bibfnamefont {R.~M.}\ \bibnamefont
  {Konik}}\ and\ \bibinfo {author} {\bibfnamefont {Y.}~\bibnamefont {Adamov}},\
  }\bibfield  {title} {\bibinfo {title} {Numerical renormalization group for
  continuum one-dimensional systems},\ }\href
  {https://doi.org/10.1103/PhysRevLett.98.147205} {\bibfield  {journal}
  {\bibinfo  {journal} {Phys. Rev. Lett.}\ }\textbf {\bibinfo {volume} {98}},\
  \bibinfo {pages} {147205} (\bibinfo {year} {2007})}\BibitemShut {NoStop}%
\bibitem [{\citenamefont {Panfil}\ and\ \citenamefont
  {Caux}(2014)}]{PhysRevA.89.033605}%
  \BibitemOpen
  \bibfield  {author} {\bibinfo {author} {\bibfnamefont {M.}~\bibnamefont
  {Panfil}}\ and\ \bibinfo {author} {\bibfnamefont {J.-S.}\ \bibnamefont
  {Caux}},\ }\bibfield  {title} {\bibinfo {title} {Finite-temperature
  correlations in the lieb-liniger one-dimensional bose gas},\ }\href
  {https://doi.org/10.1103/PhysRevA.89.033605} {\bibfield  {journal} {\bibinfo
  {journal} {Phys. Rev. A}\ }\textbf {\bibinfo {volume} {89}},\ \bibinfo
  {pages} {033605} (\bibinfo {year} {2014})}\BibitemShut {NoStop}%
\bibitem [{\citenamefont {Caux}(2016)}]{Caux_2016}%
  \BibitemOpen
  \bibfield  {author} {\bibinfo {author} {\bibfnamefont {J.-S.}\ \bibnamefont
  {Caux}},\ }\bibfield  {title} {\bibinfo {title} {The quench action},\ }\href
  {https://doi.org/10.1088/1742-5468/2016/06/064006} {\bibfield  {journal}
  {\bibinfo  {journal} {Journal of Statistical Mechanics: Theory and
  Experiment}\ }\textbf {\bibinfo {volume} {2016}},\ \bibinfo {pages} {064006}
  (\bibinfo {year} {2016})}\BibitemShut {NoStop}%
\bibitem [{\citenamefont {Kinoshita}\ \emph {et~al.}(2006)\citenamefont
  {Kinoshita}, \citenamefont {Wenger},\ and\ \citenamefont
  {Weiss}}]{kinoshita2006quantum}%
  \BibitemOpen
  \bibfield  {author} {\bibinfo {author} {\bibfnamefont {T.}~\bibnamefont
  {Kinoshita}}, \bibinfo {author} {\bibfnamefont {T.}~\bibnamefont {Wenger}},\
  and\ \bibinfo {author} {\bibfnamefont {D.~S.}\ \bibnamefont {Weiss}},\
  }\bibfield  {title} {\bibinfo {title} {A quantum newton's cradle},\ }\href
  {https://doi.org/10.1038/nature04693} {\bibfield  {journal} {\bibinfo
  {journal} {Nature}\ }\textbf {\bibinfo {volume} {440}},\ \bibinfo {pages}
  {900} (\bibinfo {year} {2006})}\BibitemShut {NoStop}%
\bibitem [{\citenamefont {Bloch}\ \emph {et~al.}(2008)\citenamefont {Bloch},
  \citenamefont {Dalibard},\ and\ \citenamefont {Zwerger}}]{RevModPhys.80.885}%
  \BibitemOpen
  \bibfield  {author} {\bibinfo {author} {\bibfnamefont {I.}~\bibnamefont
  {Bloch}}, \bibinfo {author} {\bibfnamefont {J.}~\bibnamefont {Dalibard}},\
  and\ \bibinfo {author} {\bibfnamefont {W.}~\bibnamefont {Zwerger}},\
  }\bibfield  {title} {\bibinfo {title} {Many-body physics with ultracold
  gases},\ }\href {https://doi.org/10.1103/RevModPhys.80.885} {\bibfield
  {journal} {\bibinfo  {journal} {Rev. Mod. Phys.}\ }\textbf {\bibinfo {volume}
  {80}},\ \bibinfo {pages} {885} (\bibinfo {year} {2008})}\BibitemShut
  {NoStop}%
\bibitem [{\citenamefont {Greiner}\ \emph {et~al.}(2002)\citenamefont
  {Greiner}, \citenamefont {Mandel}, \citenamefont {H{\"a}nsch},\ and\
  \citenamefont {Bloch}}]{greiner2002collapse}%
  \BibitemOpen
  \bibfield  {author} {\bibinfo {author} {\bibfnamefont {M.}~\bibnamefont
  {Greiner}}, \bibinfo {author} {\bibfnamefont {O.}~\bibnamefont {Mandel}},
  \bibinfo {author} {\bibfnamefont {T.~W.}\ \bibnamefont {H{\"a}nsch}},\ and\
  \bibinfo {author} {\bibfnamefont {I.}~\bibnamefont {Bloch}},\ }\bibfield
  {title} {\bibinfo {title} {Collapse and revival of the matter wave field of a
  bose--einstein condensate},\ }\href {https://doi.org/10.1038/nature00968}
  {\bibfield  {journal} {\bibinfo  {journal} {Nature}\ }\textbf {\bibinfo
  {volume} {419}},\ \bibinfo {pages} {51} (\bibinfo {year} {2002})}\BibitemShut
  {NoStop}%
\bibitem [{\citenamefont {Langen}\ \emph {et~al.}(2013)\citenamefont {Langen},
  \citenamefont {Geiger}, \citenamefont {Kuhnert}, \citenamefont {Rauer},\ and\
  \citenamefont {Schmiedmayer}}]{langen2013local}%
  \BibitemOpen
  \bibfield  {author} {\bibinfo {author} {\bibfnamefont {T.}~\bibnamefont
  {Langen}}, \bibinfo {author} {\bibfnamefont {R.}~\bibnamefont {Geiger}},
  \bibinfo {author} {\bibfnamefont {M.}~\bibnamefont {Kuhnert}}, \bibinfo
  {author} {\bibfnamefont {B.}~\bibnamefont {Rauer}},\ and\ \bibinfo {author}
  {\bibfnamefont {J.}~\bibnamefont {Schmiedmayer}},\ }\bibfield  {title}
  {\bibinfo {title} {Local emergence of thermal correlations in an isolated
  quantum many-body system},\ }\href {https://doi.org/10.1038/nphys2739}
  {\bibfield  {journal} {\bibinfo  {journal} {Nature Physics}\ }\textbf
  {\bibinfo {volume} {9}},\ \bibinfo {pages} {640} (\bibinfo {year}
  {2013})}\BibitemShut {NoStop}%
\bibitem [{\citenamefont {Gring}\ \emph {et~al.}(2012)\citenamefont {Gring},
  \citenamefont {Kuhnert}, \citenamefont {Langen}, \citenamefont {Kitagawa},
  \citenamefont {Rauer}, \citenamefont {Schreitl}, \citenamefont {Mazets},
  \citenamefont {Smith}, \citenamefont {Demler},\ and\ \citenamefont
  {Schmiedmayer}}]{Gring1318}%
  \BibitemOpen
  \bibfield  {author} {\bibinfo {author} {\bibfnamefont {M.}~\bibnamefont
  {Gring}}, \bibinfo {author} {\bibfnamefont {M.}~\bibnamefont {Kuhnert}},
  \bibinfo {author} {\bibfnamefont {T.}~\bibnamefont {Langen}}, \bibinfo
  {author} {\bibfnamefont {T.}~\bibnamefont {Kitagawa}}, \bibinfo {author}
  {\bibfnamefont {B.}~\bibnamefont {Rauer}}, \bibinfo {author} {\bibfnamefont
  {M.}~\bibnamefont {Schreitl}}, \bibinfo {author} {\bibfnamefont
  {I.}~\bibnamefont {Mazets}}, \bibinfo {author} {\bibfnamefont {D.~A.}\
  \bibnamefont {Smith}}, \bibinfo {author} {\bibfnamefont {E.}~\bibnamefont
  {Demler}},\ and\ \bibinfo {author} {\bibfnamefont {J.}~\bibnamefont
  {Schmiedmayer}},\ }\bibfield  {title} {\bibinfo {title} {Relaxation and
  prethermalization in an isolated quantum system},\ }\href
  {https://doi.org/10.1126/science.1224953} {\bibfield  {journal} {\bibinfo
  {journal} {Science}\ }\textbf {\bibinfo {volume} {337}},\ \bibinfo {pages}
  {1318} (\bibinfo {year} {2012})}\BibitemShut {NoStop}%
\bibitem [{\citenamefont {Paredes}\ \emph {et~al.}(2004)\citenamefont
  {Paredes}, \citenamefont {Widera}, \citenamefont {Murg}, \citenamefont
  {Mandel}, \citenamefont {F{\"o}lling}, \citenamefont {Cirac}, \citenamefont
  {Shlyapnikov}, \citenamefont {H{\"a}nsch},\ and\ \citenamefont
  {Bloch}}]{paredes2004tonks}%
  \BibitemOpen
  \bibfield  {author} {\bibinfo {author} {\bibfnamefont {B.}~\bibnamefont
  {Paredes}}, \bibinfo {author} {\bibfnamefont {A.}~\bibnamefont {Widera}},
  \bibinfo {author} {\bibfnamefont {V.}~\bibnamefont {Murg}}, \bibinfo {author}
  {\bibfnamefont {O.}~\bibnamefont {Mandel}}, \bibinfo {author} {\bibfnamefont
  {S.}~\bibnamefont {F{\"o}lling}}, \bibinfo {author} {\bibfnamefont
  {I.}~\bibnamefont {Cirac}}, \bibinfo {author} {\bibfnamefont {G.~V.}\
  \bibnamefont {Shlyapnikov}}, \bibinfo {author} {\bibfnamefont {T.~W.}\
  \bibnamefont {H{\"a}nsch}},\ and\ \bibinfo {author} {\bibfnamefont
  {I.}~\bibnamefont {Bloch}},\ }\bibfield  {title} {\bibinfo {title}
  {Tonks--girardeau gas of ultracold atoms in an optical lattice},\ }\href
  {https://doi.org/10.1038/nature02530} {\bibfield  {journal} {\bibinfo
  {journal} {Nature}\ }\textbf {\bibinfo {volume} {429}},\ \bibinfo {pages}
  {277} (\bibinfo {year} {2004})}\BibitemShut {NoStop}%
\bibitem [{\citenamefont {Meinert}\ \emph {et~al.}(2015)\citenamefont
  {Meinert}, \citenamefont {Panfil}, \citenamefont {Mark}, \citenamefont
  {Lauber}, \citenamefont {Caux},\ and\ \citenamefont
  {N\"agerl}}]{PhysRevLett.115.085301}%
  \BibitemOpen
  \bibfield  {author} {\bibinfo {author} {\bibfnamefont {F.}~\bibnamefont
  {Meinert}}, \bibinfo {author} {\bibfnamefont {M.}~\bibnamefont {Panfil}},
  \bibinfo {author} {\bibfnamefont {M.~J.}\ \bibnamefont {Mark}}, \bibinfo
  {author} {\bibfnamefont {K.}~\bibnamefont {Lauber}}, \bibinfo {author}
  {\bibfnamefont {J.-S.}\ \bibnamefont {Caux}},\ and\ \bibinfo {author}
  {\bibfnamefont {H.-C.}\ \bibnamefont {N\"agerl}},\ }\bibfield  {title}
  {\bibinfo {title} {Probing the excitations of a lieb-liniger gas from weak to
  strong coupling},\ }\href {https://doi.org/10.1103/PhysRevLett.115.085301}
  {\bibfield  {journal} {\bibinfo  {journal} {Phys. Rev. Lett.}\ }\textbf
  {\bibinfo {volume} {115}},\ \bibinfo {pages} {085301} (\bibinfo {year}
  {2015})}\BibitemShut {NoStop}%
\bibitem [{\citenamefont {Meinert}\ \emph {et~al.}(2013)\citenamefont
  {Meinert}, \citenamefont {Mark}, \citenamefont {Kirilov}, \citenamefont
  {Lauber}, \citenamefont {Weinmann}, \citenamefont {Daley},\ and\
  \citenamefont {N\"agerl}}]{PhysRevLett.111.053003}%
  \BibitemOpen
  \bibfield  {author} {\bibinfo {author} {\bibfnamefont {F.}~\bibnamefont
  {Meinert}}, \bibinfo {author} {\bibfnamefont {M.~J.}\ \bibnamefont {Mark}},
  \bibinfo {author} {\bibfnamefont {E.}~\bibnamefont {Kirilov}}, \bibinfo
  {author} {\bibfnamefont {K.}~\bibnamefont {Lauber}}, \bibinfo {author}
  {\bibfnamefont {P.}~\bibnamefont {Weinmann}}, \bibinfo {author}
  {\bibfnamefont {A.~J.}\ \bibnamefont {Daley}},\ and\ \bibinfo {author}
  {\bibfnamefont {H.-C.}\ \bibnamefont {N\"agerl}},\ }\bibfield  {title}
  {\bibinfo {title} {Quantum quench in an atomic one-dimensional ising chain},\
  }\href {https://doi.org/10.1103/PhysRevLett.111.053003} {\bibfield  {journal}
  {\bibinfo  {journal} {Phys. Rev. Lett.}\ }\textbf {\bibinfo {volume} {111}},\
  \bibinfo {pages} {053003} (\bibinfo {year} {2013})}\BibitemShut {NoStop}%
\bibitem [{\citenamefont {Schweigler}\ \emph {et~al.}(2017)\citenamefont
  {Schweigler}, \citenamefont {Kasper}, \citenamefont {Erne}, \citenamefont
  {Mazets}, \citenamefont {Rauer}, \citenamefont {Cataldini}, \citenamefont
  {Langen}, \citenamefont {Gasenzer}, \citenamefont {Berges},\ and\
  \citenamefont {Schmiedmayer}}]{schweigler2017experimental}%
  \BibitemOpen
  \bibfield  {author} {\bibinfo {author} {\bibfnamefont {T.}~\bibnamefont
  {Schweigler}}, \bibinfo {author} {\bibfnamefont {V.}~\bibnamefont {Kasper}},
  \bibinfo {author} {\bibfnamefont {S.}~\bibnamefont {Erne}}, \bibinfo {author}
  {\bibfnamefont {I.}~\bibnamefont {Mazets}}, \bibinfo {author} {\bibfnamefont
  {B.}~\bibnamefont {Rauer}}, \bibinfo {author} {\bibfnamefont
  {F.}~\bibnamefont {Cataldini}}, \bibinfo {author} {\bibfnamefont
  {T.}~\bibnamefont {Langen}}, \bibinfo {author} {\bibfnamefont
  {T.}~\bibnamefont {Gasenzer}}, \bibinfo {author} {\bibfnamefont
  {J.}~\bibnamefont {Berges}},\ and\ \bibinfo {author} {\bibfnamefont
  {J.}~\bibnamefont {Schmiedmayer}},\ }\bibfield  {title} {\bibinfo {title}
  {Experimental characterization of a quantum many-body system via higher-order
  correlations},\ }\href {https://doi.org/10.1038/nature22310} {\bibfield
  {journal} {\bibinfo  {journal} {Nature}\ }\textbf {\bibinfo {volume} {545}},\
  \bibinfo {pages} {323} (\bibinfo {year} {2017})}\BibitemShut {NoStop}%
\bibitem [{\citenamefont {Fabbri}\ \emph {et~al.}(2015)\citenamefont {Fabbri},
  \citenamefont {Panfil}, \citenamefont {Cl\'ement}, \citenamefont {Fallani},
  \citenamefont {Inguscio}, \citenamefont {Fort},\ and\ \citenamefont
  {Caux}}]{Fabbri2015Dynamical}%
  \BibitemOpen
  \bibfield  {author} {\bibinfo {author} {\bibfnamefont {N.}~\bibnamefont
  {Fabbri}}, \bibinfo {author} {\bibfnamefont {M.}~\bibnamefont {Panfil}},
  \bibinfo {author} {\bibfnamefont {D.}~\bibnamefont {Cl\'ement}}, \bibinfo
  {author} {\bibfnamefont {L.}~\bibnamefont {Fallani}}, \bibinfo {author}
  {\bibfnamefont {M.}~\bibnamefont {Inguscio}}, \bibinfo {author}
  {\bibfnamefont {C.}~\bibnamefont {Fort}},\ and\ \bibinfo {author}
  {\bibfnamefont {J.-S.}\ \bibnamefont {Caux}},\ }\bibfield  {title} {\bibinfo
  {title} {Dynamical structure factor of one-dimensional bose gases:
  Experimental signatures of beyond-luttinger-liquid physics},\ }\href
  {https://doi.org/10.1103/PhysRevA.91.043617} {\bibfield  {journal} {\bibinfo
  {journal} {Phys. Rev. A}\ }\textbf {\bibinfo {volume} {91}},\ \bibinfo
  {pages} {043617} (\bibinfo {year} {2015})}\BibitemShut {NoStop}%
\bibitem [{\citenamefont {Langen}\ \emph {et~al.}(2015)\citenamefont {Langen},
  \citenamefont {Erne}, \citenamefont {Geiger}, \citenamefont {Rauer},
  \citenamefont {Schweigler}, \citenamefont {Kuhnert}, \citenamefont
  {Rohringer}, \citenamefont {Mazets}, \citenamefont {Gasenzer},\ and\
  \citenamefont {Schmiedmayer}}]{Langen207}%
  \BibitemOpen
  \bibfield  {author} {\bibinfo {author} {\bibfnamefont {T.}~\bibnamefont
  {Langen}}, \bibinfo {author} {\bibfnamefont {S.}~\bibnamefont {Erne}},
  \bibinfo {author} {\bibfnamefont {R.}~\bibnamefont {Geiger}}, \bibinfo
  {author} {\bibfnamefont {B.}~\bibnamefont {Rauer}}, \bibinfo {author}
  {\bibfnamefont {T.}~\bibnamefont {Schweigler}}, \bibinfo {author}
  {\bibfnamefont {M.}~\bibnamefont {Kuhnert}}, \bibinfo {author} {\bibfnamefont
  {W.}~\bibnamefont {Rohringer}}, \bibinfo {author} {\bibfnamefont {I.~E.}\
  \bibnamefont {Mazets}}, \bibinfo {author} {\bibfnamefont {T.}~\bibnamefont
  {Gasenzer}},\ and\ \bibinfo {author} {\bibfnamefont {J.}~\bibnamefont
  {Schmiedmayer}},\ }\bibfield  {title} {\bibinfo {title} {Experimental
  observation of a generalized gibbs ensemble},\ }\href
  {https://doi.org/10.1126/science.1257026} {\bibfield  {journal} {\bibinfo
  {journal} {Science}\ }\textbf {\bibinfo {volume} {348}},\ \bibinfo {pages}
  {207} (\bibinfo {year} {2015})}\BibitemShut {NoStop}%
\bibitem [{\citenamefont {Schemmer}\ \emph {et~al.}(2019)\citenamefont
  {Schemmer}, \citenamefont {Bouchoule}, \citenamefont {Doyon},\ and\
  \citenamefont {Dubail}}]{schemmer2019generalized}%
  \BibitemOpen
  \bibfield  {author} {\bibinfo {author} {\bibfnamefont {M.}~\bibnamefont
  {Schemmer}}, \bibinfo {author} {\bibfnamefont {I.}~\bibnamefont {Bouchoule}},
  \bibinfo {author} {\bibfnamefont {B.}~\bibnamefont {Doyon}},\ and\ \bibinfo
  {author} {\bibfnamefont {J.}~\bibnamefont {Dubail}},\ }\bibfield  {title}
  {\bibinfo {title} {Generalized hydrodynamics on an atom chip},\ }\href
  {https://doi.org/10.1103/PhysRevLett.122.090601} {\bibfield  {journal}
  {\bibinfo  {journal} {Physical review letters}\ }\textbf {\bibinfo {volume}
  {122}},\ \bibinfo {pages} {090601} (\bibinfo {year} {2019})}\BibitemShut
  {NoStop}%
\bibitem [{\citenamefont {Polkovnikov}\ \emph {et~al.}(2011)\citenamefont
  {Polkovnikov}, \citenamefont {Sengupta}, \citenamefont {Silva},\ and\
  \citenamefont {Vengalattore}}]{Polkovnikov2011}%
  \BibitemOpen
  \bibfield  {author} {\bibinfo {author} {\bibfnamefont {A.}~\bibnamefont
  {Polkovnikov}}, \bibinfo {author} {\bibfnamefont {K.}~\bibnamefont
  {Sengupta}}, \bibinfo {author} {\bibfnamefont {A.}~\bibnamefont {Silva}},\
  and\ \bibinfo {author} {\bibfnamefont {M.}~\bibnamefont {Vengalattore}},\
  }\bibfield  {title} {\bibinfo {title} {Colloquium: Nonequilibrium dynamics of
  closed interacting quantum systems},\ }\href
  {https://doi.org/10.1103/RevModPhys.83.863} {\bibfield  {journal} {\bibinfo
  {journal} {Rev. Mod. Phys.}\ }\textbf {\bibinfo {volume} {83}},\ \bibinfo
  {pages} {863} (\bibinfo {year} {2011})}\BibitemShut {NoStop}%
\bibitem [{\citenamefont {Schweigler}\ \emph {et~al.}(2020)\citenamefont
  {Schweigler}, \citenamefont {Gluza}, \citenamefont {Tajik}, \citenamefont
  {Sotiriadis}, \citenamefont {Cataldini}, \citenamefont {Ji}, \citenamefont
  {M{\o}ller}, \citenamefont {Sabino}, \citenamefont {Rauer}, \citenamefont
  {Eisert} \emph {et~al.}}]{schweigler2020decay}%
  \BibitemOpen
  \bibfield  {author} {\bibinfo {author} {\bibfnamefont {T.}~\bibnamefont
  {Schweigler}}, \bibinfo {author} {\bibfnamefont {M.}~\bibnamefont {Gluza}},
  \bibinfo {author} {\bibfnamefont {M.}~\bibnamefont {Tajik}}, \bibinfo
  {author} {\bibfnamefont {S.}~\bibnamefont {Sotiriadis}}, \bibinfo {author}
  {\bibfnamefont {F.}~\bibnamefont {Cataldini}}, \bibinfo {author}
  {\bibfnamefont {S.-C.}\ \bibnamefont {Ji}}, \bibinfo {author} {\bibfnamefont
  {F.~S.}\ \bibnamefont {M{\o}ller}}, \bibinfo {author} {\bibfnamefont
  {J.}~\bibnamefont {Sabino}}, \bibinfo {author} {\bibfnamefont
  {B.}~\bibnamefont {Rauer}}, \bibinfo {author} {\bibfnamefont
  {J.}~\bibnamefont {Eisert}}, \emph {et~al.},\ }\bibfield  {title} {\bibinfo
  {title} {Decay and recurrence of non-gaussian correlations in a quantum
  many-body system},\ }\href@noop {} {\bibfield  {journal} {\bibinfo  {journal}
  {arXiv preprint arXiv:2003.01808}\ } (\bibinfo {year} {2020})}\BibitemShut
  {NoStop}%
\bibitem [{\citenamefont {Kinoshita}\ \emph {et~al.}(2004)\citenamefont
  {Kinoshita}, \citenamefont {Wenger},\ and\ \citenamefont
  {Weiss}}]{Kinoshita1125}%
  \BibitemOpen
  \bibfield  {author} {\bibinfo {author} {\bibfnamefont {T.}~\bibnamefont
  {Kinoshita}}, \bibinfo {author} {\bibfnamefont {T.}~\bibnamefont {Wenger}},\
  and\ \bibinfo {author} {\bibfnamefont {D.~S.}\ \bibnamefont {Weiss}},\
  }\bibfield  {title} {\bibinfo {title} {Observation of a one-dimensional
  tonks-girardeau gas},\ }\href {https://doi.org/10.1126/science.1100700}
  {\bibfield  {journal} {\bibinfo  {journal} {Science}\ }\textbf {\bibinfo
  {volume} {305}},\ \bibinfo {pages} {1125} (\bibinfo {year}
  {2004})}\BibitemShut {NoStop}%
\bibitem [{\citenamefont {Li}\ \emph {et~al.}(2020)\citenamefont {Li},
  \citenamefont {Zhou}, \citenamefont {Mazets}, \citenamefont {Stimming},
  \citenamefont {M{\o}ller}, \citenamefont {Zhu}, \citenamefont {Zhai},
  \citenamefont {Xiong}, \citenamefont {Zhou}, \citenamefont {Chen},\ and\
  \citenamefont {Schmiedmayer}}]{Li2018}%
  \BibitemOpen
  \bibfield  {author} {\bibinfo {author} {\bibfnamefont {C.}~\bibnamefont
  {Li}}, \bibinfo {author} {\bibfnamefont {T.}~\bibnamefont {Zhou}}, \bibinfo
  {author} {\bibfnamefont {I.}~\bibnamefont {Mazets}}, \bibinfo {author}
  {\bibfnamefont {H.-P.}\ \bibnamefont {Stimming}}, \bibinfo {author}
  {\bibfnamefont {F.~S.}\ \bibnamefont {M{\o}ller}}, \bibinfo {author}
  {\bibfnamefont {Z.}~\bibnamefont {Zhu}}, \bibinfo {author} {\bibfnamefont
  {Y.}~\bibnamefont {Zhai}}, \bibinfo {author} {\bibfnamefont {W.}~\bibnamefont
  {Xiong}}, \bibinfo {author} {\bibfnamefont {X.}~\bibnamefont {Zhou}},
  \bibinfo {author} {\bibfnamefont {X.}~\bibnamefont {Chen}},\ and\ \bibinfo
  {author} {\bibfnamefont {J.}~\bibnamefont {Schmiedmayer}},\ }\bibfield
  {title} {\bibinfo {title} {Relaxation of bosons in one dimension and the
  onset of dimensional crossover},\ }\href
  {https://doi.org/10.21468/SciPostPhys.9.4.058} {\bibfield  {journal}
  {\bibinfo  {journal} {SciPost Phys.}\ }\textbf {\bibinfo {volume} {9}},\
  \bibinfo {pages} {58} (\bibinfo {year} {2020})}\BibitemShut {NoStop}%
\bibitem [{\citenamefont {Haller}\ \emph {et~al.}(2009)\citenamefont {Haller},
  \citenamefont {Gustavsson}, \citenamefont {Mark}, \citenamefont {Danzl},
  \citenamefont {Hart}, \citenamefont {Pupillo},\ and\ \citenamefont
  {N\"{a}gerl}}]{Haller2009}%
  \BibitemOpen
  \bibfield  {author} {\bibinfo {author} {\bibfnamefont {E.}~\bibnamefont
  {Haller}}, \bibinfo {author} {\bibfnamefont {M.}~\bibnamefont {Gustavsson}},
  \bibinfo {author} {\bibfnamefont {M.~J.}\ \bibnamefont {Mark}}, \bibinfo
  {author} {\bibfnamefont {J.~G.}\ \bibnamefont {Danzl}}, \bibinfo {author}
  {\bibfnamefont {R.}~\bibnamefont {Hart}}, \bibinfo {author} {\bibfnamefont
  {G.}~\bibnamefont {Pupillo}},\ and\ \bibinfo {author} {\bibfnamefont {H.-C.}\
  \bibnamefont {N\"{a}gerl}},\ }\bibfield  {title} {\bibinfo {title}
  {Realization of an excited, strongly correlated quantum gas phase},\ }\href
  {https://doi.org/10.1126/science.1175850} {\bibfield  {journal} {\bibinfo
  {journal} {Science}\ }\textbf {\bibinfo {volume} {325}},\ \bibinfo {pages}
  {1224–1227} (\bibinfo {year} {2009})}\BibitemShut {NoStop}%
\bibitem [{\citenamefont {Rigol}\ \emph {et~al.}(2007)\citenamefont {Rigol},
  \citenamefont {Dunjko}, \citenamefont {Yurovsky},\ and\ \citenamefont
  {Olshanii}}]{rigol2007relaxation}%
  \BibitemOpen
  \bibfield  {author} {\bibinfo {author} {\bibfnamefont {M.}~\bibnamefont
  {Rigol}}, \bibinfo {author} {\bibfnamefont {V.}~\bibnamefont {Dunjko}},
  \bibinfo {author} {\bibfnamefont {V.}~\bibnamefont {Yurovsky}},\ and\
  \bibinfo {author} {\bibfnamefont {M.}~\bibnamefont {Olshanii}},\ }\bibfield
  {title} {\bibinfo {title} {Relaxation in a completely integrable many-body
  quantum system: An ab initio study of the dynamics of the highly excited
  states of 1d lattice hard-core bosons},\ }\href
  {https://doi.org/10.1103/PhysRevLett.98.050405} {\bibfield  {journal}
  {\bibinfo  {journal} {Phys. Rev. Lett.}\ }\textbf {\bibinfo {volume} {98}},\
  \bibinfo {pages} {050405} (\bibinfo {year} {2007})}\BibitemShut {NoStop}%
\bibitem [{\citenamefont {Rigol}(2009)}]{rigol2009breakdown}%
  \BibitemOpen
  \bibfield  {author} {\bibinfo {author} {\bibfnamefont {M.}~\bibnamefont
  {Rigol}},\ }\bibfield  {title} {\bibinfo {title} {Breakdown of thermalization
  in finite one-dimensional systems},\ }\href@noop {} {\bibfield  {journal}
  {\bibinfo  {journal} {Physical review letters}\ }\textbf {\bibinfo {volume}
  {103}},\ \bibinfo {pages} {100403} (\bibinfo {year} {2009})}\BibitemShut
  {NoStop}%
\bibitem [{\citenamefont {Rigol}\ \emph {et~al.}(2008)\citenamefont {Rigol},
  \citenamefont {Dunjko},\ and\ \citenamefont {Olshanii}}]{Rigol2008}%
  \BibitemOpen
  \bibfield  {author} {\bibinfo {author} {\bibfnamefont {M.}~\bibnamefont
  {Rigol}}, \bibinfo {author} {\bibfnamefont {V.}~\bibnamefont {Dunjko}},\ and\
  \bibinfo {author} {\bibfnamefont {M.}~\bibnamefont {Olshanii}},\ }\bibfield
  {title} {\bibinfo {title} {{Thermalization and its mechanism for generic
  isolated quantum systems}},\ }\href {https://doi.org/10.1038/nature06838}
  {\bibfield  {journal} {\bibinfo  {journal} {Nature (London)}\ }\textbf
  {\bibinfo {volume} {452}},\ \bibinfo {pages} {854} (\bibinfo {year}
  {2008})}\BibitemShut {NoStop}%
\bibitem [{\citenamefont {Gogolin}\ and\ \citenamefont
  {Eisert}(2016)}]{Gogolin2016}%
  \BibitemOpen
  \bibfield  {author} {\bibinfo {author} {\bibfnamefont {C.}~\bibnamefont
  {Gogolin}}\ and\ \bibinfo {author} {\bibfnamefont {J.}~\bibnamefont
  {Eisert}},\ }\bibfield  {title} {\bibinfo {title} {Equilibration,
  thermalisation, and the emergence of statistical mechanics in closed quantum
  systems},\ }\href {http://stacks.iop.org/0034-4885/79/i=5/a=056001}
  {\bibfield  {journal} {\bibinfo  {journal} {Reports on Progress in Physics}\
  }\textbf {\bibinfo {volume} {79}},\ \bibinfo {pages} {056001} (\bibinfo
  {year} {2016})}\BibitemShut {NoStop}%
\bibitem [{\citenamefont {Caux}\ and\ \citenamefont
  {Essler}(2013)}]{PhysRevLett.110.257203}%
  \BibitemOpen
  \bibfield  {author} {\bibinfo {author} {\bibfnamefont {J.-S.}\ \bibnamefont
  {Caux}}\ and\ \bibinfo {author} {\bibfnamefont {F.~H.~L.}\ \bibnamefont
  {Essler}},\ }\bibfield  {title} {\bibinfo {title} {Time evolution of local
  observables after quenching to an integrable model},\ }\href
  {https://doi.org/10.1103/PhysRevLett.110.257203} {\bibfield  {journal}
  {\bibinfo  {journal} {Phys. Rev. Lett.}\ }\textbf {\bibinfo {volume} {110}},\
  \bibinfo {pages} {257203} (\bibinfo {year} {2013})}\BibitemShut {NoStop}%
\bibitem [{\citenamefont {Wigner}(1955)}]{wigner1955scattering}%
  \BibitemOpen
  \bibfield  {author} {\bibinfo {author} {\bibfnamefont {E.~P.}\ \bibnamefont
  {Wigner}},\ }\bibfield  {title} {\bibinfo {title} {Lower limit for the energy
  derivative of the scattering phase shift},\ }\href
  {https://doi.org/10.1103/PhysRev.98.145} {\bibfield  {journal} {\bibinfo
  {journal} {Phys. Rev.}\ }\textbf {\bibinfo {volume} {98}},\ \bibinfo {pages}
  {145} (\bibinfo {year} {1955})}\BibitemShut {NoStop}%
\bibitem [{\citenamefont {Boldrighini}\ \emph {et~al.}(1983)\citenamefont
  {Boldrighini}, \citenamefont {Dobrushin},\ and\ \citenamefont
  {Sukhov}}]{boldrighini1983one}%
  \BibitemOpen
  \bibfield  {author} {\bibinfo {author} {\bibfnamefont {C.}~\bibnamefont
  {Boldrighini}}, \bibinfo {author} {\bibfnamefont {R.}~\bibnamefont
  {Dobrushin}},\ and\ \bibinfo {author} {\bibfnamefont {Y.~M.}\ \bibnamefont
  {Sukhov}},\ }\bibfield  {title} {\bibinfo {title} {One-dimensional hard rod
  caricature of hydrodynamics},\ }\href@noop {} {\bibfield  {journal} {\bibinfo
   {journal} {Journal of Statistical Physics}\ }\textbf {\bibinfo {volume}
  {31}},\ \bibinfo {pages} {577} (\bibinfo {year} {1983})}\BibitemShut
  {NoStop}%
\bibitem [{\citenamefont
  {Mazets}(2011{\natexlab{a}})}]{mazets2011integrability}%
  \BibitemOpen
  \bibfield  {author} {\bibinfo {author} {\bibfnamefont {I.~E.}\ \bibnamefont
  {Mazets}},\ }\bibfield  {title} {\bibinfo {title} {Integrability breakdown in
  longitudinaly trapped, one-dimensional bosonic gases},\ }\href@noop {}
  {\bibfield  {journal} {\bibinfo  {journal} {The European Physical Journal D}\
  }\textbf {\bibinfo {volume} {65}},\ \bibinfo {pages} {43} (\bibinfo {year}
  {2011}{\natexlab{a}})}\BibitemShut {NoStop}%
\bibitem [{\citenamefont {Castro-Alvaredo}\ \emph {et~al.}(2016)\citenamefont
  {Castro-Alvaredo}, \citenamefont {Doyon},\ and\ \citenamefont
  {Yoshimura}}]{castro2016emergent}%
  \BibitemOpen
  \bibfield  {author} {\bibinfo {author} {\bibfnamefont {O.~A.}\ \bibnamefont
  {Castro-Alvaredo}}, \bibinfo {author} {\bibfnamefont {B.}~\bibnamefont
  {Doyon}},\ and\ \bibinfo {author} {\bibfnamefont {T.}~\bibnamefont
  {Yoshimura}},\ }\bibfield  {title} {\bibinfo {title} {Emergent hydrodynamics
  in integrable quantum systems out of equilibrium},\ }\href
  {https://doi.org/10.1103/PhysRevX.6.041065} {\bibfield  {journal} {\bibinfo
  {journal} {Physical Review X}\ }\textbf {\bibinfo {volume} {6}},\ \bibinfo
  {pages} {041065} (\bibinfo {year} {2016})}\BibitemShut {NoStop}%
\bibitem [{\citenamefont {Bertini}\ \emph {et~al.}(2016)\citenamefont
  {Bertini}, \citenamefont {Collura}, \citenamefont {De~Nardis},\ and\
  \citenamefont {Fagotti}}]{bertini2016transport}%
  \BibitemOpen
  \bibfield  {author} {\bibinfo {author} {\bibfnamefont {B.}~\bibnamefont
  {Bertini}}, \bibinfo {author} {\bibfnamefont {M.}~\bibnamefont {Collura}},
  \bibinfo {author} {\bibfnamefont {J.}~\bibnamefont {De~Nardis}},\ and\
  \bibinfo {author} {\bibfnamefont {M.}~\bibnamefont {Fagotti}},\ }\bibfield
  {title} {\bibinfo {title} {Transport in out-of-equilibrium x x z chains:
  Exact profiles of charges and currents},\ }\href
  {https://doi.org/10.1103/PhysRevLett.117.207201} {\bibfield  {journal}
  {\bibinfo  {journal} {Physical review letters}\ }\textbf {\bibinfo {volume}
  {117}},\ \bibinfo {pages} {207201} (\bibinfo {year} {2016})}\BibitemShut
  {NoStop}%
\bibitem [{\citenamefont {Bastianello}\ \emph {et~al.}(2019)\citenamefont
  {Bastianello}, \citenamefont {Alba},\ and\ \citenamefont
  {Caux}}]{bastianello2019inhomogeneous}%
  \BibitemOpen
  \bibfield  {author} {\bibinfo {author} {\bibfnamefont {A.}~\bibnamefont
  {Bastianello}}, \bibinfo {author} {\bibfnamefont {V.}~\bibnamefont {Alba}},\
  and\ \bibinfo {author} {\bibfnamefont {J.-S.}\ \bibnamefont {Caux}},\
  }\bibfield  {title} {\bibinfo {title} {Generalized hydrodynamics with
  space-time inhomogeneous interactions},\ }\href
  {https://doi.org/10.1103/PhysRevLett.123.130602} {\bibfield  {journal}
  {\bibinfo  {journal} {Phys. Rev. Lett.}\ }\textbf {\bibinfo {volume} {123}},\
  \bibinfo {pages} {130602} (\bibinfo {year} {2019})}\BibitemShut {NoStop}%
\bibitem [{\citenamefont {Doyon}(2018)}]{doyon2018exact}%
  \BibitemOpen
  \bibfield  {author} {\bibinfo {author} {\bibfnamefont {B.}~\bibnamefont
  {Doyon}},\ }\bibfield  {title} {\bibinfo {title} {{Exact large-scale
  correlations in integrable systems out of equilibrium}},\ }\href
  {https://doi.org/10.21468/SciPostPhys.5.5.054} {\bibfield  {journal}
  {\bibinfo  {journal} {SciPost Phys.}\ }\textbf {\bibinfo {volume} {5}},\
  \bibinfo {pages} {54} (\bibinfo {year} {2018})}\BibitemShut {NoStop}%
\bibitem [{\citenamefont {Bastianello}\ \emph {et~al.}(2018)\citenamefont
  {Bastianello}, \citenamefont {Piroli},\ and\ \citenamefont
  {Calabrese}}]{bastianello2018exact}%
  \BibitemOpen
  \bibfield  {author} {\bibinfo {author} {\bibfnamefont {A.}~\bibnamefont
  {Bastianello}}, \bibinfo {author} {\bibfnamefont {L.}~\bibnamefont
  {Piroli}},\ and\ \bibinfo {author} {\bibfnamefont {P.}~\bibnamefont
  {Calabrese}},\ }\bibfield  {title} {\bibinfo {title} {Exact local
  correlations and full counting statistics for arbitrary states of the
  one-dimensional interacting bose gas},\ }\href
  {https://doi.org/10.1103/PhysRevLett.120.190601} {\bibfield  {journal}
  {\bibinfo  {journal} {Physical review letters}\ }\textbf {\bibinfo {volume}
  {120}},\ \bibinfo {pages} {190601} (\bibinfo {year} {2018})}\BibitemShut
  {NoStop}%
\bibitem [{\citenamefont {Bastianello}\ and\ \citenamefont
  {Piroli}(2018)}]{bastianello2018sinh}%
  \BibitemOpen
  \bibfield  {author} {\bibinfo {author} {\bibfnamefont {A.}~\bibnamefont
  {Bastianello}}\ and\ \bibinfo {author} {\bibfnamefont {L.}~\bibnamefont
  {Piroli}},\ }\bibfield  {title} {\bibinfo {title} {From the sinh-gordon field
  theory to the one-dimensional bose gas: exact local correlations and full
  counting statistics},\ }\href {https://doi.org/10.1088/1742-5468/aaeb48}
  {\bibfield  {journal} {\bibinfo  {journal} {Journal of Statistical Mechanics:
  Theory and Experiment}\ }\textbf {\bibinfo {volume} {2018}},\ \bibinfo
  {pages} {113104} (\bibinfo {year} {2018})}\BibitemShut {NoStop}%
\bibitem [{\citenamefont {Doyon}\ and\ \citenamefont
  {Myers}(2020)}]{doyon2020fluctuations}%
  \BibitemOpen
  \bibfield  {author} {\bibinfo {author} {\bibfnamefont {B.}~\bibnamefont
  {Doyon}}\ and\ \bibinfo {author} {\bibfnamefont {J.}~\bibnamefont {Myers}},\
  }\bibfield  {title} {\bibinfo {title} {Fluctuations in ballistic transport
  from euler hydrodynamics},\ }in\ \href@noop {} {\emph {\bibinfo {booktitle}
  {Annales Henri Poincar{\'e}}}},\ Vol.~\bibinfo {volume} {21}\ (\bibinfo
  {organization} {Springer},\ \bibinfo {year} {2020})\ pp.\ \bibinfo {pages}
  {255--302}\BibitemShut {NoStop}%
\bibitem [{\citenamefont {Møller}\ \emph {et~al.}(2020)\citenamefont
  {Møller}, \citenamefont {Perfetto}, \citenamefont {Doyon},\ and\
  \citenamefont {Schmiedmayer}}]{mller2020eulerscale}%
  \BibitemOpen
  \bibfield  {author} {\bibinfo {author} {\bibfnamefont {F.~S.}\ \bibnamefont
  {Møller}}, \bibinfo {author} {\bibfnamefont {G.}~\bibnamefont {Perfetto}},
  \bibinfo {author} {\bibfnamefont {B.}~\bibnamefont {Doyon}},\ and\ \bibinfo
  {author} {\bibfnamefont {J.}~\bibnamefont {Schmiedmayer}},\ }\href@noop {}
  {\bibinfo {title} {Euler-scale dynamical correlations in integrable systems
  with fluid motion}} (\bibinfo {year} {2020}),\ \Eprint
  {https://arxiv.org/abs/2007.00527} {arXiv:2007.00527 [cond-mat.stat-mech]}
  \BibitemShut {NoStop}%
\bibitem [{\citenamefont {Ilievski}\ and\ \citenamefont
  {De~Nardis}(2017{\natexlab{a}})}]{PhysRevLett.119.020602}%
  \BibitemOpen
  \bibfield  {author} {\bibinfo {author} {\bibfnamefont {E.}~\bibnamefont
  {Ilievski}}\ and\ \bibinfo {author} {\bibfnamefont {J.}~\bibnamefont
  {De~Nardis}},\ }\bibfield  {title} {\bibinfo {title} {Microscopic origin of
  ideal conductivity in integrable quantum models},\ }\href
  {https://doi.org/10.1103/PhysRevLett.119.020602} {\bibfield  {journal}
  {\bibinfo  {journal} {Phys. Rev. Lett.}\ }\textbf {\bibinfo {volume} {119}},\
  \bibinfo {pages} {020602} (\bibinfo {year} {2017}{\natexlab{a}})}\BibitemShut
  {NoStop}%
\bibitem [{\citenamefont {Bulchandani}\ \emph {et~al.}(2018)\citenamefont
  {Bulchandani}, \citenamefont {Vasseur}, \citenamefont {Karrasch},\ and\
  \citenamefont {Moore}}]{PhysRevB.97.045407}%
  \BibitemOpen
  \bibfield  {author} {\bibinfo {author} {\bibfnamefont {V.~B.}\ \bibnamefont
  {Bulchandani}}, \bibinfo {author} {\bibfnamefont {R.}~\bibnamefont
  {Vasseur}}, \bibinfo {author} {\bibfnamefont {C.}~\bibnamefont {Karrasch}},\
  and\ \bibinfo {author} {\bibfnamefont {J.~E.}\ \bibnamefont {Moore}},\
  }\bibfield  {title} {\bibinfo {title} {Bethe-boltzmann hydrodynamics and spin
  transport in the xxz chain},\ }\href
  {https://doi.org/10.1103/PhysRevB.97.045407} {\bibfield  {journal} {\bibinfo
  {journal} {Phys. Rev. B}\ }\textbf {\bibinfo {volume} {97}},\ \bibinfo
  {pages} {045407} (\bibinfo {year} {2018})}\BibitemShut {NoStop}%
\bibitem [{\citenamefont {Doyon}\ and\ \citenamefont
  {Spohn}(2017)}]{SciPostPhys.3.6.039}%
  \BibitemOpen
  \bibfield  {author} {\bibinfo {author} {\bibfnamefont {B.}~\bibnamefont
  {Doyon}}\ and\ \bibinfo {author} {\bibfnamefont {H.}~\bibnamefont {Spohn}},\
  }\bibfield  {title} {\bibinfo {title} {{Drude Weight for the Lieb-Liniger
  Bose Gas}},\ }\href {https://doi.org/10.21468/SciPostPhys.3.6.039} {\bibfield
   {journal} {\bibinfo  {journal} {SciPost Phys.}\ }\textbf {\bibinfo {volume}
  {3}},\ \bibinfo {pages} {039} (\bibinfo {year} {2017})}\BibitemShut {NoStop}%
\bibitem [{\citenamefont {Ilievski}\ and\ \citenamefont
  {De~Nardis}(2017{\natexlab{b}})}]{PhysRevB.96.081118}%
  \BibitemOpen
  \bibfield  {author} {\bibinfo {author} {\bibfnamefont {E.}~\bibnamefont
  {Ilievski}}\ and\ \bibinfo {author} {\bibfnamefont {J.}~\bibnamefont
  {De~Nardis}},\ }\bibfield  {title} {\bibinfo {title} {Ballistic transport in
  the one-dimensional hubbard model: The hydrodynamic approach},\ }\href
  {https://doi.org/10.1103/PhysRevB.96.081118} {\bibfield  {journal} {\bibinfo
  {journal} {Phys. Rev. B}\ }\textbf {\bibinfo {volume} {96}},\ \bibinfo
  {pages} {081118(R)} (\bibinfo {year} {2017}{\natexlab{b}})}\BibitemShut
  {NoStop}%
\bibitem [{\citenamefont {De~Nardis}\ \emph {et~al.}(2018)\citenamefont
  {De~Nardis}, \citenamefont {Bernard},\ and\ \citenamefont
  {Doyon}}]{PhysRevLett.121.160603}%
  \BibitemOpen
  \bibfield  {author} {\bibinfo {author} {\bibfnamefont {J.}~\bibnamefont
  {De~Nardis}}, \bibinfo {author} {\bibfnamefont {D.}~\bibnamefont {Bernard}},\
  and\ \bibinfo {author} {\bibfnamefont {B.}~\bibnamefont {Doyon}},\ }\bibfield
   {title} {\bibinfo {title} {Hydrodynamic diffusion in integrable systems},\
  }\href {https://doi.org/10.1103/PhysRevLett.121.160603} {\bibfield  {journal}
  {\bibinfo  {journal} {Phys. Rev. Lett.}\ }\textbf {\bibinfo {volume} {121}},\
  \bibinfo {pages} {160603} (\bibinfo {year} {2018})}\BibitemShut {NoStop}%
\bibitem [{\citenamefont {Gopalakrishnan}\ \emph {et~al.}(2018)\citenamefont
  {Gopalakrishnan}, \citenamefont {Huse}, \citenamefont {Khemani},\ and\
  \citenamefont {Vasseur}}]{PhysRevB.98.220303}%
  \BibitemOpen
  \bibfield  {author} {\bibinfo {author} {\bibfnamefont {S.}~\bibnamefont
  {Gopalakrishnan}}, \bibinfo {author} {\bibfnamefont {D.~A.}\ \bibnamefont
  {Huse}}, \bibinfo {author} {\bibfnamefont {V.}~\bibnamefont {Khemani}},\ and\
  \bibinfo {author} {\bibfnamefont {R.}~\bibnamefont {Vasseur}},\ }\bibfield
  {title} {\bibinfo {title} {Hydrodynamics of operator spreading and
  quasiparticle diffusion in interacting integrable systems},\ }\href
  {https://doi.org/10.1103/PhysRevB.98.220303} {\bibfield  {journal} {\bibinfo
  {journal} {Phys. Rev. B}\ }\textbf {\bibinfo {volume} {98}},\ \bibinfo
  {pages} {220303(R)} (\bibinfo {year} {2018})}\BibitemShut {NoStop}%
\bibitem [{\citenamefont {Nardis}\ \emph {et~al.}(2019)\citenamefont {Nardis},
  \citenamefont {Bernard},\ and\ \citenamefont
  {Doyon}}]{10.21468/SciPostPhys.6.4.049}%
  \BibitemOpen
  \bibfield  {author} {\bibinfo {author} {\bibfnamefont {J.~D.}\ \bibnamefont
  {Nardis}}, \bibinfo {author} {\bibfnamefont {D.}~\bibnamefont {Bernard}},\
  and\ \bibinfo {author} {\bibfnamefont {B.}~\bibnamefont {Doyon}},\ }\bibfield
   {title} {\bibinfo {title} {{Diffusion in generalized hydrodynamics and
  quasiparticle scattering}},\ }\href
  {https://doi.org/10.21468/SciPostPhys.6.4.049} {\bibfield  {journal}
  {\bibinfo  {journal} {SciPost Phys.}\ }\textbf {\bibinfo {volume} {6}},\
  \bibinfo {pages} {49} (\bibinfo {year} {2019})}\BibitemShut {NoStop}%
\bibitem [{\citenamefont {Gopalakrishnan}\ and\ \citenamefont
  {Vasseur}(2019)}]{PhysRevLett.122.127202}%
  \BibitemOpen
  \bibfield  {author} {\bibinfo {author} {\bibfnamefont {S.}~\bibnamefont
  {Gopalakrishnan}}\ and\ \bibinfo {author} {\bibfnamefont {R.}~\bibnamefont
  {Vasseur}},\ }\bibfield  {title} {\bibinfo {title} {Kinetic theory of spin
  diffusion and superdiffusion in $xxz$ spin chains},\ }\href
  {https://doi.org/10.1103/PhysRevLett.122.127202} {\bibfield  {journal}
  {\bibinfo  {journal} {Phys. Rev. Lett.}\ }\textbf {\bibinfo {volume} {122}},\
  \bibinfo {pages} {127202} (\bibinfo {year} {2019})}\BibitemShut {NoStop}%
\bibitem [{\citenamefont {Bouchoule}\ \emph {et~al.}(2020)\citenamefont
  {Bouchoule}, \citenamefont {Doyon},\ and\ \citenamefont
  {Dubail}}]{bouchoule2020effect}%
  \BibitemOpen
  \bibfield  {author} {\bibinfo {author} {\bibfnamefont {I.}~\bibnamefont
  {Bouchoule}}, \bibinfo {author} {\bibfnamefont {B.}~\bibnamefont {Doyon}},\
  and\ \bibinfo {author} {\bibfnamefont {J.}~\bibnamefont {Dubail}},\
  }\bibfield  {title} {\bibinfo {title} {{The effect of atom losses on the
  distribution of rapidities in the one-dimensional Bose gas}},\ }\href
  {https://doi.org/10.21468/SciPostPhys.9.4.044} {\bibfield  {journal}
  {\bibinfo  {journal} {SciPost Phys.}\ }\textbf {\bibinfo {volume} {9}},\
  \bibinfo {pages} {44} (\bibinfo {year} {2020})}\BibitemShut {NoStop}%
\bibitem [{\citenamefont {Bastianello}\ \emph
  {et~al.}(2020{\natexlab{a}})\citenamefont {Bastianello}, \citenamefont
  {De~Nardis},\ and\ \citenamefont {De~Luca}}]{bastianello2020generalised}%
  \BibitemOpen
  \bibfield  {author} {\bibinfo {author} {\bibfnamefont {A.}~\bibnamefont
  {Bastianello}}, \bibinfo {author} {\bibfnamefont {J.}~\bibnamefont
  {De~Nardis}},\ and\ \bibinfo {author} {\bibfnamefont {A.}~\bibnamefont
  {De~Luca}},\ }\bibfield  {title} {\bibinfo {title} {Generalised hydrodynamics
  with dephasing noise},\ }\href@noop {} {\bibfield  {journal} {\bibinfo
  {journal} {arXiv preprint arXiv:2003.01702}\ } (\bibinfo {year}
  {2020}{\natexlab{a}})}\BibitemShut {NoStop}%
\bibitem [{\citenamefont {Bastianello}\ \emph
  {et~al.}(2020{\natexlab{b}})\citenamefont {Bastianello}, \citenamefont
  {Luca}, \citenamefont {Doyon},\ and\ \citenamefont
  {Nardis}}]{bastianello2020thermalisation}%
  \BibitemOpen
  \bibfield  {author} {\bibinfo {author} {\bibfnamefont {A.}~\bibnamefont
  {Bastianello}}, \bibinfo {author} {\bibfnamefont {A.~D.}\ \bibnamefont
  {Luca}}, \bibinfo {author} {\bibfnamefont {B.}~\bibnamefont {Doyon}},\ and\
  \bibinfo {author} {\bibfnamefont {J.~D.}\ \bibnamefont {Nardis}},\
  }\href@noop {} {\bibinfo {title} {Thermalisation of a trapped one-dimensional
  bose gas via diffusion}} (\bibinfo {year} {2020}{\natexlab{b}}),\ \Eprint
  {https://arxiv.org/abs/2007.04861} {arXiv:2007.04861 [cond-mat.quant-gas]}
  \BibitemShut {NoStop}%
\bibitem [{\citenamefont {Mazets}\ \emph {et~al.}(2008)\citenamefont {Mazets},
  \citenamefont {Schumm},\ and\ \citenamefont {Schmiedmayer}}]{Mazets2008}%
  \BibitemOpen
  \bibfield  {author} {\bibinfo {author} {\bibfnamefont {I.~E.}\ \bibnamefont
  {Mazets}}, \bibinfo {author} {\bibfnamefont {T.}~\bibnamefont {Schumm}},\
  and\ \bibinfo {author} {\bibfnamefont {J.}~\bibnamefont {Schmiedmayer}},\
  }\bibfield  {title} {\bibinfo {title} {Breakdown of integrability in a
  quasi-1d ultracold bosonic gas},\ }\href
  {https://doi.org/10.1103/PhysRevLett.100.210403} {\bibfield  {journal}
  {\bibinfo  {journal} {Phys. Rev. Lett.}\ }\textbf {\bibinfo {volume} {100}},\
  \bibinfo {pages} {210403} (\bibinfo {year} {2008})}\BibitemShut {NoStop}%
\bibitem [{\citenamefont {Gerbier}\ and\ \citenamefont
  {Castin}(2010)}]{Gerbier2010}%
  \BibitemOpen
  \bibfield  {author} {\bibinfo {author} {\bibfnamefont {F.}~\bibnamefont
  {Gerbier}}\ and\ \bibinfo {author} {\bibfnamefont {Y.}~\bibnamefont
  {Castin}},\ }\bibfield  {title} {\bibinfo {title} {{Heating rates for an atom
  in a far-detuned optical lattice}},\ }\href
  {https://doi.org/10.1103/PhysRevA.82.013615} {\bibfield  {journal} {\bibinfo
  {journal} {Physical Review A}\ }\textbf {\bibinfo {volume} {82}},\ \bibinfo
  {pages} {013615} (\bibinfo {year} {2010})}\BibitemShut {NoStop}%
\bibitem [{\citenamefont {Pichler}\ \emph {et~al.}(2010)\citenamefont
  {Pichler}, \citenamefont {Daley},\ and\ \citenamefont
  {Zoller}}]{Pichler2010}%
  \BibitemOpen
  \bibfield  {author} {\bibinfo {author} {\bibfnamefont {H.}~\bibnamefont
  {Pichler}}, \bibinfo {author} {\bibfnamefont {A.~J.}\ \bibnamefont {Daley}},\
  and\ \bibinfo {author} {\bibfnamefont {P.}~\bibnamefont {Zoller}},\
  }\bibfield  {title} {\bibinfo {title} {{Nonequilibrium dynamics of bosonic
  atoms in optical lattices: Decoherence of many-body states due to spontaneous
  emission}},\ }\href {https://doi.org/10.1103/PhysRevA.82.063605} {\bibfield
  {journal} {\bibinfo  {journal} {Physical Review A}\ }\textbf {\bibinfo
  {volume} {82}},\ \bibinfo {pages} {063605} (\bibinfo {year}
  {2010})}\BibitemShut {NoStop}%
\bibitem [{\citenamefont {Mazets}(2011{\natexlab{b}})}]{Mazets2011}%
  \BibitemOpen
  \bibfield  {author} {\bibinfo {author} {\bibfnamefont {I.~E.}\ \bibnamefont
  {Mazets}},\ }\bibfield  {title} {\bibinfo {title} {{Dynamics and kinetics of
  quasiparticle decay in a nearly-one-dimensional degenerate Bose gas}},\
  }\href {https://doi.org/10.1103/PhysRevA.83.043625} {\bibfield  {journal}
  {\bibinfo  {journal} {Physical Review A}\ }\textbf {\bibinfo {volume} {83}},\
  \bibinfo {pages} {043625} (\bibinfo {year} {2011}{\natexlab{b}})}\BibitemShut
  {NoStop}%
\bibitem [{\citenamefont {Riou}\ \emph {et~al.}(2012)\citenamefont {Riou},
  \citenamefont {Reinhard}, \citenamefont {Zundel},\ and\ \citenamefont
  {Weiss}}]{Riou2012}%
  \BibitemOpen
  \bibfield  {author} {\bibinfo {author} {\bibfnamefont {J.-F.}\ \bibnamefont
  {Riou}}, \bibinfo {author} {\bibfnamefont {A.}~\bibnamefont {Reinhard}},
  \bibinfo {author} {\bibfnamefont {L.~A.}\ \bibnamefont {Zundel}},\ and\
  \bibinfo {author} {\bibfnamefont {D.~S.}\ \bibnamefont {Weiss}},\ }\bibfield
  {title} {\bibinfo {title} {{Spontaneous-emission-induced transition rates
  between atomic states in optical lattices}},\ }\href
  {https://doi.org/10.1103/PhysRevA.86.033412} {\bibfield  {journal} {\bibinfo
  {journal} {Physical Review A}\ }\textbf {\bibinfo {volume} {86}},\ \bibinfo
  {pages} {033412} (\bibinfo {year} {2012})}\BibitemShut {NoStop}%
\bibitem [{\citenamefont {Riou}\ \emph {et~al.}(2014)\citenamefont {Riou},
  \citenamefont {Zundel}, \citenamefont {Reinhard},\ and\ \citenamefont
  {Weiss}}]{Riou2014}%
  \BibitemOpen
  \bibfield  {author} {\bibinfo {author} {\bibfnamefont {J.-F.}\ \bibnamefont
  {Riou}}, \bibinfo {author} {\bibfnamefont {L.~A.}\ \bibnamefont {Zundel}},
  \bibinfo {author} {\bibfnamefont {A.}~\bibnamefont {Reinhard}},\ and\
  \bibinfo {author} {\bibfnamefont {D.~S.}\ \bibnamefont {Weiss}},\ }\bibfield
  {title} {\bibinfo {title} {Effect of optical-lattice heating on the momentum
  distribution of a one-dimensional bose gas},\ }\href
  {https://doi.org/10.1103/PhysRevA.90.033401} {\bibfield  {journal} {\bibinfo
  {journal} {Phys. Rev. A}\ }\textbf {\bibinfo {volume} {90}},\ \bibinfo
  {pages} {033401} (\bibinfo {year} {2014})}\BibitemShut {NoStop}%
\bibitem [{\citenamefont {Tang}\ \emph {et~al.}(2018)\citenamefont {Tang},
  \citenamefont {Kao}, \citenamefont {Li}, \citenamefont {Seo}, \citenamefont
  {Mallayya}, \citenamefont {Rigol}, \citenamefont {Gopalakrishnan},\ and\
  \citenamefont {Lev}}]{tang2018cradle}%
  \BibitemOpen
  \bibfield  {author} {\bibinfo {author} {\bibfnamefont {Y.}~\bibnamefont
  {Tang}}, \bibinfo {author} {\bibfnamefont {W.}~\bibnamefont {Kao}}, \bibinfo
  {author} {\bibfnamefont {K.-Y.}\ \bibnamefont {Li}}, \bibinfo {author}
  {\bibfnamefont {S.}~\bibnamefont {Seo}}, \bibinfo {author} {\bibfnamefont
  {K.}~\bibnamefont {Mallayya}}, \bibinfo {author} {\bibfnamefont
  {M.}~\bibnamefont {Rigol}}, \bibinfo {author} {\bibfnamefont
  {S.}~\bibnamefont {Gopalakrishnan}},\ and\ \bibinfo {author} {\bibfnamefont
  {B.~L.}\ \bibnamefont {Lev}},\ }\bibfield  {title} {\bibinfo {title}
  {Thermalization near integrability in a dipolar quantum newton's cradle},\
  }\href {https://doi.org/10.1103/PhysRevX.8.021030} {\bibfield  {journal}
  {\bibinfo  {journal} {Phys. Rev. X}\ }\textbf {\bibinfo {volume} {8}},\
  \bibinfo {pages} {021030} (\bibinfo {year} {2018})}\BibitemShut {NoStop}%
\bibitem [{\citenamefont {Zundel}\ \emph {et~al.}(2019)\citenamefont {Zundel},
  \citenamefont {Wilson}, \citenamefont {Malvania}, \citenamefont {Xia},
  \citenamefont {Riou},\ and\ \citenamefont {Weiss}}]{Zundel2019}%
  \BibitemOpen
  \bibfield  {author} {\bibinfo {author} {\bibfnamefont {L.~A.}\ \bibnamefont
  {Zundel}}, \bibinfo {author} {\bibfnamefont {J.~M.}\ \bibnamefont {Wilson}},
  \bibinfo {author} {\bibfnamefont {N.}~\bibnamefont {Malvania}}, \bibinfo
  {author} {\bibfnamefont {L.}~\bibnamefont {Xia}}, \bibinfo {author}
  {\bibfnamefont {J.-F.}\ \bibnamefont {Riou}},\ and\ \bibinfo {author}
  {\bibfnamefont {D.~S.}\ \bibnamefont {Weiss}},\ }\bibfield  {title} {\bibinfo
  {title} {Energy-dependent three-body loss in 1d bose gases},\ }\href
  {https://doi.org/10.1103/PhysRevLett.122.013402} {\bibfield  {journal}
  {\bibinfo  {journal} {Phys. Rev. Lett.}\ }\textbf {\bibinfo {volume} {122}},\
  \bibinfo {pages} {013402} (\bibinfo {year} {2019})}\BibitemShut {NoStop}%
\bibitem [{\citenamefont {Caux}\ \emph {et~al.}(2019)\citenamefont {Caux},
  \citenamefont {Doyon}, \citenamefont {Dubail}, \citenamefont {Konik},\ and\
  \citenamefont {Yoshimura}}]{caux2019cradle}%
  \BibitemOpen
  \bibfield  {author} {\bibinfo {author} {\bibfnamefont {J.-S.}\ \bibnamefont
  {Caux}}, \bibinfo {author} {\bibfnamefont {B.}~\bibnamefont {Doyon}},
  \bibinfo {author} {\bibfnamefont {J.}~\bibnamefont {Dubail}}, \bibinfo
  {author} {\bibfnamefont {R.}~\bibnamefont {Konik}},\ and\ \bibinfo {author}
  {\bibfnamefont {T.}~\bibnamefont {Yoshimura}},\ }\bibfield  {title} {\bibinfo
  {title} {{Hydrodynamics of the interacting Bose gas in the Quantum Newton
  Cradle setup}},\ }\href {https://doi.org/10.21468/SciPostPhys.6.6.070}
  {\bibfield  {journal} {\bibinfo  {journal} {SciPost Phys.}\ }\textbf
  {\bibinfo {volume} {6}},\ \bibinfo {pages} {70} (\bibinfo {year}
  {2019})}\BibitemShut {NoStop}%
\bibitem [{\citenamefont {Mallayya}\ \emph {et~al.}(2019)\citenamefont
  {Mallayya}, \citenamefont {Rigol},\ and\ \citenamefont
  {De~Roeck}}]{PhysRevX.9.021027}%
  \BibitemOpen
  \bibfield  {author} {\bibinfo {author} {\bibfnamefont {K.}~\bibnamefont
  {Mallayya}}, \bibinfo {author} {\bibfnamefont {M.}~\bibnamefont {Rigol}},\
  and\ \bibinfo {author} {\bibfnamefont {W.}~\bibnamefont {De~Roeck}},\
  }\bibfield  {title} {\bibinfo {title} {Prethermalization and thermalization
  in isolated quantum systems},\ }\href
  {https://doi.org/10.1103/PhysRevX.9.021027} {\bibfield  {journal} {\bibinfo
  {journal} {Phys. Rev. X}\ }\textbf {\bibinfo {volume} {9}},\ \bibinfo {pages}
  {021027} (\bibinfo {year} {2019})}\BibitemShut {NoStop}%
\bibitem [{\citenamefont {Friedman}\ \emph {et~al.}(2020)\citenamefont
  {Friedman}, \citenamefont {Gopalakrishnan},\ and\ \citenamefont
  {Vasseur}}]{PhysRevB.101.180302}%
  \BibitemOpen
  \bibfield  {author} {\bibinfo {author} {\bibfnamefont {A.~J.}\ \bibnamefont
  {Friedman}}, \bibinfo {author} {\bibfnamefont {S.}~\bibnamefont
  {Gopalakrishnan}},\ and\ \bibinfo {author} {\bibfnamefont {R.}~\bibnamefont
  {Vasseur}},\ }\bibfield  {title} {\bibinfo {title} {Diffusive hydrodynamics
  from integrability breaking},\ }\href
  {https://doi.org/10.1103/PhysRevB.101.180302} {\bibfield  {journal} {\bibinfo
   {journal} {Phys. Rev. B}\ }\textbf {\bibinfo {volume} {101}},\ \bibinfo
  {pages} {180302(R)} (\bibinfo {year} {2020})}\BibitemShut {NoStop}%
\bibitem [{\citenamefont {Durnin}\ \emph {et~al.}(2020)\citenamefont {Durnin},
  \citenamefont {Bhaseen},\ and\ \citenamefont {Doyon}}]{durnin2020non}%
  \BibitemOpen
  \bibfield  {author} {\bibinfo {author} {\bibfnamefont {J.}~\bibnamefont
  {Durnin}}, \bibinfo {author} {\bibfnamefont {M.}~\bibnamefont {Bhaseen}},\
  and\ \bibinfo {author} {\bibfnamefont {B.}~\bibnamefont {Doyon}},\ }\bibfield
   {title} {\bibinfo {title} {Non-equilibrium dynamics and weakly broken
  integrability},\ }\href@noop {} {\bibfield  {journal} {\bibinfo  {journal}
  {arXiv preprint arXiv:2004.11030}\ } (\bibinfo {year} {2020})}\BibitemShut
  {NoStop}%
\bibitem [{\citenamefont {Lopez-Piqueres}\ \emph {et~al.}(2020)\citenamefont
  {Lopez-Piqueres}, \citenamefont {Ware}, \citenamefont {Gopalakrishnan},\ and\
  \citenamefont {Vasseur}}]{lopez2020hydrodynamics}%
  \BibitemOpen
  \bibfield  {author} {\bibinfo {author} {\bibfnamefont {J.}~\bibnamefont
  {Lopez-Piqueres}}, \bibinfo {author} {\bibfnamefont {B.}~\bibnamefont
  {Ware}}, \bibinfo {author} {\bibfnamefont {S.}~\bibnamefont
  {Gopalakrishnan}},\ and\ \bibinfo {author} {\bibfnamefont {R.}~\bibnamefont
  {Vasseur}},\ }\bibfield  {title} {\bibinfo {title} {Hydrodynamics of
  non-integrable systems from relaxation-time approximation},\ }\href@noop {}
  {\bibfield  {journal} {\bibinfo  {journal} {arXiv preprint arXiv:2005.13546}\
  } (\bibinfo {year} {2020})}\BibitemShut {NoStop}%
\bibitem [{\citenamefont {Bland}\ \emph {et~al.}(2018)\citenamefont {Bland},
  \citenamefont {.Parker}, \citenamefont {Proukakis},\ and\ \citenamefont
  {Malomed}}]{Bland_2018}%
  \BibitemOpen
  \bibfield  {author} {\bibinfo {author} {\bibfnamefont {T.}~\bibnamefont
  {Bland}}, \bibinfo {author} {\bibfnamefont {N.~G.}\ \bibnamefont {.Parker}},
  \bibinfo {author} {\bibfnamefont {N.~P.}\ \bibnamefont {Proukakis}},\ and\
  \bibinfo {author} {\bibfnamefont {B.~A.}\ \bibnamefont {Malomed}},\
  }\bibfield  {title} {\bibinfo {title} {Probing quasi-integrability of the
  gross-pitaevskii equation in a harmonic-oscillator potential},\ }\href
  {https://doi.org/10.1088/1361-6455/aae0ba} {\bibfield  {journal} {\bibinfo
  {journal} {Journal of Physics B: Atomic, Molecular and Optical Physics}\
  }\textbf {\bibinfo {volume} {51}},\ \bibinfo {pages} {205303} (\bibinfo
  {year} {2018})}\BibitemShut {NoStop}%
\bibitem [{\citenamefont {Pitaevskii}\ and\ \citenamefont
  {Lifshitz}(2012)}]{pitaevskii2012physical}%
  \BibitemOpen
  \bibfield  {author} {\bibinfo {author} {\bibfnamefont {L.}~\bibnamefont
  {Pitaevskii}}\ and\ \bibinfo {author} {\bibfnamefont {E.}~\bibnamefont
  {Lifshitz}},\ }\href@noop {} {\emph {\bibinfo {title} {Physical Kinetics:
  Volume 10}}},\ Vol.~\bibinfo {volume} {10}\ (\bibinfo  {publisher}
  {Butterworth-Heinemann},\ \bibinfo {year} {2012})\BibitemShut {NoStop}%
\bibitem [{\citenamefont {Yang}(1967)}]{PhysRevLett.19.1312}%
  \BibitemOpen
  \bibfield  {author} {\bibinfo {author} {\bibfnamefont {C.~N.}\ \bibnamefont
  {Yang}},\ }\bibfield  {title} {\bibinfo {title} {Some exact results for the
  many-body problem in one dimension with repulsive delta-function
  interaction},\ }\href {https://doi.org/10.1103/PhysRevLett.19.1312}
  {\bibfield  {journal} {\bibinfo  {journal} {Phys. Rev. Lett.}\ }\textbf
  {\bibinfo {volume} {19}},\ \bibinfo {pages} {1312} (\bibinfo {year}
  {1967})}\BibitemShut {NoStop}%
\bibitem [{\citenamefont {Yang}\ and\ \citenamefont
  {Yang}(1969)}]{doi:10.1063/1.1664947}%
  \BibitemOpen
  \bibfield  {author} {\bibinfo {author} {\bibfnamefont {C.~N.}\ \bibnamefont
  {Yang}}\ and\ \bibinfo {author} {\bibfnamefont {C.~P.}\ \bibnamefont
  {Yang}},\ }\bibfield  {title} {\bibinfo {title} {Thermodynamics of a
  one‐dimensional system of bosons with repulsive delta‐function
  interaction},\ }\href {https://doi.org/10.1063/1.1664947} {\bibfield
  {journal} {\bibinfo  {journal} {Journal of Mathematical Physics}\ }\textbf
  {\bibinfo {volume} {10}},\ \bibinfo {pages} {1115} (\bibinfo {year}
  {1969})}\BibitemShut {NoStop}%
\bibitem [{\citenamefont {Lieb}\ and\ \citenamefont
  {Liniger}(1963)}]{lieb1963exact}%
  \BibitemOpen
  \bibfield  {author} {\bibinfo {author} {\bibfnamefont {E.~H.}\ \bibnamefont
  {Lieb}}\ and\ \bibinfo {author} {\bibfnamefont {W.}~\bibnamefont {Liniger}},\
  }\bibfield  {title} {\bibinfo {title} {Exact analysis of an interacting bose
  gas. i. the general solution and the ground state},\ }\href
  {https://doi.org/10.1103/PhysRev.130.1605} {\bibfield  {journal} {\bibinfo
  {journal} {Physical Review}\ }\textbf {\bibinfo {volume} {130}},\ \bibinfo
  {pages} {1605} (\bibinfo {year} {1963})}\BibitemShut {NoStop}%
\bibitem [{\citenamefont {Lieb}(2004)}]{lieb2004exact}%
  \BibitemOpen
  \bibfield  {author} {\bibinfo {author} {\bibfnamefont {E.~H.}\ \bibnamefont
  {Lieb}},\ }\bibfield  {title} {\bibinfo {title} {Exact analysis of an
  interacting bose gas. ii. the excitation spectrum},\ }in\ \href
  {https://doi.org/10.1007/978-3-662-06390-3_37} {\emph {\bibinfo {booktitle}
  {Condensed Matter Physics and Exactly Soluble Models}}}\ (\bibinfo
  {publisher} {Springer},\ \bibinfo {year} {2004})\ pp.\ \bibinfo {pages}
  {617--625}\BibitemShut {NoStop}%
\bibitem [{\citenamefont {Takahashi}(2005)}]{takahashi2005thermodynamics}%
  \BibitemOpen
  \bibfield  {author} {\bibinfo {author} {\bibfnamefont {M.}~\bibnamefont
  {Takahashi}},\ }\href@noop {} {\emph {\bibinfo {title} {Thermodynamics of
  one-dimensional solvable models}}}\ (\bibinfo  {publisher} {Cambridge
  University Press},\ \bibinfo {year} {2005})\BibitemShut {NoStop}%
\bibitem [{\citenamefont {Doyon}\ \emph {et~al.}(2018)\citenamefont {Doyon},
  \citenamefont {Yoshimura},\ and\ \citenamefont
  {Caux}}]{PhysRevLett.120.045301}%
  \BibitemOpen
  \bibfield  {author} {\bibinfo {author} {\bibfnamefont {B.}~\bibnamefont
  {Doyon}}, \bibinfo {author} {\bibfnamefont {T.}~\bibnamefont {Yoshimura}},\
  and\ \bibinfo {author} {\bibfnamefont {J.-S.}\ \bibnamefont {Caux}},\
  }\bibfield  {title} {\bibinfo {title} {Soliton gases and generalized
  hydrodynamics},\ }\href {https://doi.org/10.1103/PhysRevLett.120.045301}
  {\bibfield  {journal} {\bibinfo  {journal} {Phys. Rev. Lett.}\ }\textbf
  {\bibinfo {volume} {120}},\ \bibinfo {pages} {045301} (\bibinfo {year}
  {2018})}\BibitemShut {NoStop}%
\bibitem [{\citenamefont {Doyon}\ and\ \citenamefont
  {Yoshimura}(2017)}]{Doyon2017note}%
  \BibitemOpen
  \bibfield  {author} {\bibinfo {author} {\bibfnamefont {B.}~\bibnamefont
  {Doyon}}\ and\ \bibinfo {author} {\bibfnamefont {T.}~\bibnamefont
  {Yoshimura}},\ }\bibfield  {title} {\bibinfo {title} {{A note on generalized
  hydrodynamics: inhomogeneous fields and other concepts}},\ }\href
  {https://doi.org/10.21468/SciPostPhys.2.2.014} {\bibfield  {journal}
  {\bibinfo  {journal} {SciPost Phys.}\ }\textbf {\bibinfo {volume} {2}},\
  \bibinfo {pages} {014} (\bibinfo {year} {2017})}\BibitemShut {NoStop}%
\bibitem [{\citenamefont {Klauser}\ and\ \citenamefont
  {Caux}(2011)}]{klauser2011}%
  \BibitemOpen
  \bibfield  {author} {\bibinfo {author} {\bibfnamefont {A.}~\bibnamefont
  {Klauser}}\ and\ \bibinfo {author} {\bibfnamefont {J.-S.}\ \bibnamefont
  {Caux}},\ }\bibfield  {title} {\bibinfo {title} {Equilibrium thermodynamic
  properties of interacting two-component bosons in one dimension},\ }\href
  {https://doi.org/10.1103/PhysRevA.84.033604} {\bibfield  {journal} {\bibinfo
  {journal} {Phys. Rev. A}\ }\textbf {\bibinfo {volume} {84}},\ \bibinfo
  {pages} {033604} (\bibinfo {year} {2011})}\BibitemShut {NoStop}%
\bibitem [{\citenamefont {Sutherland}(1968)}]{Sutherland1968}%
  \BibitemOpen
  \bibfield  {author} {\bibinfo {author} {\bibfnamefont {B.}~\bibnamefont
  {Sutherland}},\ }\bibfield  {title} {\bibinfo {title} {Further results for
  the many-body problem in one dimension},\ }\href
  {https://doi.org/10.1103/PhysRevLett.20.98} {\bibfield  {journal} {\bibinfo
  {journal} {Phys. Rev. Lett.}\ }\textbf {\bibinfo {volume} {20}},\ \bibinfo
  {pages} {98} (\bibinfo {year} {1968})}\BibitemShut {NoStop}%
\bibitem [{Note1()}]{Note1}%
  \BibitemOpen
  \bibinfo {note} {See Supplemental Material for further details.}\BibitemShut
  {Stop}%
\bibitem [{Note2()}]{Note2}%
  \BibitemOpen
  \bibinfo {note} {Unlike for collisions, transitions caused by external
  heating do not abide to parity conservation. Thus, the heating will typically
  cause an atom to jump one transverse level~\cite {Li2018}. It is therefore
  reasonable to assume that the rate of atoms transferred via heating to the
  second excited state is proportional to the population of the first one,
  hence $\gamma _2 = \gamma _1 \nu _1$.}\BibitemShut {Stop}%
\bibitem [{Note3()}]{Note3}%
  \BibitemOpen
  \bibinfo {note} {See Supplemental Material for a detailed construction of the
  collision integral, which includes Refs.~\cite {Olshanii1998}}\BibitemShut
  {NoStop}%
\bibitem [{\citenamefont {Salasnich}\ \emph {et~al.}(2002)\citenamefont
  {Salasnich}, \citenamefont {Parola},\ and\ \citenamefont
  {Reatto}}]{PhysRevA.65.043614}%
  \BibitemOpen
  \bibfield  {author} {\bibinfo {author} {\bibfnamefont {L.}~\bibnamefont
  {Salasnich}}, \bibinfo {author} {\bibfnamefont {A.}~\bibnamefont {Parola}},\
  and\ \bibinfo {author} {\bibfnamefont {L.}~\bibnamefont {Reatto}},\
  }\bibfield  {title} {\bibinfo {title} {Effective wave equations for the
  dynamics of cigar-shaped and disk-shaped bose condensates},\ }\href
  {https://doi.org/10.1103/PhysRevA.65.043614} {\bibfield  {journal} {\bibinfo
  {journal} {Phys. Rev. A}\ }\textbf {\bibinfo {volume} {65}},\ \bibinfo
  {pages} {043614} (\bibinfo {year} {2002})}\BibitemShut {NoStop}%
\bibitem [{\citenamefont {Mateo}\ and\ \citenamefont
  {Delgado}(2008)}]{PhysRevA.77.013617}%
  \BibitemOpen
  \bibfield  {author} {\bibinfo {author} {\bibfnamefont {A.~M.}\ \bibnamefont
  {Mateo}}\ and\ \bibinfo {author} {\bibfnamefont {V.}~\bibnamefont
  {Delgado}},\ }\bibfield  {title} {\bibinfo {title} {Effective mean-field
  equations for cigar-shaped and disk-shaped bose-einstein condensates},\
  }\href {https://doi.org/10.1103/PhysRevA.77.013617} {\bibfield  {journal}
  {\bibinfo  {journal} {Phys. Rev. A}\ }\textbf {\bibinfo {volume} {77}},\
  \bibinfo {pages} {013617} (\bibinfo {year} {2008})}\BibitemShut {NoStop}%
\bibitem [{\citenamefont {Adhikari}\ and\ \citenamefont
  {Malomed}(2009)}]{adhikari2009gap}%
  \BibitemOpen
  \bibfield  {author} {\bibinfo {author} {\bibfnamefont {S.~K.}\ \bibnamefont
  {Adhikari}}\ and\ \bibinfo {author} {\bibfnamefont {B.~A.}\ \bibnamefont
  {Malomed}},\ }\bibfield  {title} {\bibinfo {title} {Gap solitons in a model
  of a superfluid fermion gas in optical lattices},\ }\href
  {https://doi.org/10.1016/j.physd.2008.07.025} {\bibfield  {journal} {\bibinfo
   {journal} {Physica D: Nonlinear Phenomena}\ }\textbf {\bibinfo {volume}
  {238}},\ \bibinfo {pages} {1402} (\bibinfo {year} {2009})}\BibitemShut
  {NoStop}%
\bibitem [{\citenamefont {Mazets}\ and\ \citenamefont
  {Schmiedmayer}(2010)}]{Mazets_2010}%
  \BibitemOpen
  \bibfield  {author} {\bibinfo {author} {\bibfnamefont {I.~E.}\ \bibnamefont
  {Mazets}}\ and\ \bibinfo {author} {\bibfnamefont {J.}~\bibnamefont
  {Schmiedmayer}},\ }\bibfield  {title} {\bibinfo {title} {Thermalization in a
  quasi-one-dimensional ultracold bosonic gas},\ }\href
  {https://doi.org/10.1088/1367-2630/12/5/055023} {\bibfield  {journal}
  {\bibinfo  {journal} {New Journal of Physics}\ }\textbf {\bibinfo {volume}
  {12}},\ \bibinfo {pages} {055023} (\bibinfo {year} {2010})}\BibitemShut
  {NoStop}%
\bibitem [{\citenamefont {van~den Berg}\ \emph {et~al.}(2016)\citenamefont
  {van~den Berg}, \citenamefont {Wouters}, \citenamefont {Eli\"ens},
  \citenamefont {De~Nardis}, \citenamefont {Konik},\ and\ \citenamefont
  {Caux}}]{berg2016separation}%
  \BibitemOpen
  \bibfield  {author} {\bibinfo {author} {\bibfnamefont {R.}~\bibnamefont
  {van~den Berg}}, \bibinfo {author} {\bibfnamefont {B.}~\bibnamefont
  {Wouters}}, \bibinfo {author} {\bibfnamefont {S.}~\bibnamefont {Eli\"ens}},
  \bibinfo {author} {\bibfnamefont {J.}~\bibnamefont {De~Nardis}}, \bibinfo
  {author} {\bibfnamefont {R.~M.}\ \bibnamefont {Konik}},\ and\ \bibinfo
  {author} {\bibfnamefont {J.-S.}\ \bibnamefont {Caux}},\ }\bibfield  {title}
  {\bibinfo {title} {Separation of time scales in a quantum newton's cradle},\
  }\href {https://doi.org/10.1103/PhysRevLett.116.225302} {\bibfield  {journal}
  {\bibinfo  {journal} {Phys. Rev. Lett.}\ }\textbf {\bibinfo {volume} {116}},\
  \bibinfo {pages} {225302} (\bibinfo {year} {2016})}\BibitemShut {NoStop}%
\bibitem [{Note4()}]{Note4}%
  \BibitemOpen
  \bibinfo {note} {See Supplemental Material for a detailed description of the
  measurement of each parameter, which includes Refs.~\cite {Cazalilla2004,
  PhysRevLett.87.050404, PhysRevLett.91.010405, PhysRevA.67.051602,
  PhysRevA.83.031604, PhysRevLett.121.220402}}\BibitemShut {NoStop}%
\bibitem [{\citenamefont {M{\o}ller}\ and\ \citenamefont
  {Schmiedmayer}(2020)}]{iFluid}%
  \BibitemOpen
  \bibfield  {author} {\bibinfo {author} {\bibfnamefont {F.~S.}\ \bibnamefont
  {M{\o}ller}}\ and\ \bibinfo {author} {\bibfnamefont {J.}~\bibnamefont
  {Schmiedmayer}},\ }\bibfield  {title} {\bibinfo {title} {{Introducing iFluid:
  a numerical framework for solving hydrodynamical equations in integrable
  models}},\ }\href {https://doi.org/10.21468/SciPostPhys.8.3.041} {\bibfield
  {journal} {\bibinfo  {journal} {SciPost Phys.}\ }\textbf {\bibinfo {volume}
  {8}},\ \bibinfo {pages} {41} (\bibinfo {year} {2020})}\BibitemShut {NoStop}%
\bibitem [{Note5()}]{Note5}%
  \BibitemOpen
  \bibinfo {note} {A similar phenomenon can also be seen in Ref. \cite
  {caux2019cradle} at long times, albeit not as clearly.}\BibitemShut {Stop}%
\bibitem [{Note6()}]{Note6}%
  \BibitemOpen
  \bibinfo {note} {Averaging the profiles over one periods also reduces the
  difference between the RDFs and MDFs.}\BibitemShut {Stop}%
\bibitem [{\citenamefont {Wilson}\ \emph {et~al.}(2020)\citenamefont {Wilson},
  \citenamefont {Malvania}, \citenamefont {Le}, \citenamefont {Zhang},
  \citenamefont {Rigol},\ and\ \citenamefont {Weiss}}]{Wilson2020}%
  \BibitemOpen
  \bibfield  {author} {\bibinfo {author} {\bibfnamefont {J.~M.}\ \bibnamefont
  {Wilson}}, \bibinfo {author} {\bibfnamefont {N.}~\bibnamefont {Malvania}},
  \bibinfo {author} {\bibfnamefont {Y.}~\bibnamefont {Le}}, \bibinfo {author}
  {\bibfnamefont {Y.}~\bibnamefont {Zhang}}, \bibinfo {author} {\bibfnamefont
  {M.}~\bibnamefont {Rigol}},\ and\ \bibinfo {author} {\bibfnamefont {D.~S.}\
  \bibnamefont {Weiss}},\ }\bibfield  {title} {\bibinfo {title} {{Observation
  of dynamical fermionization}},\ }\href
  {https://doi.org/10.1126/science.aaz0242} {\bibfield  {journal} {\bibinfo
  {journal} {Science}\ }\textbf {\bibinfo {volume} {367}},\ \bibinfo {pages}
  {1461} (\bibinfo {year} {2020})}\BibitemShut {NoStop}%
\bibitem [{\citenamefont {Giamarchi}\ and\ \citenamefont
  {Press}(2004)}]{giamarchi2004quantum}%
  \BibitemOpen
  \bibfield  {author} {\bibinfo {author} {\bibfnamefont {T.}~\bibnamefont
  {Giamarchi}}\ and\ \bibinfo {author} {\bibfnamefont {O.~U.}\ \bibnamefont
  {Press}},\ }\href {https://books.google.at/books?id=1MwTDAAAQBAJ} {\emph
  {\bibinfo {title} {Quantum Physics in One Dimension}}},\ International Series
  of Monogr\ (\bibinfo  {publisher} {Clarendon Press},\ \bibinfo {year}
  {2004})\BibitemShut {NoStop}%
\bibitem [{Note7()}]{Note7}%
  \BibitemOpen
  \bibinfo {note} {Note, the initial increase of $\protect \mathcal {T}(t)$
  observed in figure \ref {fig:thermalization_a} for the GHD simulations stems
  from the initial depletion of quasiparticles at low rapidity (which can also
  be seen in figure \ref {fig:theta}). This is caused by the interactions in
  standard GHD and is not a product of our extended model.}\BibitemShut {Stop}%
\bibitem [{Note8()}]{Note8}%
  \BibitemOpen
  \bibinfo {note} {Note, the experimental band-mapping technique used to
  extract $\nu _1$ and $\nu _2$ requires a sufficient number of atoms in the
  transverse excited states in order to overcome the measurement noise. Since
  the second transverse state is only sparsely populated, the resulting
  errorbars on $\nu _2$ are quite large. See Ref. \cite {Li2018} for more
  details.}\BibitemShut {Stop}%
\bibitem [{Note9()}]{Note9}%
  \BibitemOpen
  \bibinfo {note} {See Supplemental Material for a detailed discussion of the
  various methods applicable in different regimes of the Lieb-Liniger phase
  diagram, which includes Refs.~\cite
  {PhysRevA.67.053615,Bland_2018,Tonks1936,Girardeau1960,Collura2014,Krauth2006,PhysRevLett.120.045301,Li2018,schemmer2019generalized,pitaevskii2012physical}}\BibitemShut
  {NoStop}%
\bibitem [{\citenamefont {van Amerongen}\ \emph {et~al.}(2008)\citenamefont
  {van Amerongen}, \citenamefont {van Es}, \citenamefont {Wicke}, \citenamefont
  {Kheruntsyan},\ and\ \citenamefont {van Druten}}]{Amerongen2008yang}%
  \BibitemOpen
  \bibfield  {author} {\bibinfo {author} {\bibfnamefont {A.~H.}\ \bibnamefont
  {van Amerongen}}, \bibinfo {author} {\bibfnamefont {J.~J.~P.}\ \bibnamefont
  {van Es}}, \bibinfo {author} {\bibfnamefont {P.}~\bibnamefont {Wicke}},
  \bibinfo {author} {\bibfnamefont {K.~V.}\ \bibnamefont {Kheruntsyan}},\ and\
  \bibinfo {author} {\bibfnamefont {N.~J.}\ \bibnamefont {van Druten}},\
  }\bibfield  {title} {\bibinfo {title} {Yang-yang thermodynamics on an atom
  chip},\ }\href {https://doi.org/10.1103/PhysRevLett.100.090402} {\bibfield
  {journal} {\bibinfo  {journal} {Phys. Rev. Lett.}\ }\textbf {\bibinfo
  {volume} {100}},\ \bibinfo {pages} {090402} (\bibinfo {year}
  {2008})}\BibitemShut {NoStop}%
\bibitem [{\citenamefont {Davis}\ \emph {et~al.}(2012)\citenamefont {Davis},
  \citenamefont {Blakie}, \citenamefont {van Amerongen}, \citenamefont {van
  Druten},\ and\ \citenamefont {Kheruntsyan}}]{davis2012thermometry}%
  \BibitemOpen
  \bibfield  {author} {\bibinfo {author} {\bibfnamefont {M.~J.}\ \bibnamefont
  {Davis}}, \bibinfo {author} {\bibfnamefont {P.~B.}\ \bibnamefont {Blakie}},
  \bibinfo {author} {\bibfnamefont {A.~H.}\ \bibnamefont {van Amerongen}},
  \bibinfo {author} {\bibfnamefont {N.~J.}\ \bibnamefont {van Druten}},\ and\
  \bibinfo {author} {\bibfnamefont {K.~V.}\ \bibnamefont {Kheruntsyan}},\
  }\bibfield  {title} {\bibinfo {title} {Yang-yang thermometry and momentum
  distribution of a trapped one-dimensional bose gas},\ }\href
  {https://doi.org/10.1103/PhysRevA.85.031604} {\bibfield  {journal} {\bibinfo
  {journal} {Phys. Rev. A}\ }\textbf {\bibinfo {volume} {85}},\ \bibinfo
  {pages} {031604(R)} (\bibinfo {year} {2012})}\BibitemShut {NoStop}%
\bibitem [{\citenamefont {Jacqmin}\ \emph {et~al.}(2011)\citenamefont
  {Jacqmin}, \citenamefont {Armijo}, \citenamefont {Berrada}, \citenamefont
  {Kheruntsyan},\ and\ \citenamefont {Bouchoule}}]{jacqmin2011fluctuations}%
  \BibitemOpen
  \bibfield  {author} {\bibinfo {author} {\bibfnamefont {T.}~\bibnamefont
  {Jacqmin}}, \bibinfo {author} {\bibfnamefont {J.}~\bibnamefont {Armijo}},
  \bibinfo {author} {\bibfnamefont {T.}~\bibnamefont {Berrada}}, \bibinfo
  {author} {\bibfnamefont {K.~V.}\ \bibnamefont {Kheruntsyan}},\ and\ \bibinfo
  {author} {\bibfnamefont {I.}~\bibnamefont {Bouchoule}},\ }\bibfield  {title}
  {\bibinfo {title} {Sub-poissonian fluctuations in a 1d bose gas: From the
  quantum quasicondensate to the strongly interacting regime},\ }\href
  {https://doi.org/10.1103/PhysRevLett.106.230405} {\bibfield  {journal}
  {\bibinfo  {journal} {Phys. Rev. Lett.}\ }\textbf {\bibinfo {volume} {106}},\
  \bibinfo {pages} {230405} (\bibinfo {year} {2011})}\BibitemShut {NoStop}%
\bibitem [{\citenamefont {Jacqmin}\ \emph {et~al.}(2012)\citenamefont
  {Jacqmin}, \citenamefont {Fang}, \citenamefont {Berrada}, \citenamefont
  {Roscilde},\ and\ \citenamefont {Bouchoule}}]{jacqmin2012momentum}%
  \BibitemOpen
  \bibfield  {author} {\bibinfo {author} {\bibfnamefont {T.}~\bibnamefont
  {Jacqmin}}, \bibinfo {author} {\bibfnamefont {B.}~\bibnamefont {Fang}},
  \bibinfo {author} {\bibfnamefont {T.}~\bibnamefont {Berrada}}, \bibinfo
  {author} {\bibfnamefont {T.}~\bibnamefont {Roscilde}},\ and\ \bibinfo
  {author} {\bibfnamefont {I.}~\bibnamefont {Bouchoule}},\ }\bibfield  {title}
  {\bibinfo {title} {Momentum distribution of one-dimensional bose gases at the
  quasicondensation crossover: Theoretical and experimental investigation},\
  }\href {https://doi.org/10.1103/PhysRevA.86.043626} {\bibfield  {journal}
  {\bibinfo  {journal} {Phys. Rev. A}\ }\textbf {\bibinfo {volume} {86}},\
  \bibinfo {pages} {043626} (\bibinfo {year} {2012})}\BibitemShut {NoStop}%
\bibitem [{\citenamefont {Armijo}\ \emph {et~al.}(2011)\citenamefont {Armijo},
  \citenamefont {Jacqmin}, \citenamefont {Kheruntsyan},\ and\ \citenamefont
  {Bouchoule}}]{PhysRevA.83.021605}%
  \BibitemOpen
  \bibfield  {author} {\bibinfo {author} {\bibfnamefont {J.}~\bibnamefont
  {Armijo}}, \bibinfo {author} {\bibfnamefont {T.}~\bibnamefont {Jacqmin}},
  \bibinfo {author} {\bibfnamefont {K.}~\bibnamefont {Kheruntsyan}},\ and\
  \bibinfo {author} {\bibfnamefont {I.}~\bibnamefont {Bouchoule}},\ }\bibfield
  {title} {\bibinfo {title} {Mapping out the quasicondensate transition through
  the dimensional crossover from one to three dimensions},\ }\href
  {https://doi.org/10.1103/PhysRevA.83.021605} {\bibfield  {journal} {\bibinfo
  {journal} {Phys. Rev. A}\ }\textbf {\bibinfo {volume} {83}},\ \bibinfo
  {pages} {021605(R)} (\bibinfo {year} {2011})}\BibitemShut {NoStop}%
\bibitem [{\citenamefont {Tonks}(1936)}]{Tonks1936}%
  \BibitemOpen
  \bibfield  {author} {\bibinfo {author} {\bibfnamefont {L.}~\bibnamefont
  {Tonks}},\ }\bibfield  {title} {\bibinfo {title} {The complete equation of
  state of one, two and three-dimensional gases of hard elastic spheres},\
  }\href {https://doi.org/10.1103/PhysRev.50.955} {\bibfield  {journal}
  {\bibinfo  {journal} {Phys. Rev.}\ }\textbf {\bibinfo {volume} {50}},\
  \bibinfo {pages} {955} (\bibinfo {year} {1936})}\BibitemShut {NoStop}%
\bibitem [{\citenamefont {Girardeau}(1960)}]{Girardeau1960}%
  \BibitemOpen
  \bibfield  {author} {\bibinfo {author} {\bibfnamefont {M.}~\bibnamefont
  {Girardeau}},\ }\bibfield  {title} {\bibinfo {title} {Relationship between
  systems of impenetrable bosons and fermions in one dimension},\ }\href
  {https://doi.org/10.1063/1.1703687} {\bibfield  {journal} {\bibinfo
  {journal} {Journal of Mathematical Physics}\ }\textbf {\bibinfo {volume}
  {1}},\ \bibinfo {pages} {516} (\bibinfo {year} {1960})},\ \Eprint
  {https://arxiv.org/abs/https://doi.org/10.1063/1.1703687}
  {https://doi.org/10.1063/1.1703687} \BibitemShut {NoStop}%
\bibitem [{\citenamefont {Collura}\ and\ \citenamefont
  {Karevski}(2014)}]{Collura2014}%
  \BibitemOpen
  \bibfield  {author} {\bibinfo {author} {\bibfnamefont {M.}~\bibnamefont
  {Collura}}\ and\ \bibinfo {author} {\bibfnamefont {D.}~\bibnamefont
  {Karevski}},\ }\bibfield  {title} {\bibinfo {title} {Quantum quench from a
  thermal tensor state: Boundary effects and generalized gibbs ensemble},\
  }\href {https://doi.org/10.1103/PhysRevB.89.214308} {\bibfield  {journal}
  {\bibinfo  {journal} {Phys. Rev. B}\ }\textbf {\bibinfo {volume} {89}},\
  \bibinfo {pages} {214308} (\bibinfo {year} {2014})}\BibitemShut {NoStop}%
\bibitem [{\citenamefont {Krauth}(2006)}]{Krauth2006}%
  \BibitemOpen
  \bibfield  {author} {\bibinfo {author} {\bibfnamefont {W.}~\bibnamefont
  {Krauth}},\ }\href@noop {} {\emph {\bibinfo {title} {Statistical mechanics:
  {A}lgorithms and computation}}}\ (\bibinfo  {publisher} {Oxford University
  Press},\ \bibinfo {address} {Oxford},\ \bibinfo {year} {2006})\BibitemShut
  {NoStop}%
\bibitem [{\citenamefont {Cazalilla}(2004)}]{Cazalilla2004}%
  \BibitemOpen
  \bibfield  {author} {\bibinfo {author} {\bibfnamefont {M.~A.}\ \bibnamefont
  {Cazalilla}},\ }\bibfield  {title} {\bibinfo {title} {Bosonizing
  one-dimensional cold atomic gases},\ }\href
  {https://doi.org/10.1088/0953-4075/37/7/051} {\bibfield  {journal} {\bibinfo
  {journal} {Journal of Physics B: Atomic, Molecular and Optical Physics}\
  }\textbf {\bibinfo {volume} {37}},\ \bibinfo {pages} {S1} (\bibinfo {year}
  {2004})}\BibitemShut {NoStop}%
\bibitem [{\citenamefont {Petrov}\ \emph {et~al.}(2001)\citenamefont {Petrov},
  \citenamefont {Shlyapnikov},\ and\ \citenamefont
  {Walraven}}]{PhysRevLett.87.050404}%
  \BibitemOpen
  \bibfield  {author} {\bibinfo {author} {\bibfnamefont {D.~S.}\ \bibnamefont
  {Petrov}}, \bibinfo {author} {\bibfnamefont {G.~V.}\ \bibnamefont
  {Shlyapnikov}},\ and\ \bibinfo {author} {\bibfnamefont {J.~T.~M.}\
  \bibnamefont {Walraven}},\ }\bibfield  {title} {\bibinfo {title}
  {Phase-fluctuating 3d bose-einstein condensates in elongated traps},\ }\href
  {https://doi.org/10.1103/PhysRevLett.87.050404} {\bibfield  {journal}
  {\bibinfo  {journal} {Phys. Rev. Lett.}\ }\textbf {\bibinfo {volume} {87}},\
  \bibinfo {pages} {050404} (\bibinfo {year} {2001})}\BibitemShut {NoStop}%
\bibitem [{\citenamefont {Richard}\ \emph {et~al.}(2003)\citenamefont
  {Richard}, \citenamefont {Gerbier}, \citenamefont {Thywissen}, \citenamefont
  {Hugbart}, \citenamefont {Bouyer},\ and\ \citenamefont
  {Aspect}}]{PhysRevLett.91.010405}%
  \BibitemOpen
  \bibfield  {author} {\bibinfo {author} {\bibfnamefont {S.}~\bibnamefont
  {Richard}}, \bibinfo {author} {\bibfnamefont {F.}~\bibnamefont {Gerbier}},
  \bibinfo {author} {\bibfnamefont {J.~H.}\ \bibnamefont {Thywissen}}, \bibinfo
  {author} {\bibfnamefont {M.}~\bibnamefont {Hugbart}}, \bibinfo {author}
  {\bibfnamefont {P.}~\bibnamefont {Bouyer}},\ and\ \bibinfo {author}
  {\bibfnamefont {A.}~\bibnamefont {Aspect}},\ }\bibfield  {title} {\bibinfo
  {title} {Momentum spectroscopy of 1d phase fluctuations in bose-einstein
  condensates},\ }\href {https://doi.org/10.1103/PhysRevLett.91.010405}
  {\bibfield  {journal} {\bibinfo  {journal} {Phys. Rev. Lett.}\ }\textbf
  {\bibinfo {volume} {91}},\ \bibinfo {pages} {010405} (\bibinfo {year}
  {2003})}\BibitemShut {NoStop}%
\bibitem [{\citenamefont {Gerbier}\ \emph {et~al.}(2003)\citenamefont
  {Gerbier}, \citenamefont {Thywissen}, \citenamefont {Richard}, \citenamefont
  {Hugbart}, \citenamefont {Bouyer},\ and\ \citenamefont
  {Aspect}}]{PhysRevA.67.051602}%
  \BibitemOpen
  \bibfield  {author} {\bibinfo {author} {\bibfnamefont {F.}~\bibnamefont
  {Gerbier}}, \bibinfo {author} {\bibfnamefont {J.~H.}\ \bibnamefont
  {Thywissen}}, \bibinfo {author} {\bibfnamefont {S.}~\bibnamefont {Richard}},
  \bibinfo {author} {\bibfnamefont {M.}~\bibnamefont {Hugbart}}, \bibinfo
  {author} {\bibfnamefont {P.}~\bibnamefont {Bouyer}},\ and\ \bibinfo {author}
  {\bibfnamefont {A.}~\bibnamefont {Aspect}},\ }\bibfield  {title} {\bibinfo
  {title} {Momentum distribution and correlation function of quasicondensates
  in elongated traps},\ }\href {https://doi.org/10.1103/PhysRevA.67.051602}
  {\bibfield  {journal} {\bibinfo  {journal} {Phys. Rev. A}\ }\textbf {\bibinfo
  {volume} {67}},\ \bibinfo {pages} {051602} (\bibinfo {year}
  {2003})}\BibitemShut {NoStop}%
\bibitem [{\citenamefont {Fabbri}\ \emph {et~al.}(2011)\citenamefont {Fabbri},
  \citenamefont {Cl\'ement}, \citenamefont {Fallani}, \citenamefont {Fort},\
  and\ \citenamefont {Inguscio}}]{PhysRevA.83.031604}%
  \BibitemOpen
  \bibfield  {author} {\bibinfo {author} {\bibfnamefont {N.}~\bibnamefont
  {Fabbri}}, \bibinfo {author} {\bibfnamefont {D.}~\bibnamefont {Cl\'ement}},
  \bibinfo {author} {\bibfnamefont {L.}~\bibnamefont {Fallani}}, \bibinfo
  {author} {\bibfnamefont {C.}~\bibnamefont {Fort}},\ and\ \bibinfo {author}
  {\bibfnamefont {M.}~\bibnamefont {Inguscio}},\ }\bibfield  {title} {\bibinfo
  {title} {Momentum-resolved study of an array of one-dimensional strongly
  phase-fluctuating bose gases},\ }\href
  {https://doi.org/10.1103/PhysRevA.83.031604} {\bibfield  {journal} {\bibinfo
  {journal} {Phys. Rev. A}\ }\textbf {\bibinfo {volume} {83}},\ \bibinfo
  {pages} {031604} (\bibinfo {year} {2011})}\BibitemShut {NoStop}%
\bibitem [{\citenamefont {Yao}\ \emph {et~al.}(2018)\citenamefont {Yao},
  \citenamefont {Cl\'ement}, \citenamefont {Minguzzi}, \citenamefont
  {Vignolo},\ and\ \citenamefont {Sanchez-Palencia}}]{PhysRevLett.121.220402}%
  \BibitemOpen
  \bibfield  {author} {\bibinfo {author} {\bibfnamefont {H.}~\bibnamefont
  {Yao}}, \bibinfo {author} {\bibfnamefont {D.}~\bibnamefont {Cl\'ement}},
  \bibinfo {author} {\bibfnamefont {A.}~\bibnamefont {Minguzzi}}, \bibinfo
  {author} {\bibfnamefont {P.}~\bibnamefont {Vignolo}},\ and\ \bibinfo {author}
  {\bibfnamefont {L.}~\bibnamefont {Sanchez-Palencia}},\ }\bibfield  {title}
  {\bibinfo {title} {Tan's contact for trapped lieb-liniger bosons at finite
  temperature},\ }\href {https://doi.org/10.1103/PhysRevLett.121.220402}
  {\bibfield  {journal} {\bibinfo  {journal} {Phys. Rev. Lett.}\ }\textbf
  {\bibinfo {volume} {121}},\ \bibinfo {pages} {220402} (\bibinfo {year}
  {2018})}\BibitemShut {NoStop}%
\bibitem [{Note10()}]{Note10}%
  \BibitemOpen
  \bibinfo {note} {Note, that the cited paper deals mainly with 1D Fermi
  systems; the result for bosons is only briefly presented}\BibitemShut
  {NoStop}%
\bibitem [{\citenamefont {Olshanii}(1998)}]{Olshanii1998}%
  \BibitemOpen
  \bibfield  {author} {\bibinfo {author} {\bibfnamefont {M.}~\bibnamefont
  {Olshanii}},\ }\bibfield  {title} {\bibinfo {title} {Atomic scattering in the
  presence of an external confinement and a gas of impenetrable bosons},\
  }\href {https://doi.org/10.1103/PhysRevLett.81.938} {\bibfield  {journal}
  {\bibinfo  {journal} {Phys. Rev. Lett.}\ }\textbf {\bibinfo {volume} {81}},\
  \bibinfo {pages} {938} (\bibinfo {year} {1998})}\BibitemShut {NoStop}%
\end{thebibliography}%

\end{document}